\UseRawInputEncoding      
\RequirePackage{fix-cm}
\documentclass[natbib,smallcondensed]{svjour3}    
\smartqed  
\usepackage[graphicx]{realboxes}   	
\usepackage{slashbox} 							
\usepackage{color}           
\usepackage[colorlinks=true,citecolor=blue,urlcolor=blue,linkcolor=blue]{hyperref}
\usepackage{amstext}
\usepackage{amssymb}
\usepackage{enumerate}
\usepackage{upgreek}
\usepackage{rotating}        
\usepackage{lscape}
\usepackage{longtable}       
\usepackage{multirow}

\usepackage{url}             
\usepackage{breakurl}        

\graphicspath{{figs/}}	

\journalname{my journal}

\newcommand{\adsurl}[1]{\href{http://adsabs.harvard.edu/abs/#1}{ADS}}
\newcommand{\doiurl}[1]{\href{http://dx.doi.org/#1}{DOI}}

\newcommand{\eg}{e.g.,{\ }}
\newcommand{\ie}{i.e.,{\ }}
\newcommand{\SC}{spacecraft}
\newcommand{\los}{line-of-sight{\ }}
\newcommand{\losnosp}{line-of-sight}
\newcommand{\loss}{lines-of-sight{\ }}
\newcommand{\fov}{field of view{\ }}
\newcommand{\fovnosp}{field of view}
\newcommand{\fovs}{fields of view}

\newcommand{\degr}{{^\circ}}
\renewcommand{\deg}{$^\circ$}
\newcommand{\Rsun}{\,R$_\odot$}  									
\newcommand{\Bsun}{\,$\overline{{\rm B}_\odot}${\ }} 
\newcommand{\Bsunnosp}{\,$\overline{{\rm B}_\odot}$} 
\newcommand{\SZL}{S10$_\odot${\ }}
\newcommand{\SZLnosp}{S10$_\odot$}

\begin{document}

\title{Observations of the Solar F-corona from Space}


\author{P. L. Lamy  \and
        H. Gilardy \and
				A. Llebaria
}


\institute{P.L. Lamy \at 
				Laboratoire Atmosph\`eres, Milieux et Observations Spatiales, CNRS \& 
				Universit\'e de Versailles Saint-Quentin-en-Yvelines, 
				11 Bd d'Alembert, 78280 Guyancourt, France
				\email{philippe.lamy@latmos.ipsl.fr}
				\and
				H. Gilardy \at
				Laboratoire Atmosph\`eres, Milieux et Observations Spatiales, CNRS \& 
				Universit\'e de Versailles Saint-Quentin-en-Yvelines, 
				11 Bd d'Alembert, 78280 Guyancourt, France	
				\email{hugo.gilardy@latmos.ipsl.fr}
				\and
				A. Llebaria \at
				Laboratoire Atmosph\`eres, Milieux et Observations Spatiales, CNRS \& 
				Universit\'e de Versailles Saint-Quentin-en-Yvelines, 
				11 Bd d'Alembert, 78280 Guyancourt, France					
        \email{antoine.llebaria@latmos.ipsl.fr} 
}

\date{Received: 2022-03-23 / Accepted: }

\maketitle

\noindent {\bf Corresponding author: P.L. Lamy (philippe.lamy@latmos.ipsl.fr)}

\bigskip

\begin{abstract}

We present a review of the observations of the solar F-corona from space with a special emphasis of the 25 years of continuous monitoring achieved by the LASCO-C2 and C3 coronagraphs. 
Our work includes images obtained by the navigation cameras of the \textit{Clementine} spacecraft, the SECCHI/HI-1A heliospheric imager onboard STEREO-A, and the Wide Field Imager for Solar Probe onboard the \textit{Parker Solar Probe}.
The connection to the zodiacal light is considered based on ground- and space-based observations, prominently from the past \textit{Helios}, IRAS, COBE, and IRAKI missions.
The characteristic radiance profiles along the two symmetry axis of the ``elliptically'' shaped F-corona (aka equatorial and polar directions) follow power laws in the 5\deg\,--\,50\deg\ range of elongation, with constant power exponents of -2.33 and -2.55.
Both profiles connect extremely well to the corresponding standard profiles of the zodiacal light.
The LASCO equatorial profile exhibits a shoulder implying a $\approx$17\,\% decrease of the radiance within  $\approx$\,10\Rsun\ that may be explained by the disappearance of organic materials within 0.3 AU. 
LASCO detected for the first time a secular variation of the F-corona, an increase at a rate of 0.46\,\% per year of the integrated radiance in the LASCO-C3 \fovnosp. This is likely the first observational evidence of the role of collisions in the inner zodiacal cloud.
The temporal evolution of the integrated radiance in the LASCO-C2 \fov is more complex suggesting possible additional processes.
Whereas it is well established that the F-corona is slightly redder than the Sun, the spectral variation of its color index is not yet well established.
A composite of C2 and C3 images produced the LASCO reference map of the radiance of the F-corona from 2 to 30\Rsun\ and, by combining with ground-based measurements, the LASCO extended map from 1 to 6\Rsun.
An upper limit of 0.03\,R$_\odot$ is obtained for the offset between the center of the Sun and that of the F-corona with a most likely value of zero.
The flattening index of the F-corona starts from zero at an elongation of 0.5\deg\ $\pm$ 0.01\deg\ (1.9\Rsun) and increases linearly with the logarithm of the elongation to connect to that of the zodiacal light with however a small hump related to the shoulder in the equatorial profile.
The shape of the isophotes is best described by super-ellipses with an exponent linked to the flattening index.
An ellipsoid model of the spatial density of interplanetary dust is solely capable of reproducing this shape, thus rejecting other classical models such as fan, and cosine.
The plane of symmetry of the inner zodiacal cloud is strongly warped, its inclination increasing towards the planes of the inner planets and ultimately the solar equator. 
In contrast, its longitude of ascending node is found to be constant and equal to 87.6\deg.
LASCO did not detect any small scale structures such as putative rings occasionally reported during solar eclipses.
The outer border of the depletion zone where interplanetary dust particles start to be affected by sublimation appears well constrained at $\approx$\,19\Rsun. 
This zone extends down to $\approx$\,5\Rsun, thus defining the boundary of the dust-free zone where the most refractory materials -- likely moderately absorbing silicates -- disappear.


\end{abstract}

\section{Introduction}
\label{sec:intro}
The optical manifestation of the interplanetary dust cloud has been traditionally split into two regions according to their angular extension from the Sun, F-corona for the inner region and the zodiacal light for the outer region, essentially for practical observational reasons since the former has been traditionally observed during solar eclipses and the latter at night. 
In one of his pioneering work, \cite{VdH1947} realized that the F-corona ``is just the extension of the zodiacal light'' as illustrated by his Figure~2, and he demonstrated the key role of diffraction in the light scattering process responsible for the F-corona. 
In addition to the dominating role of large dust particles (sizes larger than visible wavelengths), several of his premonitory insights are worth recalling: i) removing particles within 0.1~AU from the Sun (\ie introducing a dust-free zone using present terminology) changes the brightness of the F-corona only slightly, ii) its color should be slightly redder than the Sun, and iii) the albedo of the particles is about 1\,\% or their space density increases a little toward the Sun.
However, from an observational point of view, the dichotomy between F-corona and zodiacal light resulted in a gap in photometric measurements between them that has persisted for a long time (\eg Figure~2 of \cite{VdH1947}) although it was reduced by \cite{Blackwell1955} see his Figure~10.
It is quite revealing that the latest revision of the famous Allen's Astrophysical Quantities by \cite{Cox2000} presents tabulated values of the brightness of the zodiacal light as a function of helio-ecliptic longitude and latitude (\ie a coarse map at solar elongations $\epsilon$ $\geq$ 15\deg) expressed in units of $S_{10}(V)$ in the section ``Solar System Small Bodies'' and only two photometric profiles, equatorial and polar, of the F-corona expressed in units of mean solar brightness $\overline{B_\odot}$ in the section ``Sun''.
Indeed, although eclipse observations have produced many images of the solar corona, the superposition of the K and F components requires a rigorous separation and we are not aware of any photometrically calibrated images of the F-corona in the literature except those recently published by us \citep{Llebaria2021}. 

The latest thorough review of the properties of the F-corona remains that of \cite{KoutchmyLamy1985} and their synthetic equatorial and polar profiles, hereafter referred to as the Koutchmy--Lamy or K--L model, connect reasonably well to those of the zodiacal light \citep{Lamy1986}. 
\cite{Kimura1998} discussed the brightness of the F-corona, but they mostly concentrated on the near-infrared peaks occasionally observed at solar eclipses and their possible connection to a dust ring. 
The review of dust near the Sun by \cite{Mann2004} was prominently concerned by the physical and dynamical processes at work in the inner solar system and the sources and sinks of circum-solar dust. 
Among their conclusions, they stated ``that under present conditions no prominent dust ring exists near the Sun'' which is particularly relevant to the present study.

Although not strictly connected to the F-corona, several space observations are worth considering, notably in the perspective of unifying F-corona and zodiacal light and bridging the gap mentioned above.
The two \textit{Helios} space probes continuously scanned the heliosphere with their photometers pointing at three constant ecliptic latitudes of $\pm$16\deg, $\pm31$\deg, and $\pm90$\deg while traveling from 1 to 0.3\,AU  \citep{Porsche1981}.
Although \textit{Helios~1} launched in December 1974 was active during seven years and \textit{Helios~2} launched in January 1976 during four years, only the first two years of each \SC\ operation with good data coverage and nearly constant spin axis orientation were analyzed in detail \citep{Leinert1981}.
Among the most important results, the radiance (improperly called ``intensity'' by these authors) was found to increase towards the Sun as a function of the heliocentric distance $d$ of the observer following the law  $d^{-2.3 \pm 0.05}$, the upper and lower limits being recorded at small and large elongations from the Sun, respectively.

\begin{figure*}[!htpb]
\centering
\includegraphics[width=0.465\textwidth]{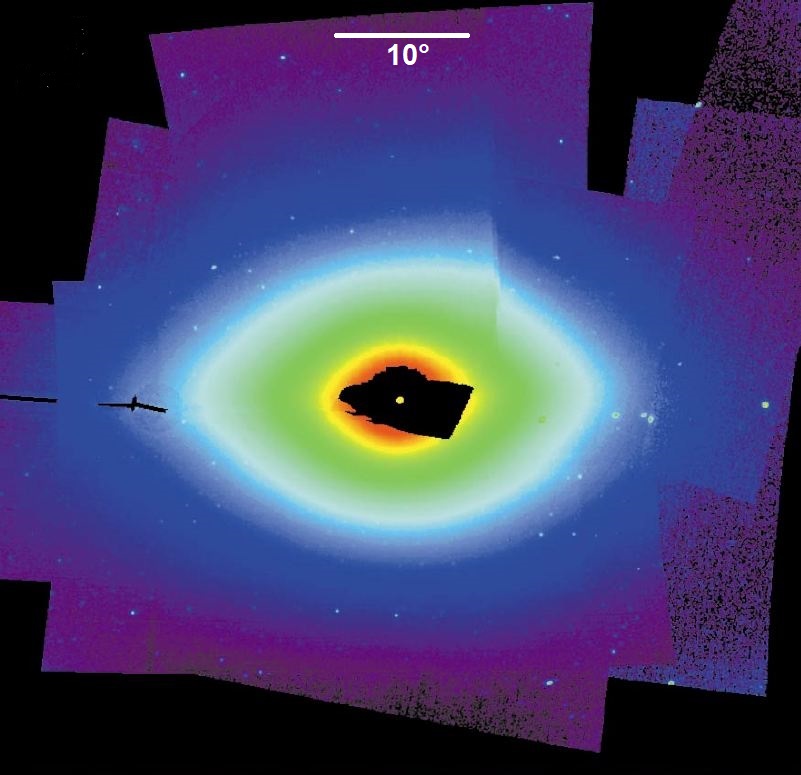}
\includegraphics[width=0.45\textwidth]{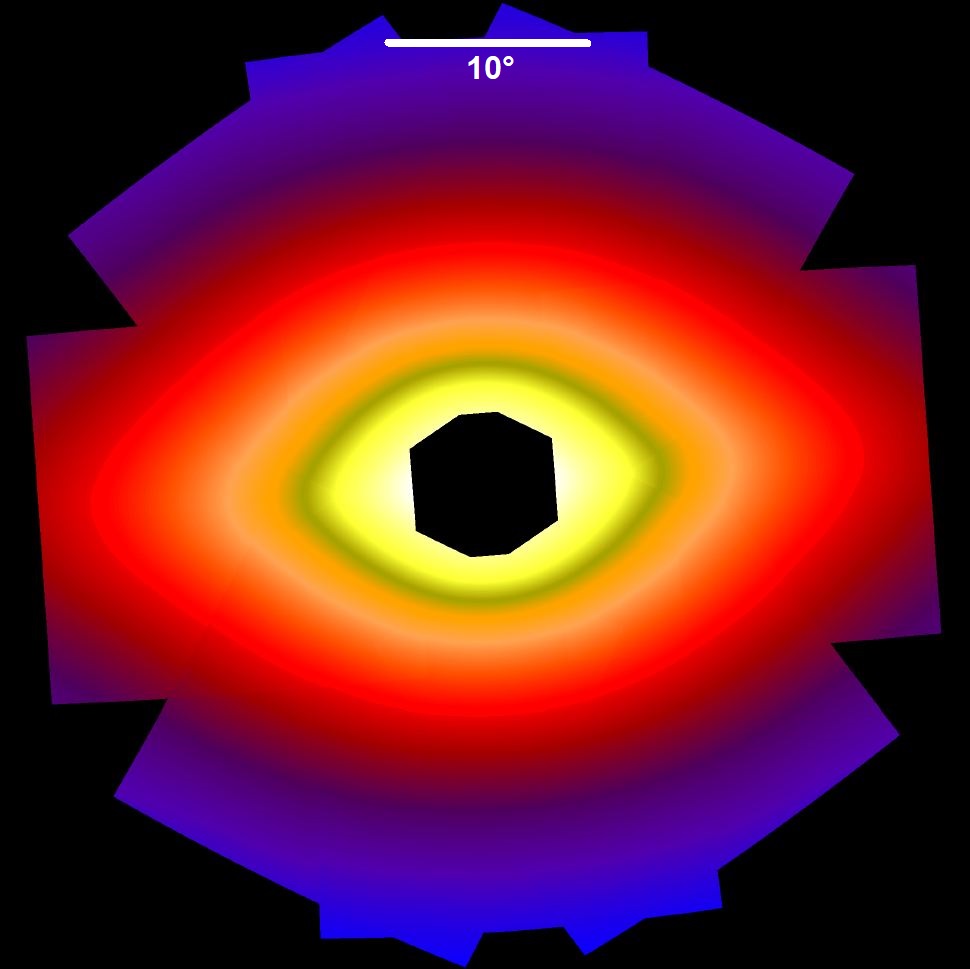}
\includegraphics[width=0.465\textwidth]{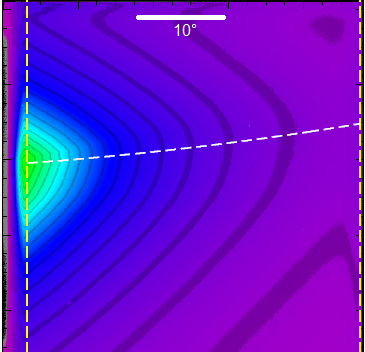}
\includegraphics[width=0.45\textwidth]{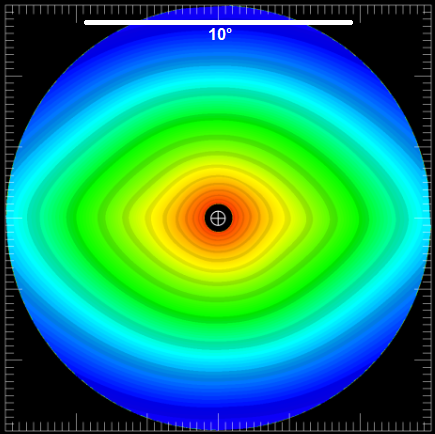}
\caption{Upper left panel: mosaic of seven fields of the inner zodiacal light observed by the \textit{Clementine} star tracker camera in March-April 1994.
The field of view is 60\deg $\times$ 60\deg and the Sun is drawn to scale at the center of the mosaic.
This is a reproduction of Figure~5 of \cite{Hahn2002} where further detail may be found.
Upper right panel: mosaic of eight fields of view of the inner zodiacal light observed by the SECCHI/HI-1 heliospheric imager on 10 March 2009.
The field of view is 48\deg $\times$ 48\deg.
This is a reproduction of Figure~1 of \cite{Stenborg2018} where further detail may be found.
Lower left panel: WISPR image of the inner zodiacal light taken on 5 November 2018 in orbit 1 of the \textit{Parker Solar Probe}.
The two dashed yellow lines delimit a band extending from 16\deg\ to 53\deg\ from the Sun and the dashed white line shows the location of the symmetry axis.
This is a reproduction of Figure~1 of \cite{Stenborg2021} where further detail may be found.
Lower right panel: composite of LASCO C2+C3 images of the F-corona on 21 December 1997. 
The field of view is 16\deg $\times$ 16\deg and the Sun is drawn to scale as a white circle.}
\label{fig:F_examples}
\end{figure*}

The next generation space missions secured images of the F-corona/zodiacal light thanks to the implementation of bi-dimensional CCD detectors, but from a heliocentric distance of or close to 1\,AU, and from or within a few degrees of the ecliptic plane.
Four examples are presented in Figure~\ref{fig:F_examples} coming from the \textit{Clementine}, SoHO, STEREO, and \textit{Parker Solar Probe} missions and they illustrate the continuity of the two phenomena so that the historical distinction between F-corona and zodiacal light is less and less justified.

The navigation cameras of the \textit{Clementine} spacecraft allowed building a mosaic of seven fields of the inner zodiacal light obtained in March-April 1994 that covers a \fov ranging from 5\deg to 30\deg \citep{Hahn2002}. 
Although affected by stray light from Venus and the Moon (used as external occulter) in a couple of angular sectors, this image clearly reveals the geometry of the inner interplanetary dust cloud which, to first order, appears globally axi-symmetric.
However slight asymmetries of $\approx$\,10\,\% were detected in the east--west and north--south directions.

The SECCHI/HI-1A heliospheric imager \citep{Howard2008} onboard the \textit{Solar Terrestrial Relationships Observatory} Ahead (STEREO-A) spacecraft observed the eastern side of the Sun between 5\deg and 24\deg elongations from December 2007 to March 2017.
Several roll maneuvers allowed reconstructing the full 360\deg view of the inner zodiacal light sixteen times and the case of 10 March 2009 is displayed in Figure~1 of \cite{Stenborg2018} reproduced here in Figure~\ref{fig:F_examples}.
It can be directly compared with the \textit{Clementine} image as they cover roughly the same \fovnosp\ and the geometry of the isophotes are in overall agreement.
\cite{Stenborg2018} found that the radial profiles of the radiance follow power laws of the solar elongation with power exponents ranging from $-2.31$ to $-2.35$ along the east--west direction and from $-2.45$ to $-2.53$ along the north--south direction.
In addition, several interesting features emerged: i) an east--west asymmetry suggesting that the projected center of the zodiacal cloud is offset from the Sun’s center by $\approx$\,0.4--0.5\Rsun, and ii) a subtle secular variation which appears to be driven by the combined gravitational forces exerted by the major planets on the cloud.

The LASCO-C2 and C3 coronagraphs \citep{Brueckner1995} of the \textit{Solar and Heliospheric Observatory} (SoHO) mission launched in September 1995, have been (and are still) continuously imaging the corona over the past 25 years [1996\,--\,2020] over a useful circular \fov extending from 0.6\deg (2.2\Rsun) to 8\deg (30\Rsun). 
This unprecedented coverage of the F-corona has so far received limited attention, a probable explanation lying in the difficulty of correctly separating the K and F components recently resolved by the thorough analysis performed by our team (\cite{Lamy2020}; \cite{Lamy2021}; \cite{Llebaria2021}).
However, a preliminary investigation of the long-term evolution of F-corona suggested that it remained stable until $\approx$\,2003, but that beyond, both its general radiance and its geometry (ellipticity) appeared to have progressively changed \citep{Gardes2013}.
\cite{Morgan2007} considered the question of the effects of coronal mass ejections on the circum-solar dust cloud following a theoretical investigation by \cite{Ragot2003}.
They searched for a variation in the F-corona by comparing LASCO-C2 observations taken at the minimum and maximum of Solar Cycle 23, but found none and therefore concluded that its radiance, at heights of 3 to 6\Rsun\ in the visible, remains very stable. 
It is one of the purpose of the present article to extend this kind of investigation to 25 years of LASCO-C2 and C3 observations.

The Wide Field Imager for Solar Probe (WISPR) onboard the \textit{Parker Solar Probe} (PSP) provides visible images of the west side of the Sun between 13.5\deg and 108\deg elongation thanks to its two telescopes with overlapping \fovs\ at approximately 50\deg \citep{Vourlidas2016}.
PSP has already completed ten perihelion passes and the WISPR images obtained during the orbit inbound to the first perihelion and during the first five solar encounters (excluding encounter 3) have been analyzed by \cite{Howard2019} and \cite{Stenborg2021}, respectively.
From measurements along the axis of symmetry of the zodiacal light, both works concluded on a power exponent of $-2.3$ for the variation of the radiance with elongation, but as PSP approaches a perihelion distance of 28\Rsun, this variation becomes less steep starting at about 19\Rsun.
This was interpreted as the signature of the existence of a depletion in circum-solar dust density within that heliocentric distance.
 
The \textit{Solar Orbiter Mission} (SolO) \cite{Muller2020}) has already reached its first perihelion at 0.51\,AU in June 2020, but the images of the METIS coronagraph \citep{Antonucci2020} have not yet been processed to the point of restoring the F-corona.

An in-depth review of these past results complemented by a detailed presentation of the LASCO observation benefiting from its remarkable photometric stability and unprecedented time coverage (25 years) is timely to offer an up-to-date characterization of the F-corona.
It will further provide the framework for the analysis and interpretation of the new views from different vantage points in both heliocentric distance and inclination offered by PSP and SoLO and by other forthcoming space missions as well.

This review is structured according to the different properties of the F-corona that can be deduced from its observation.
Its organization follows that implemented in a similar review of coronal mass ejections \citep{Lamy2019}: past results are first critically assessed to provide the background for the new results from the LASCO observations.
The preamble is composed of two sections, a general introduction (Section~\ref{sec:intro}) and an overview of the global properties of the F-corona (Section~\ref{sec:overview}).
In Section~\ref{sec:SOHOLASCO}, we briefly summarize the operations of SoHO and LASCO relevant to the observations of the F-corona, as well as the LASCO images and their analysis.
The photometric properties of the F-corona are the subject of Section~\ref{sec:photometry} from which we construct a standard model in Section~\ref{sec:model}.
The next three sections pursue the characterization of the F-corona: geometric properties (Section~\ref{sec:geometry}), plane of symmetry (Section~\ref{sec:plane}), and stability (Section~\ref{sec:stability}).
The discussion and interpretation of the results are carried out in Section~\ref{sec:discussion} and we summarize our results in Section~\ref{sec:summary}.
Appendix A presents an updated version of the volume scattering function of \cite{Lamy1986} in both tabular and graphical forms.
Appendices B and C are devoted to technical aspects concerning the correction of the C3 images.
Appendix D presents a tabular form of the LASCO reference model of the F-corona.

It is worth clarifying the units to be used in this article as they have been traditionally different for the corona and the zodiacal light.
Solar elongations will be preferably expressed in units of degree since, unlike the angular radius of the Sun, this makes elongations independent of the heliocentric distance.
The units of \Rsun, when used as an angular unit as already employed above, strictly corresponds to the angular extent of the solar radius at Earth mean distance (1\,AU), \ie 1\Rsun = 959.63 arcsec = 0.2666\deg. 
This corresponds to a physical dimension of the solar radius of 695,990 km. 
We are aware that in 2016, the International Astronomical Union adopted a new definition of the solar radius based on seismic data determined by the turnaround point for intermediate and high-I oscillations. 
This turnaround point is located below the photosphere, depending on the specific I value, and consequently the newly adopted radius of 695,700 km is substantially smaller than the classical photospheric value of 695,990 km. 
It is understood that a correction must be applied to reconcile the seismic and photospheric values of the solar radius \citep{Haberreiter2008}.
The radiance, commonly called ``brightness'' or incorrectly ``intensity'', will be expressed in units of mean solar brightness $\overline{B_\odot}$, or in practice, in units of $10^{-10}$\Bsun as more appropriate to coronal values.

\section{Overview of the Global Properties of the F-corona}
\label{sec:overview}
Ground- and space-based images of the F-corona (\eg Figure~\ref{fig:F_examples}) and of the zodiacal light exhibit an approximately elliptical shape that led to the description of the zodiacal cloud of interplanetary dust particles (IDPs) as a spheroid centered at the Sun, symmetric about the plane of maximum density and axially symmetric about an axis perpendicular to this plane, at least to first approximation.
Consequently, the three-dimensional density distribution of IDPs is generally modeled by functions of heliocentric distance and either latitude or height with respect to this plane (\eg \cite{Giese1986}; \cite{Giese1989}).
This plane of symmetry of the zodiacal cloud, hereafter abbreviated as ``PSZC'', is characterized by its inclination $i$ with respect to the ecliptic plane, typically a few degrees, and the longitude of its ascending node $\Omega_A$. 

The F-corona/zodiacal light as seen at visible wavelengths results from a double integration of the photospheric light scattered by individual IDP, one along \loss defined by an observer and the viewing directions, and the other over their size distribution.
This has several immediate consequences. 
First, the geometry and the radiance perceived by an observer very much depend upon its location within the cloud. 
In particular, the F-corona/zodiacal light appears symmetric with respect to the PSZC only when observed from this plane.
For an Earth-based observer or in orbit around it, this occurs twice per year when the PSZC intersects the ecliptic plane, \ie at the so-called nodes of late June and December. 
Second and whereas the size distribution of IDPs ranges from sub-micron to centimeter, the size range that really contributes to the visible radiance is much narrower. 
In-situ measurements of their flux with impact detectors once converted to distribution of cross sectional area and plotted on a linear scale as a function of the logarithm of the radius of dust particles (Figure~5.19 of \cite{Leinert1990}) exhibits a bell shape peaking at 30\,$\mu$m.  
Its full width at half maximum extends from radii of $\approx$\,10 to $\approx$\,100\,$\mu$m, thus delimiting the range of IDP size that contributes to the bulk of the F-corona/zodiacal light at visible wavelengths. 
This in turn has far reaching implications.

The light scattering problem may then be simply treated by the combination of independent processes, forward diffraction, Fresnel reflection, nonpolarized reflection and transmission as proposed by \cite{Giese1977}, and not necessarily with the Mie formalism with its severe restrictions (\eg spherical shape). 
Even the classical approximation of the superposition of a diffraction peak and isotropic scattering \citep{VdH1947} is applicable (\eg \cite{Leinert1975}; \cite{Mann1992}).
More elaborate and robust solutions are available that handle large (\ie radius $>>$ wavelength) rough particles such as the high-energy approximation (\cite{Chiappetta1980}; \cite{Perrin1983}) or the eikonal model \citep{Perrin1986}.

However, the preferred approach nowadays makes use of the volume scattering function (VSF) introduced by \cite{Dumont1973} and which characterizes the scattering phase function of a unit volume of interplanetary dust (\eg \cite{Leinert1975}; \cite{Mann1992}; \cite{Hahn2002}). 
Directly determined from the observations themselves by inversion with a limited set of assumptions, it has the advantage of by-passing the integral over the dust size distribution, thus avoiding the difficulties of having to specify the physical properties of the dust (\eg size, composition, roughness, and optical properties). 
Different determinations of the VSF have been published, but the most elaborate one has been obtained by \cite{Lamy1986} since they made use of ground as well as space-based observations (\textit{Helios, Pioneer}). 
However, their nominal VSF $\Psi_0$ at the reference heliocentric distance of 1\,AU was not specified below 5\deg\ and given only in graphical form.
We presently remedy these shortcomings by presenting in Appendix A its detailed extension below 5\deg\ in both graphical and tabulated forms.

\begin{figure}[htpb!]
\begin{center}
\includegraphics[width=\textwidth]{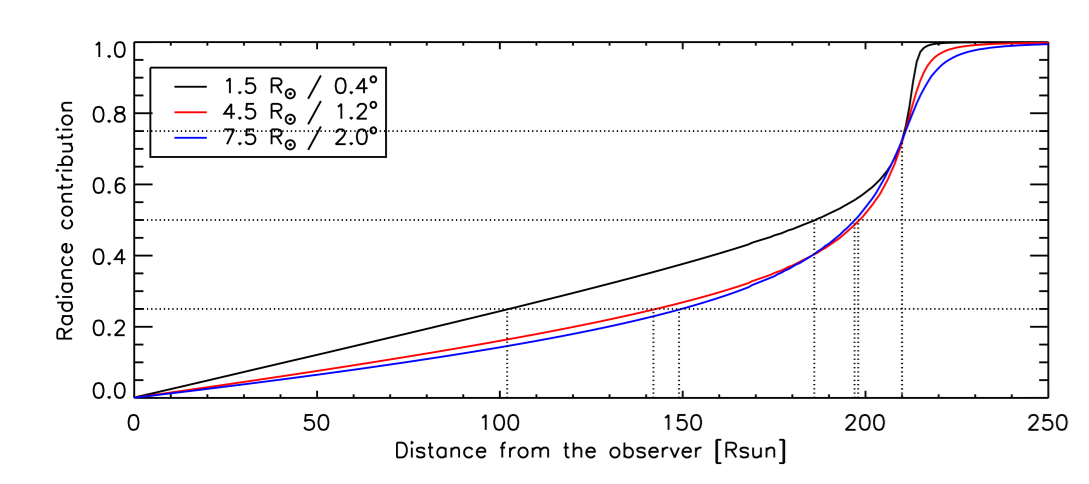}
\caption{Cumulative distributions of the radiance of the F-corona along three \loss defined by their elongation for an observer located at 1\,AU (215\Rsun).}
\label{fig:los}
\end{center} 
\end{figure}

Our immediate application of the VSF aims at completing this overview by investigating the cumulative distribution with heliocentric distance of the radiance along different \loss in order to foster our understanding and interpretation of the observations of the F-corona.
The contributions result from the interplay of the spatial density of the interplanetary dust and of the scattering function.
\cite{Mann1992} has already shown that for the F-corona observed at 1\,AU, up to $\approx$\,50\,\% of the total radiance comes from particles within 0.1\,AU around the Sun (her Table~2).
Figure~\ref{fig:los} offers a more detailed view in the form of cumulative distributions of the radiance along three lines-of-sight.
The first two at elongations of 1.2\deg\ and 2\deg\ are very close and indicate that 50\,\% of the radiance comes from particles within 0.07\,AU around the Sun, 75\,\% of the radiance comes from within $\approx$\,0.3\,AU, the remaining 25\,\% coming from the region from the observer to $\approx$\,0.7\,AU.
The third curve at an elongation of 0.4\deg\ is intended to highlight the increasing role of diffraction with decreasing elongation as the contribution of the region from the observer to $\approx$\,0.7\,AU raises to $\approx$\,40\,\%.
Altogether, these curves confirm the importance of the contribution of the innermost circum-solar zodiacal cloud to the radiance of the F-corona.
 
A final implication of the relatively large sizes of IDPs responsible for the F-corona/zodiacal light concerns their orbital evolution which is mainly controlled by gravity, radiation pressure, and the Poynting-Robertson effect.
Other forces such as electromagnetic are just too small to significantly alter the dynamics of these IDPs (\eg \cite{Leinert1990}).

\section{SoHO and LASCO Operations, LASCO Images, and their Analysis}
\label{sec:SOHOLASCO}
Detailed presentations of the SoHO and LASCO operations may be found in our past articles, notably \cite{Lamy2019} for a description of the operations, \cite{Lamy2020} and \cite{Llebaria2021}for the processing of the C2 images, and \cite{Lamy2021} for that of the C3 images. 
Let us briefly summarize those aspects which are important for the observations of the F-corona.
\begin{itemize}
	\item SoHO was launched on 2 December 1995, intermittent observations took place during the cruise phase to the L1 Lagrangian point, and the regular synoptic program started in early May 1996.
	\item The orbit of SoHO around the L1 point is slightly elliptical and lies in the ecliptic plane.
	\item SoHO and LASCO have been in quasi continuous operation except for two major interruptions which altogether resulted in a data gap from 25 June 1998 to 6 February 1999 for LASCO.
	\item Starting in June 2003, SoHO is periodically (every three months) rolled by 180$\degr$ to maximize telemetry transmission to Earth with its blocked antenna.
	\item Until 29 October 2010, the reference axis of SoHO was aligned along the sky-projected direction of the solar rotational axis resulting in solar north being up or down (in case of rolled images) on the LASCO images. 
Thereafter, the reference orientation was fixed to the perpendicular to the ecliptic plane causing the projected direction of the solar rotational axis to oscillate around the vertical direction on the LASCO images.
\end{itemize}

The present work makes use of daily images of the F-corona obtained until the end of 2020, that is an overall quasi continuous time coverage of almost 25 years, orders of magnitude longer than the aggregated eclipse time over the last century.
These C2 and C3 images taken through the same broadband ``orange'' filter having a nearly rectangular bandpass extending from $540$ to $640$ nm (central wavelength of $585$ nm) are described in the next sections.

\subsection{LASCO-C2 Images of the F-corona}
\label{sec:C2}
\begin{figure}[htpb!]
\begin{center}
\includegraphics[width=0.8\textwidth]{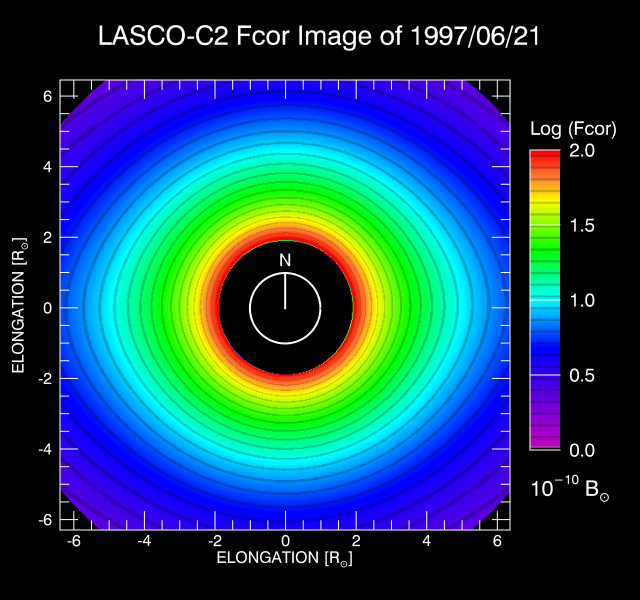}
\includegraphics[width=0.8\textwidth]{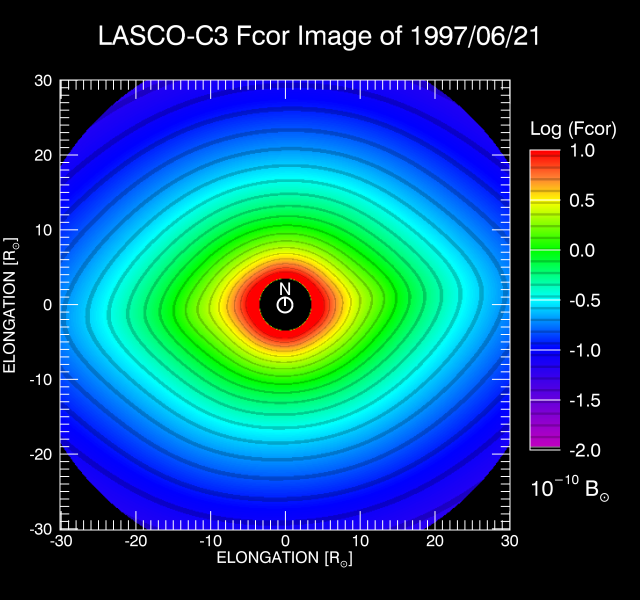}
\caption{Images of the F-corona restored from LASCO-C2 (upper panel) and C3 (lower panel) observations on 21 June 1997.
The white circle corresponds to the solar disk and the direction of solar north is indicated.
The radiance of the F-corona expressed in units of $10^{-10}$\Bsun is coded according to the respective color bars.}
\label{fig:Final_F}
\end{center} 
\end{figure}

The LASCO-C2 final images of the F-corona are the so-called ``Fcor'' images restored by \cite{Llebaria2021} following a complex procedure to eliminate the instrumental stray light (SL) from the ``F+SL'' images resulting from the polarimetric analysis which separated the K-corona \citep{Lamy2020}.
They form a time series of daily images hereafter called ``C2-Fcor'' in the format of 1024 $\times$ 1024 pixels, and their absolute radiance expressed in units of $10^{-10}$\,\Bsun $\space$ is given at the heliocentric distance of SoHO at the specified dates.
An important point is that their orientation is such that north is always up. 
This means that the images obtained when SoHO is rolled by 180\deg were counter-rotated by the same angle.
The vertical direction is aligned with the sky projected direction of the solar axis until 29 October 2010, and with the direction of the ecliptic (north) pole thereafter.
As an example, we display in Figure~\ref{fig:Final_F} the C2-Fcor image at the node of June 1997.

\subsection{LASCO-C3 Images of the F-corona}
\label{sec:C3}
The LASCO-C3 images of the F-corona come from the polarimetric analysis performed by \cite{Lamy2021}.
Strictly speaking and likewise the C2 images, this analysis left the two unpolarized components F-corona and stray light entangled, but it did show that the stray light is extremely low except for the diffraction fringe surrounding the C3 occulter.
Comparing the C2 and C3 photometric radial profiles led to the conclusion that the influence of this fringe is vanishingly small beyond $\approx$\,5\Rsun\ ensuring a nearly perfect transition between the C2 and C3 radiances at $\approx$\,5.5\Rsun. 
This led to the conclusion that a systematic application of the complex restoration procedure implemented for the C2 ``F+SL'' images by \cite{Llebaria2021} was not warranted for the C3 ``F+SL'' images which were therefore assimilated to images of the F-corona as long as the innermost region is excluded.
 
However and as we proceeded with the present in-depth analysis of the F-corona, we did notice problems that required further investigation and correction.
First, a comparison of C3 ``F+SL'' images obtained just before and just after 180\deg\ rolls of SoHO revealed that they were slightly different.
The difference between the ``before'' and the unrolled ``after'' images suggested a faint stray light ramp reminiscent of the diagonal one found in the raw images \citep{Lamy2021}, but oriented in the east--west direction, thus resulting in an artificial asymmetry.
The method implemented to characterize and correct for this secondary ramp is described in Appendix B.
Second, we found necessary to eliminate i) the diagonal streak created by the pylon of the occulter as it was perturbing several photometric analysis and ii) faint stray light residuals. 
As a reminder and in contrast with C2, the pylon of the C3 occulter obstructs its \fov along a narrow diagonal sector either in the south--east quadrant (SoHO roll angle of 0\deg) or the north--west quadrant (SoHO roll angle of 180\deg).
The method proceeded in two stages as illustrated in Figure~\ref{fig:pilcor} and is described in Appendix C.
We display in Figure~\ref{fig:Final_F} the restored C3-Fcor image at the node of June 1997.

\begin{figure}[!htpb]
\centering
\includegraphics[width=\textwidth]{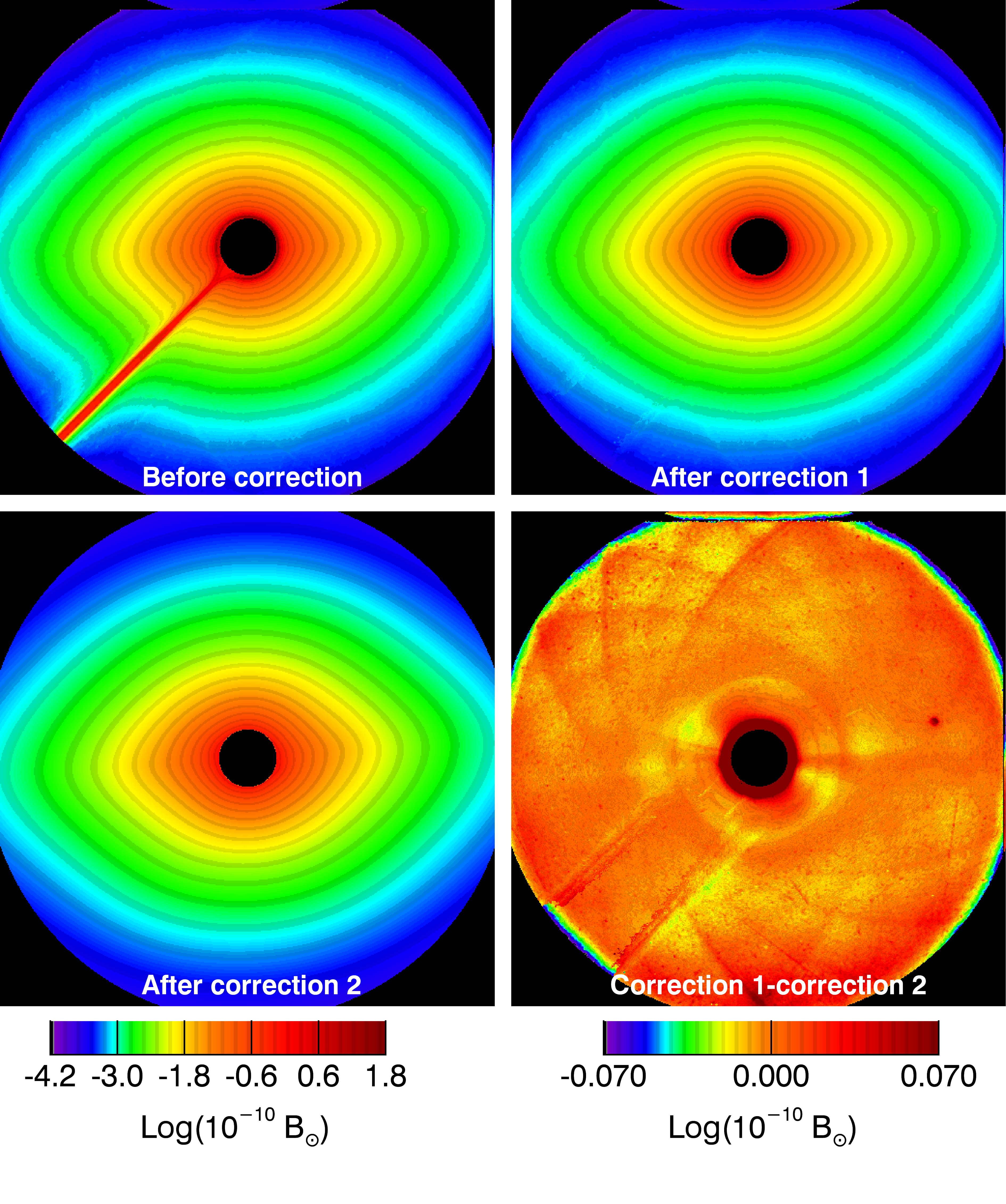}
\caption{Illustration of the two-stage procedure of the fine restoration of the C3 images of the F-corona.
Starting from the initial image (upper left panel), correction 1 removes the bulk of the contribution of the pylon (upper right panel) and correction 2 removes faint stray light residuals (lower left panel)
The logarithm of the three images are displayed according to the color bar at left.
The lower right panel displays the map of the residuals according to the color bar at right.}
\label{fig:pilcor}
\end{figure}

\subsection{Method of Analysis of the C2 and C3 images of the F-corona}
\label{sec:method}
We describe below the different procedures implemented to analyze the C2 and C3 images and to characterize the F-corona.
The first basic and classical one relies on profiles extracted along the  major and minor axes of the ``elliptically'' shaped F-corona, thus corresponding to the two symmetry axes for an observer located in the PSZC.
They are also referred to as photometric axes or axes of maximum (along the major axis) and minimum (along the minor axis) brightness.
In practice, they are conveniently named east, west, north, and south profiles and when averaged over both sides, traditionally named ``equatorial'' and ``polar'' profiles although they may not be strictly aligned with the equatorial and polar directions of the Sun. 

The second characterization implements rings centered at the center of symmetry of the F-corona (nominally the center of the Sun) and at various locations and of different widths depending upon the sought purposes. 
The mean radiances calculated in these rings then reflect the global variations of the F-corona by offsetting the influence of asymmetries resulting from observations outside the PSZC.

\begin{figure}[!htpb]
\centering
\includegraphics[width=\textwidth]{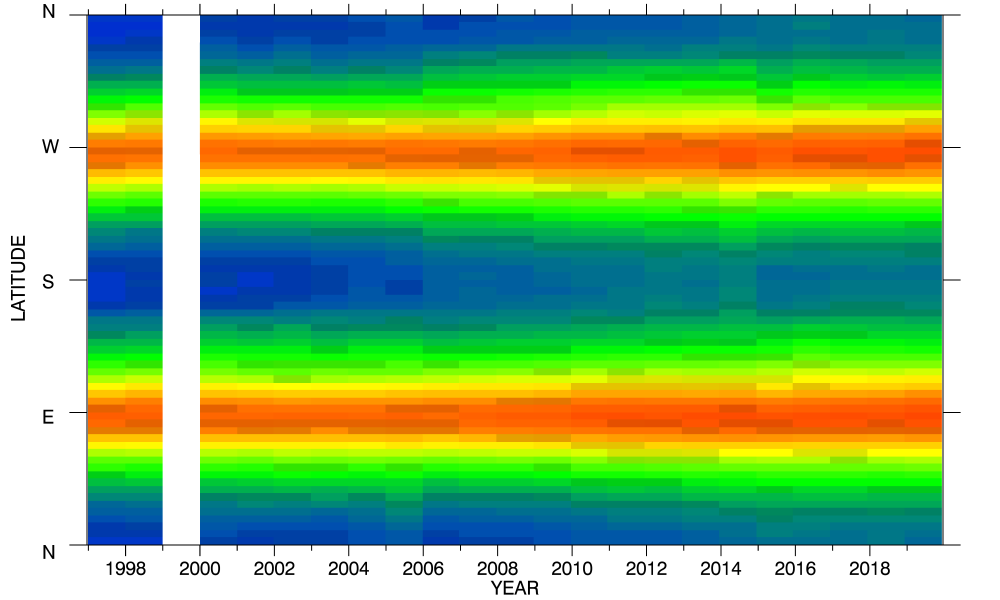}
\caption{Illustration of a stackmap calculated from C3 images at the nodes of December using a ring extending from 12 to 17\Rsun.}
\label{fig:stackmap_C3_intern}
\end{figure}

The third procedure relies on stackmaps constructed at the nodes which basically consist in low resolution heliolatitudinal maps extending over the whole mission. 
The stackmaps were constructed likewise our Carrington synoptic maps\footnote{\url{http://idoc-lasco-c2-archive.ias.u-psud.fr}} by extracting profiles along full circles at selected radii centered on the Sun and stacking them in rectangular arrays where the horizontal axis is time and the vertical axis is position angle measured counter-clockwise from solar north.
There are however specific changes adapted to the F-corona. 
\begin{itemize}
	\item 
The data set is restricted to images obtained at the nodes, either June or December, to ensure the same symmetric configuration of the F-corona.
	\item
Rather than circles, we considered rings as defined above and stacked the mean profiles.
	\item
The angular resolution is set at 5\deg and each column is duplicated five times resulting in frames of 125 $\times$ 72 pixels.
	\item
Each image was enlarged by a factor of five for better legibility.
\end{itemize}
Figure~\ref{fig:stackmap_C3_intern} gives an illustration of a stackmap constructed from the set of C3 images of the F-corona obtained at the December nodes and using a ring extending from 12 to 17\Rsun.
Stackmaps offer global views of the spatial and temporal evolution of the F-corona at given elongations or elongation ranges.

\begin{figure}[!htpb]
\centering
\includegraphics[width=0.49\textwidth]{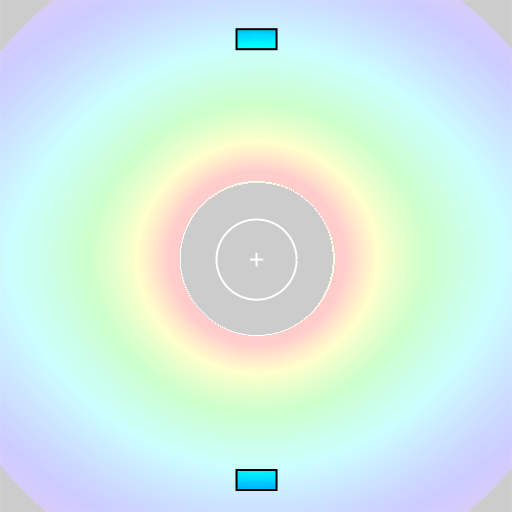}
\includegraphics[width=0.49\textwidth]{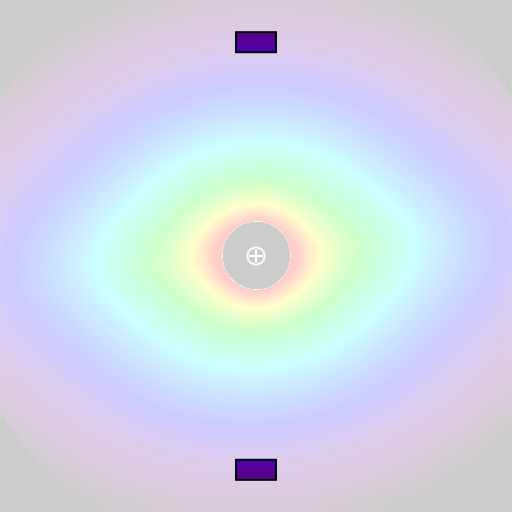}
\caption{Illustration of the two windows in which the integrated radiances were used to monitor the F-corona in the \fov of the LASCO-C2 (left panel) and C3 (right panel) coronagraphs.}
\label{fig:windows}
\end{figure}

A fourth procedure relies on the temporal monitoring of the coronal radiance in two identical small windows labelled ``north'', and ``south''.
They are centered at the same elongation on the vertical axis passing through the center of the Sun (Figure~\ref{fig:windows}).
These windows are rectangular with a common size of 20 $\times$ 40 pixels (for the image format of 512 $\times$ 512 pixels) and with their long side oriented ``tangentially''.
This allows mitigating the effect of the slight periodic oscillation or waddling of the F-corona as the polar axis of the zodiacal cloud is slightly offset from both the solar and ecliptic north pole directions.

\subsection{Comparison of the C2 and C3 Images of the F-corona}
\label{sec:C2C3}
\begin{figure*}[!htpb]
\centering
\includegraphics[width=0.49\textwidth]{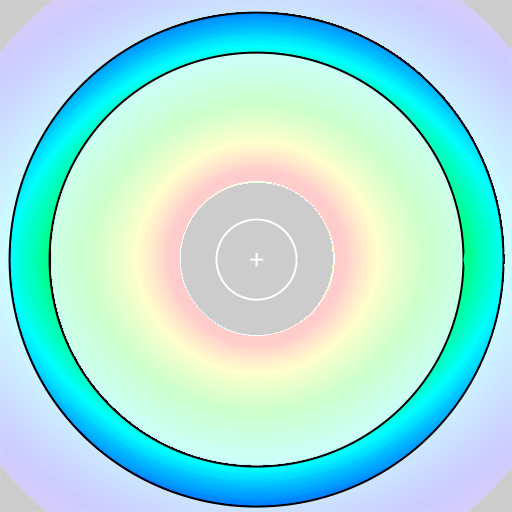}
\includegraphics[width=0.49\textwidth]{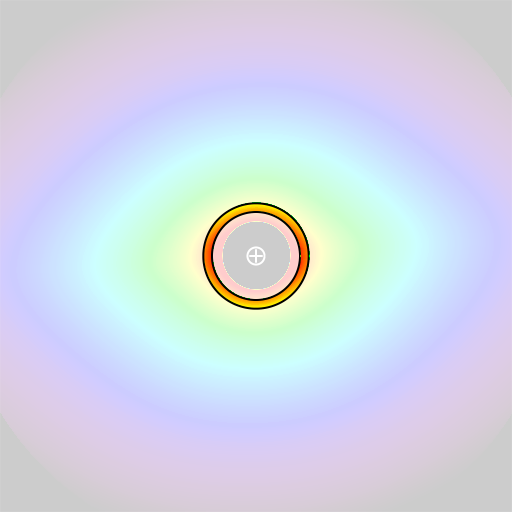}
\caption{Illustration of the common ring used to monitor the integrated radiance in the C2 (left panel) and in the C3 \fovs (right panel).
The white circles represent the solar disk with a cross at its center.} 
\label{fig:Two_rings}
\end{figure*}

The comparison of the C2 and C3 Images of the F-corona by \cite{Lamy2021} mentioned above was however limited to seven dates spanning the [1996\,--\,2019] time interval and we found necessary to strengthen it for the present analysis.
Rather than using radial and circular profiles, we made use of a specific narrow ring where the fields of view of C2 and C3 overlap and we monitored the integrated radiances as a function of time, a method already implemented by \cite{Llebaria2021}.
The inner edge of this ring is constrained by C3 in order to avoid its diffraction fringe and set to ${\rm R}_{\rm in}$(C3) = 44 pixels (5.15\Rsun). 
Taking into account the respective pixel scales of the two instruments, this corresponds to ${\rm R}_{\rm in}$(C2) = 207 pixels in the binned format of 512 $\times$ 512 pixels.
The outer edge is constrained by the C2 \fov on the one hand and by the requirement of having the C3 ring width large enough to obtain a good signal-to-noise ratio on the other hand.
We adopted ${\rm R}_{\rm out}$(C2) = 247 pixels (6.15\Rsun) and this corresponds to ${\rm R}_{\rm out}$(C3) = 52.5 pixels.
The resulting widths of the ring illustrated in Figure~\ref{fig:Two_rings} amount to 40 and 8.5 pixels on the C2 and C3 images, respectively. 
We note that the mid-points of these rings correspond to an elongation of 5.65\Rsun, close to 5.5\Rsun\ used by \cite{Lamy2021} for their comparison.

\begin{figure}[htpb!]
\begin{center}
\includegraphics[width=0.9\textwidth]{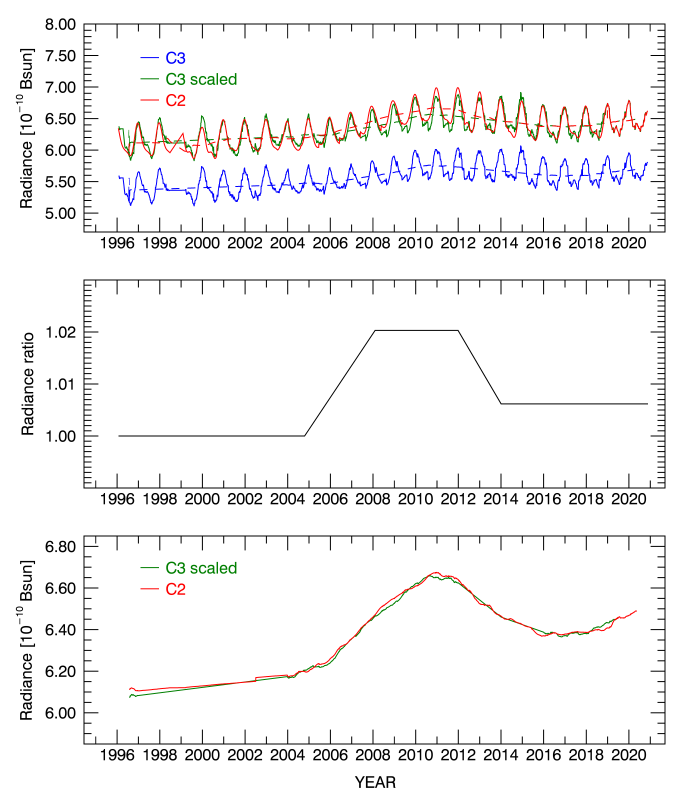}
\caption{Upper panel: daily (solid lines) and  annual (dashed lines) variations of the radiance of the F-corona integrated in a narrow ring common to C2 and C3.
The green curves labeled ``C3 scaled'' correspond to an up-scaled version of the original C3 curve (in blue) compensating for the slight difference in ring areas.
The regular periodic pattern is perturbed in late 1998 to early 1999 by the loss of SoHO.
Middle panel: Simple model of the ratio of the long-term evolutions of the C2 and of the up-scaled C3 data.
Lower panel: Comparison of the long-term evolutions of the C2 and C3 radiances after scaling and correcting the C3 data using the model ratio of the middle panel.}
\label{fig:C2C3ringphot}
\end{center} 
\end{figure}

The upper panel of Figure~\ref{fig:C2C3ringphot} displays the temporal evolutions of the integrated radiances with their characteristic quasi-sinusoidal pattern resulting only from the annual variation of the Sun--SoHO distance as the other periodic oscillation resulting from the back and forth motion of SoHO about the PSZC was eliminated by the integration in the rings.
The long-term variations were obtained by applying a running average with a window of one year thus removing the yearly pattern.
Although the two rings were designed to closely overlap in the C2 and C3 fields of view, we did not expect the C2 and C3 radiances to rigorously match due in part to different pixellisations so that the sky-projected areas of the two rings are not exactly equal.
Considering the first nine years of the mission, we found that the C3 data have to be up-scaled by a factor of 1.14 to perfectly match the C2 data.  
In other words, this means that there is a modest difference of 14\,\% between the areas of the C2 and C3 rings, thus confirming that they were properly constructed.
Whereas C2 is continuously calibrated using stars since 1996, C3 has been calibrated only until 2003 \citep{Thernisien2006}.
Based on their detailed photometric analysis, \cite{Lamy2021} found that the independent calibrations of C2 and C3 are in nearly perfect agreement and that beyond 2003, the sensitivity of C3 with respect to that of C2 has only marginally evolved.
This is well confirmed by the middle panel of Figure~\ref{fig:C2C3ringphot} which displays the temporal variation of the ratio between the C2 and the up-scaled C3 radiances.
The maximum deviation amounts to a mere 2\,\% and took place during only four years, from 2008 to 2011, a quite remarkable achievement.

\section{Photometric Properties of the F-corona}
\label{sec:photometry}
As we are observing the zodiacal cloud from within, the photometric properties of its optical manifestation, F-corona and zodiacal light, very much depends upon the location of the observer, notably its heliocentric distance and its elevation with respect to the plane of symmetry of the zodiacal cloud in the framework of a flattened axially-symmetric cloud described in Section~\ref{sec:overview}.
This consideration becomes obviously acute when comparing different observations.
It is therefore important to understand and quantify the variations resulting from observations obtained at different vantage points.
When appropriate, it will be convenient to then use a reference heliocentric distance of 1\,AU and consider symmetric configurations.
For an observer in or very near the ecliptic plane, the latter condition favors the June or December nodes when this plane crosses the PSZC.

Power laws are generally used to describe the variations of photometric quantities and the notation ``-$\nu$'' is often used for the power exponent, $\nu$ being positive.
The same notation is however sometimes used for different quantities so that we first clarify this point.
A \los is defined by its longitude with respect to that of the Sun $\lambda$-$\lambda_\odot$ and its latitude $\beta$; its elongation $\epsilon$ is then the angle between the observer--Sun line and the \losnosp.
In specific cases, we will however use the concurrent notation $R$ for the elongation to stick with traditional practices and facilitate comparisons with past results; this will be clearly stated to avoid any confusion.
If the PSZC is taken as a reference, then the ecliptic latitude $\beta$ must be replaced by $\beta_{\scriptscriptstyle\rm sym}$, the latitude with respect to this plane.
The heliocentric distance of a point in interplanetary space is denoted ``$r$'', but a specific notation ``$d$'' is introduced for the heliocentric distance of the observer such as a space probe.
\cite{Leinert1981} used the notation $R$, but we prefer ``$d$'' following \cite{Lamy1986} to emphasize the specificity of this vantage point. 
The notation ``$\nu$'' is strictly used for the exponent of the heliocentric distance in describing the spatial density: $N_{\rm d}(r) \propto r^{-\nu}$.
When the photometric profiles defined by the variation of the coronal radiance with elongation at a constant heliocentric distance of the observer follow a power law, we introduce a power exponent denoted -$\nu_{\scriptscriptstyle\rm P}$ ($\nu_{\scriptscriptstyle\rm P}>0$) so that $B_{\rm F}(\epsilon) \propto \epsilon^{-\nu_{\scriptscriptstyle\rm P}}$.
The variation of the radiance with the heliocentric distance of the observer at constant elongation was described by a power law by \cite{Leinert1981} based \textit{Helios} observations; we adopt the notation -$\nu_{\scriptscriptstyle\rm ZL}$ for the corresponding power exponent ($\nu_{\scriptscriptstyle\rm ZL}>0$) so that $B_{\rm F}(d) \propto d^{-\scalebox{1.}{$\nu$}_{\scriptscriptstyle\rm ZL}}$.
Note that \cite{Llebaria2021} used the notation $\nu_{\scriptscriptstyle\rm F}$, but these two power exponents are strictly identical and we adopt the subscript``ZL'' in the present work.

In the simple case of an axially symmetric zodiacal cloud centered at the Sun with its mid-plane in the ecliptic and with the same type of dust everywhere, it can be shown that the power exponent of the radiance profile along the symmetry axis (= ecliptic) -$\nu_{\scriptscriptstyle\rm P}$ is then related to the power exponent of the spatial density -$\nu$ via the expression $\nu_{\scriptscriptstyle\rm P}$ = $\nu$+1, see for instance \cite{Hahn2002}. 
Under the same assumptions, but for all viewing directions, the power exponent of the variation of the radiance with heliocentric distance -$\nu_{\scriptscriptstyle\rm ZL}$ is then related to that of the spatial density -$\nu$ via the expression $\nu_{\scriptscriptstyle\rm ZL}$ = $\nu$+1, see for instance \cite{Leinert1981}.
Therefore, in the case of radiance profiles along the symmetry axis, the following simple relationship holds: $\nu_{\scriptscriptstyle\rm P}$ = $\nu_{\scriptscriptstyle\rm ZL}$.

\subsection{Dependence upon the Heliocentric Distance of the Observer}
\label{sec:dependence}
\begin{figure*}[htpb!]
\centering
\includegraphics[width=0.95\textwidth]{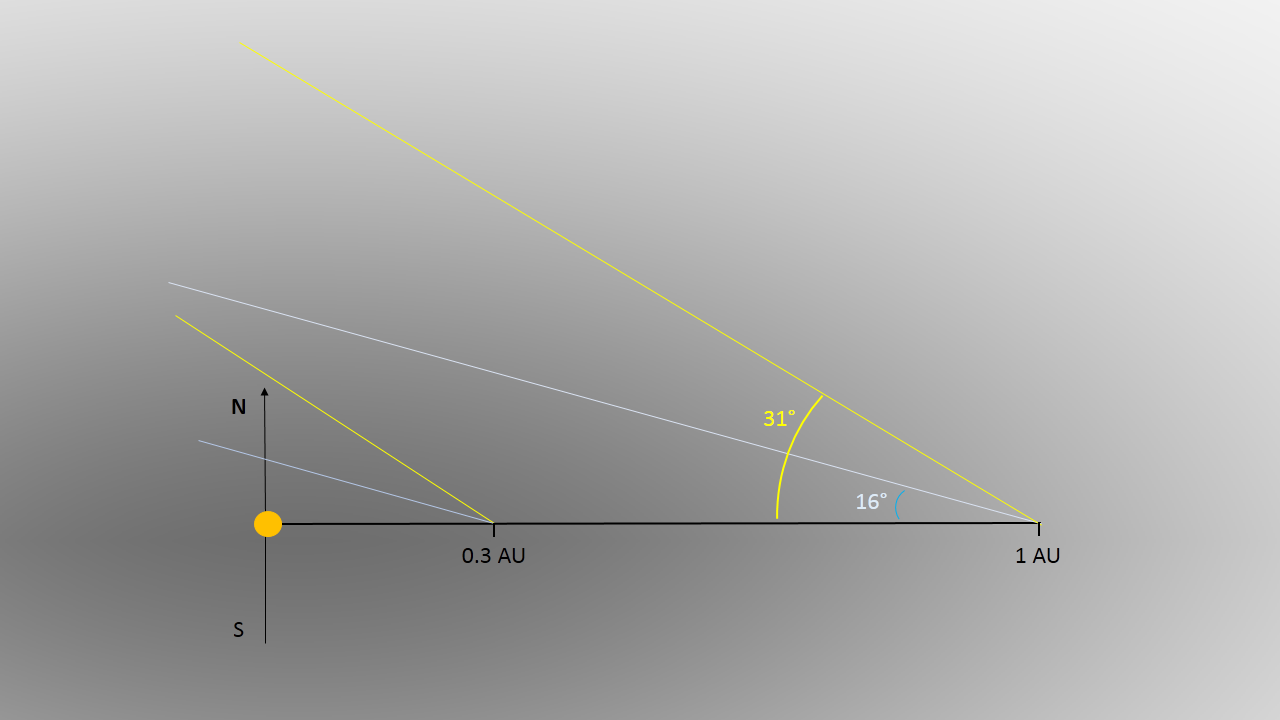}
\caption{Schematic view of the geometry of the \textit{Helios 2} observations of the zodiacal light from the two vantage points at 0.3 and 1\,AU, as seen in the meridian plane (perpendicular to the ecliptic plane).
The axially symmetric zodiacal cloud is represented by the gray background. 
The \loss of the two \textit{Helios 2}  photometers were pointed at northern latitudes of 16\deg and 31\deg.
At the two extreme heliocentric distances, the photometers probed very different regions of the zodiacal cloud.}
\label{fig:helios}
\end{figure*}

It is known from the \textit{Helios} observations that the radiance of the zodiacal light increases as the heliocentric distance $d$ of the observer decreases, at least down to $d$\,=\,0.3\,AU, and this variation follows a power law $d^{-\scalebox{1.}{$\nu$}_{\scriptscriptstyle\rm ZL}}$, with $\nu_{\scriptscriptstyle\rm ZL}$\,$\approx$\,2.3 \citep{Leinert1981}.
It is important to realize that $\nu_{\scriptscriptstyle\rm ZL}$ is determined by comparing measurements obtained at two (or several) heliocentric distances with \loss pointing to the \textit{same} direction. 
Then the \loss are parallel and therefore probe different regions of the zodiacal cloud as illustrated in Figure~\ref{fig:helios} in the case of the \textit{Helios 2} space probe observing the northern hemisphere.
In turn, this allows constraining the three-dimensional spatial distribution of dust in interplanetary space and this aspect will be addressed in the discussion.

The \textit{Helios} measurements were obtained by the twin \textit{Helios~1} and \textit{Helios~2} spacecraft using their scanning photometers mounted at fixed angles of $\pm$16\deg and $\pm$31\deg to the probe orbital plane, nominally the ecliptic plane.
Applying different corrections, \cite{Leinert1981} derived what they called ``normalized intensities'' as they would be observed by an observer in the PSZC at latitudes $\beta_{\scriptscriptstyle\rm sym}$ of 16.2\deg and 31.0\deg.
Next, comparing the results at the aphelion (1\,AU) and perihelion (0.3\,AU) distances of the \textit{Helios} orbits at the \textit{same} solar elongations , they deduced a value $\nu_{\scriptscriptstyle\rm ZL}$\,=\,2.3\,$\pm$\,0.05. 
In their tabulated summary, \cite{Leinert1982} specified that the larger value of 2.35 is more appropriate to small elongations $\epsilon\leq$\,50\deg\ and the smaller value of 2.25, at large elongations $\epsilon\geq$\,100\deg. 
Based on this result, we thus presumed that a power exponent of approximately -2.35 could be expected when observing the F-corona at different heliocentric distances.

In principle and as the orbits of the twin STEREO space probes are not strictly circular, the HI-1 heliospheric imagers could have detected a variation of radiance with heliocentric distance, but this aspect was not considered in the article of \cite{Stenborg2018} devoted to the HI-1 observations of the white-light brightness of the F-corona.
In contrast, this effect was indeed detected in the LASCO data as resulting from the eccentricity of the Earth and therefore of the SoHO orbit \citep{Llebaria2021}. 

The Wide Field Imager for Solar Probe (WISPR) is obviously the instrument of choice for this question, but here again the two articles that report on observations acquired during the first solar encounters by \cite{Howard2019} and \cite{Stenborg2021} did not address it. 
They concentrated on the variation of the radiance with solar elongation along the symmetry axis of the zodiacal cloud and their photometric normalization unfortunately concealed this key information. 
Moreover, these two articles as well as that of \cite{Stenborg2018} made the same systematic confusion between $\nu_{\scriptscriptstyle\rm ZL}$ which characterizes the variation of the radiance with the heliocentric distance of the observer at constant elongation $B_{\rm F}(d)$ and $\nu_{\scriptscriptstyle\rm P}$ which characterizes the variation of the radiance with elongation at constant heliocentric distance of the observer $B_{\rm F}(\epsilon)$. 
What is really needed to ascertain the value of $\nu_{\scriptscriptstyle\rm ZL}$ from the WISPR observations is a plot similar to that presented by \cite{Leinert1981} based on \textit{Helios} observations (their Figure~4), namely a set of curves $B_{\rm F}(\epsilon)$ at different heliocentric distances $d$ (equivalent to ``R'', the notation used by \cite{Leinert1981}) of the observer, where $\epsilon$ is expressed in degree to further avoid the confusion introduced by the variable angular extent of the solar radius \Rsun\ with $d$.

\begin{figure*}[htpb!]
\centering
\includegraphics[width=0.95\textwidth]{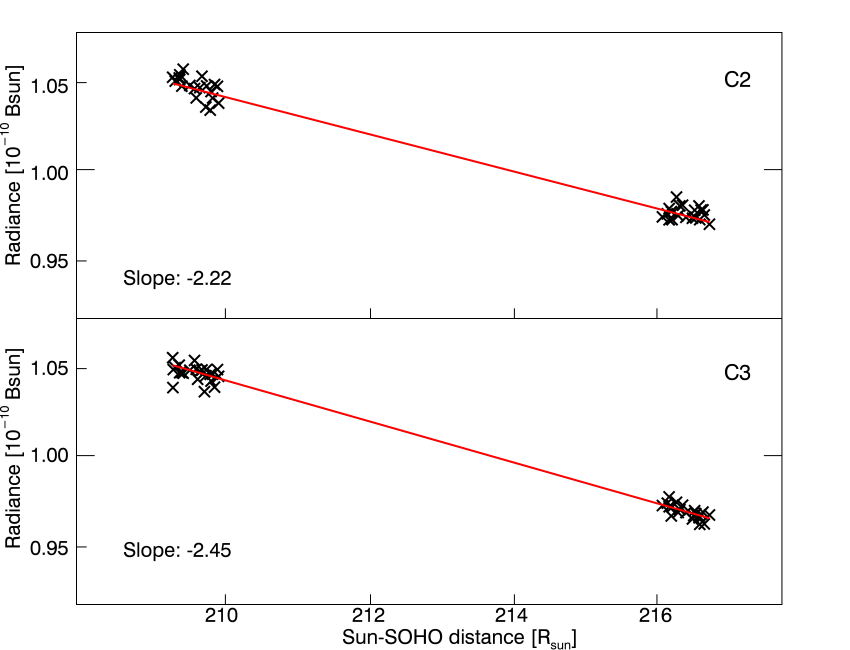}
\caption{Determination of the power exponent -$\nu_{\scriptscriptstyle\rm ZL}$ for C2 (upper panel) and C3 (lower panel) using the integrated radiances in broad rings at the two, nearly extreme, Sun--SoHO distances.}
\label{fig:C2_C3_Nu}
\end{figure*}

The determination of $\nu_{\scriptscriptstyle\rm ZL}$ from the LASCO-C2 and C3 images follows the method implemented by \cite{Llebaria2021} in their analysis of the C2 ``F+SL'' images.
In summary, the radiances were integrated in rings in order to remove the periodic north--south asymmetry affecting the F-corona when SoHO moves back and forth about the PSZC, so that their temporal variations prominently reflect the effect of the varying Sun--SoHO distance.
We specifically followed their second procedure considering only the radiance values at the nodes, exploiting the fact that they take place at nearly the extreme values of the Sun--SoHO distance.
The nodes further offer an advantage as each coronagraph sees the  \emph{same} volume of the zodiacal cloud from two opposite vantage points so that the variation of the radiance between  consecutive nodes can be safely attributed to the varying Sun--SoHO distance.
Using broad rings encompassing the \fov of each coronagraph, we obtained $\nu_{\scriptscriptstyle\rm ZL}$\,=\,2.22\,$\pm$\,0.04 for C2 and $\nu_{\scriptscriptstyle\rm ZL}$\,=\,2.45\,$\pm$\,0.13 for C3 (Figure~\ref{fig:C2_C3_Nu}).
When comparing these values with the \textit{Helios} result $\nu_{\scriptscriptstyle\rm ZL}$\,=\,2.3\,$\pm$\,0.05, several important points must be kept in mind:
\begin{itemize}
	\item 
	The very different ranges of elongation, [0.5\deg\,--\,8\deg] for C2+C3, [16\deg\,--\,160\deg] and [31\deg\,--\,147\deg] for the \textit{Helios} 16\deg\ and 31\deg\ photometers, respectively.
	\item 
	The very narrow range of heliocentric distance of SoHO, typically 0.973 to 1.009\,AU compared with 0.3 to 1\,AU for \textit{Helios}.
\end{itemize}
Owing to its rather large uncertainty, the C3 value is compatible with the \textit{Helios} result $\nu_{\scriptscriptstyle\rm ZL}$\,=\,2.35 at elongations $\epsilon\leq$50\deg.
The lower C2 value tends to suggest that the progressive increase of $\nu_{\scriptscriptstyle\rm ZL}$ with decreasing elongation may not hold down to the inner F-corona and that a turnover may take place at some elongation.
Whereas appropriate to our objective of scaling the C2 and C3 radiances to account for the varying Sun--SoHO distance, our results should not be extrapolated beyond this range in view of the limitations spelled above.
Nevertheless, the fact that LASCO could detect the influence of the varying Sun--SoHO distance over a very narrow range constitutes a noteworthy achievement.

\subsection{Radiance Profiles of the F-corona}
\label{sec:C2C3}
The classical photometric characterization of the F-corona relies on the radiance profiles along two directions, equatorial and polar. 
Many such profiles have been published in the past and for instance, Figure~1 of \cite{Kimura1998} presents a compilation in the equatorial case. 
The observer is assumed to be at 1\,AU and the F-corona to be symmetric or east--west and south--north profiles are averaged to create the above two profiles.
A sound comparison with these past results require that we consider LASCO profiles obtained under similar conditions. 
The symmetric configuration is ensured by selecting observations obtained at the nodes, either June or December. 
There are currently 46 such nodes and we obviously had to make a choice.
Based on Figure~\ref{fig:C2C3ringphot} which revealed a temporal variation of the radiance of the F-corona integrated in a ring common to C2 and C3, we selected the ``extreme'' cases of this variation.
Ultimately, we retained the December node of 1997 typical of the first few years of LASCO observations during which the radiance of the F-corona was nearly constant, and the nodes of December 2010 and 2011 when the radiance reached two consecutive and similar maxima; in practice, the images corresponding to these two nodes were averaged.
 %

\begin{figure*}[htpb!]
\centering
\includegraphics[width=0.49\textwidth]{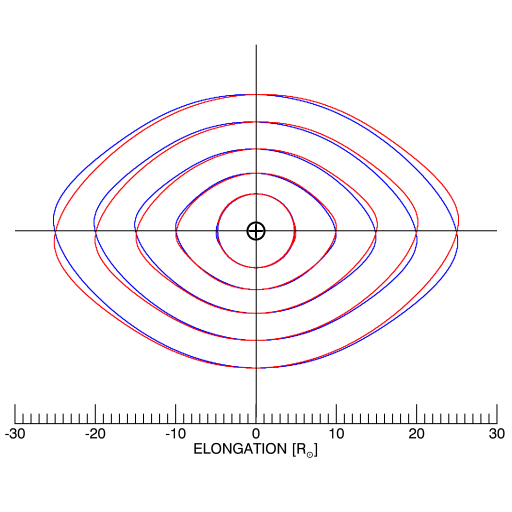}
\includegraphics[width=0.49\textwidth]{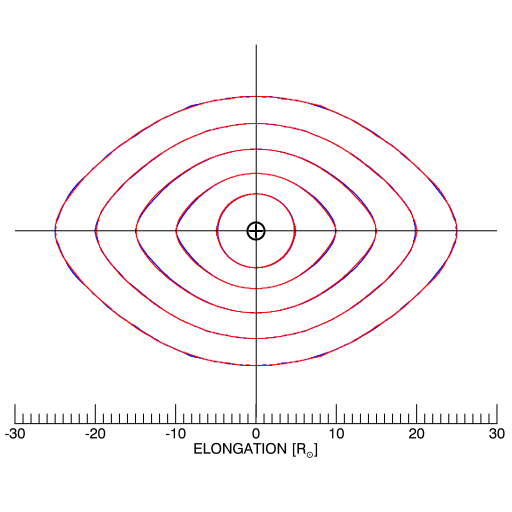}
\caption{Schematic illustration of the procedure to extract the photometric profiles of the F-corona along the major and minor axes of the ``elliptical'' isophotes.
The original (in blue) and the flipped (in red) isophotes (left panel) are rotated until coincidence is achieved (right panel).}
\label{fig:control_tilt}
\end{figure*}

We naturally used the C2-Fcor and C3-Fcor images introduced in Section~\ref{sec:SOHOLASCO} and applied two normalizations imposed by the selection of a Sun-SoHO reference distance of 1\,AU, a geometric one that redefines the pixel scale using the value of the solar radius at 1\,AU  and a photometric one using the power law $d^{-\scalebox{1.}{$\nu$}_{\scriptscriptstyle\rm ZL}}$ with $\nu_{\scriptscriptstyle\rm ZL}$\,=\,2.22 for C2 and 2.45 for C3 as determined in the above sub-section.
The profiles were extracted along the directions of the major and minor axes of the ``elliptically'' shaped F-corona.
Figure~\ref{fig:control_tilt} illustrates how we proceeded since the direction of the axes changed with time, at least until SoHO switched to ecliptic north orientation.
A selected set of isophotes extracted from a given image were plotted along with those of its mirror or flipped version with respect to the column direction.
The image was progressively rotated until the two sets of isophotes coincided ensuring that the major and minor axes were aligned with the row and column directions, respectively.
Then the desired profiles were simply extracted along these two directions.

\begin{figure*}[htpb!]
\centering
\includegraphics[width=0.95\textwidth]{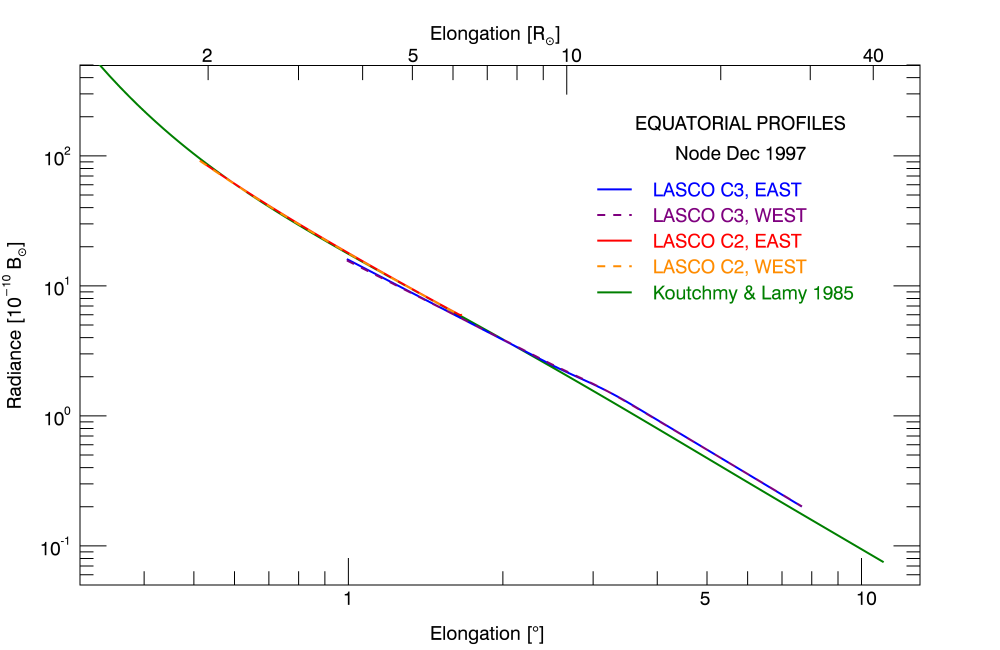}
\includegraphics[width=0.95\textwidth]{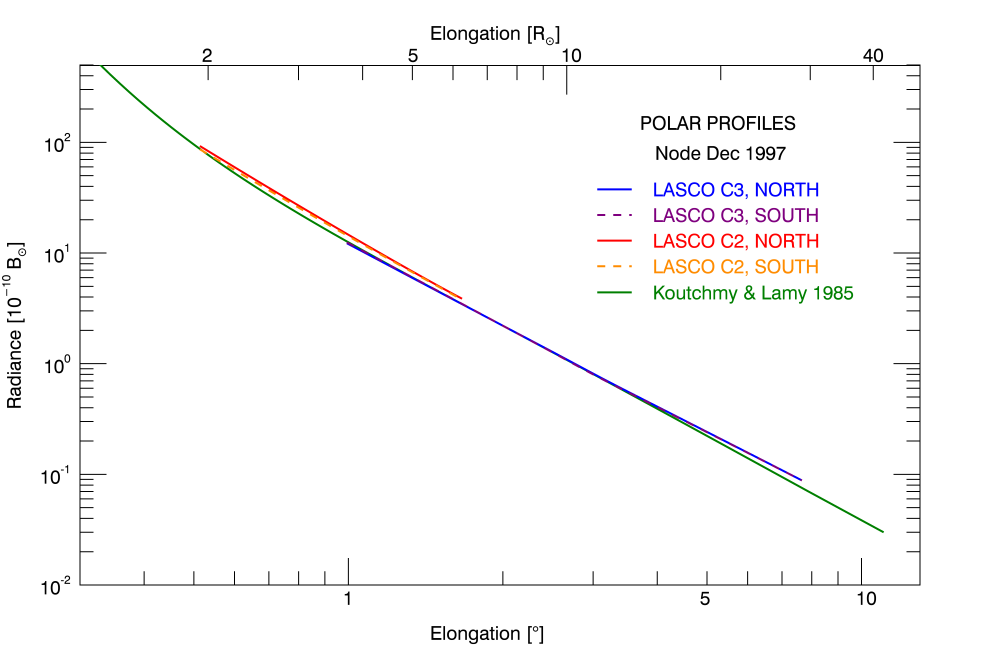}
\includegraphics[width=0.95\textwidth]{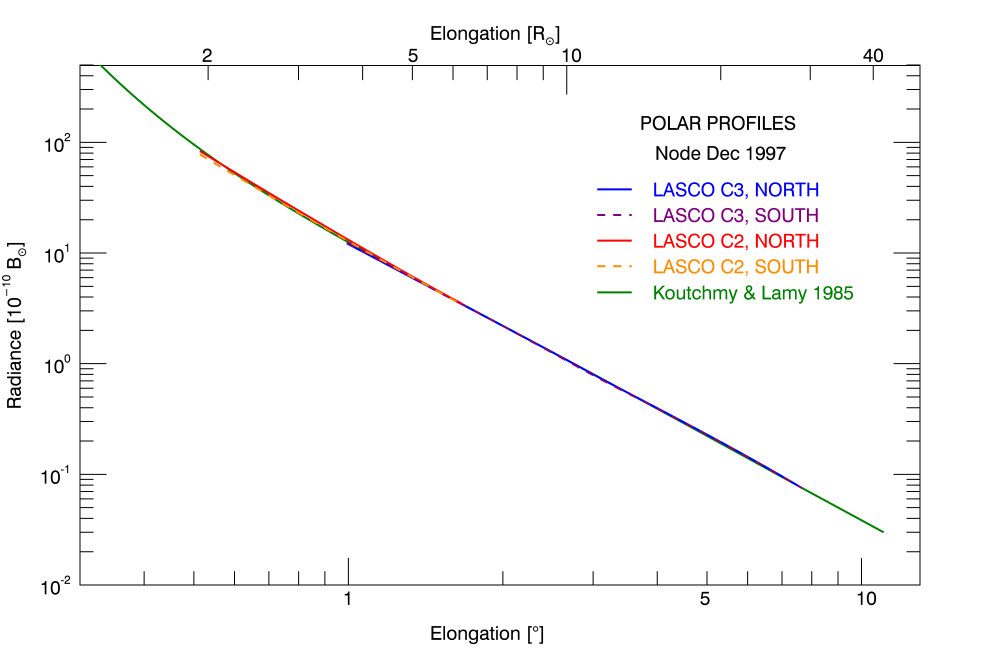}
\caption{Profiles of the radiance of the F-corona recorded by LASCO-C2 and C3 when located at 1\,AU in the plane of symmetry of the zodiacal cloud at the node of December 1997 together with the Koutchmy--Lamy model (green curves).
The upper panel displays the equatorial east and west profiles and the middle and lower panels display the polar north and south profiles.
In the lower panel, slight corrections were applied to the C2 and C3 profiles as described in the text.
Although all curves were plotted, they are sometime indistinguishable when the agreement is nearly perfect.}
\label{fig:Prof_C2C3_50803}
\end{figure*}

Figure~\ref{fig:Prof_C2C3_50803} displays the four east, west, north, and south profiles corresponding to the node of December 1997 as well as the Koutchmy--Lamy (K--L) model \citep{KoutchmyLamy1985}, nowadays considered as a reference of the F-corona.
We highlight the most striking features revealed by this figure.
\begin{itemize}
	\item 
	The nearly perfect similarity of the opposite profiles, east--west and north--south, confirming the symmetry of the F-corona when observed in the conditions stipulated above.
	\item 
	The globally excellent agreement between the LASCO profiles and the K--L model.
	\item 
	The consistency between the C2 and C3 profiles.
\end{itemize}

A closer inspection of the equatorial profiles reveals several interesting details.
First, the inner part of the C3 equatorial profiles below $\approx$\,6\Rsun\ tends to slightly depart from the C2 profile reaching a maximum deviation of $\approx$12\,\% at 4\Rsun\ a probable consequence of the subtraction of the stray light from the occulter. 
However, this slight deviation does not prevent the two profiles to smoothly connect at approximately 6\Rsun.
Second, the C3 profile exhibits a shoulder starting at $\approx$\,10\Rsun\ in contrast with the K--L model characterized by a constant slope.
Beyond $\approx$\,13\Rsun, the two profiles become quasi parallel (\ie same slope) with the C3 radiance exceeding the K--L model by $\approx$17\,\%.
This feature will be further explored when comparing with other results in the next sub-section.
Turning to the polar profiles, two points are worth mentioning.
First, the C2 profile is systematically brighter than, but quasi parallel to the C3 and K--L profiles by 10\,\% which incidentally corresponds to the uncertainty estimated by \cite{Llebaria2021} for the restoration of the K- and F-coronae.
Second, beyond $\approx$\,16\Rsun, the C3 profile progressively diverges from the K--L model and, in this case, we suspect the presence of a very faint stray light background. 
Both slight discrepancies may be efficiently corrected by down-scaling the C2 profile by a factor of 0.9 on the one hand and by subtracting a constant background of 
$1.3 \times 10^{-12}$\,\Bsun from the C3 profile on the other hand as illustrated in the lower panel of Figures~\ref{fig:Prof_C2C3_50803}.

\begin{figure*}[htpb!]
\centering
\includegraphics[width=0.95\textwidth]{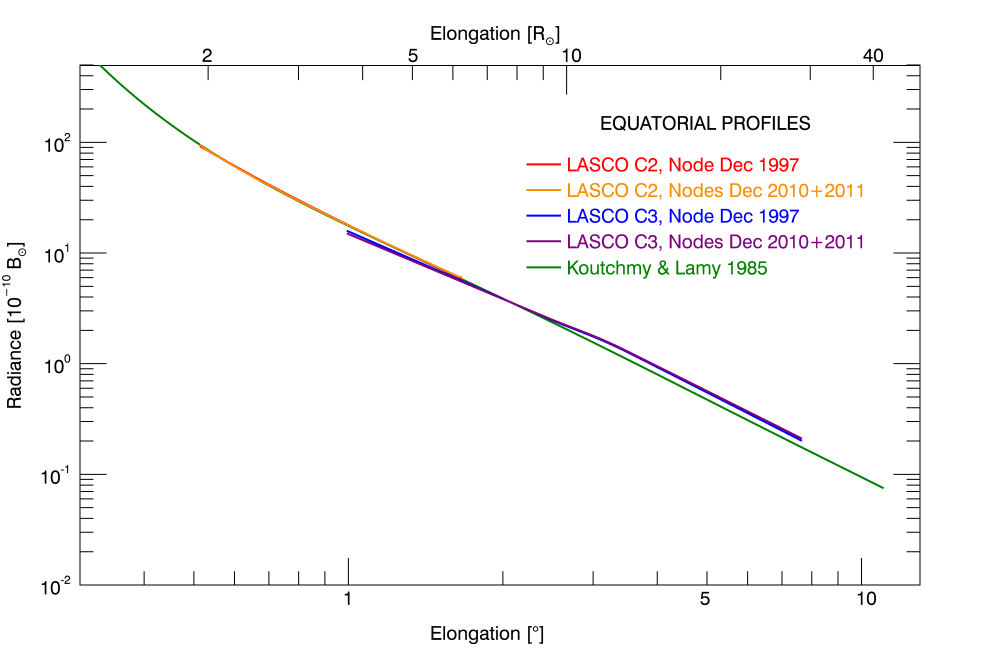}
\includegraphics[width=0.95\textwidth]{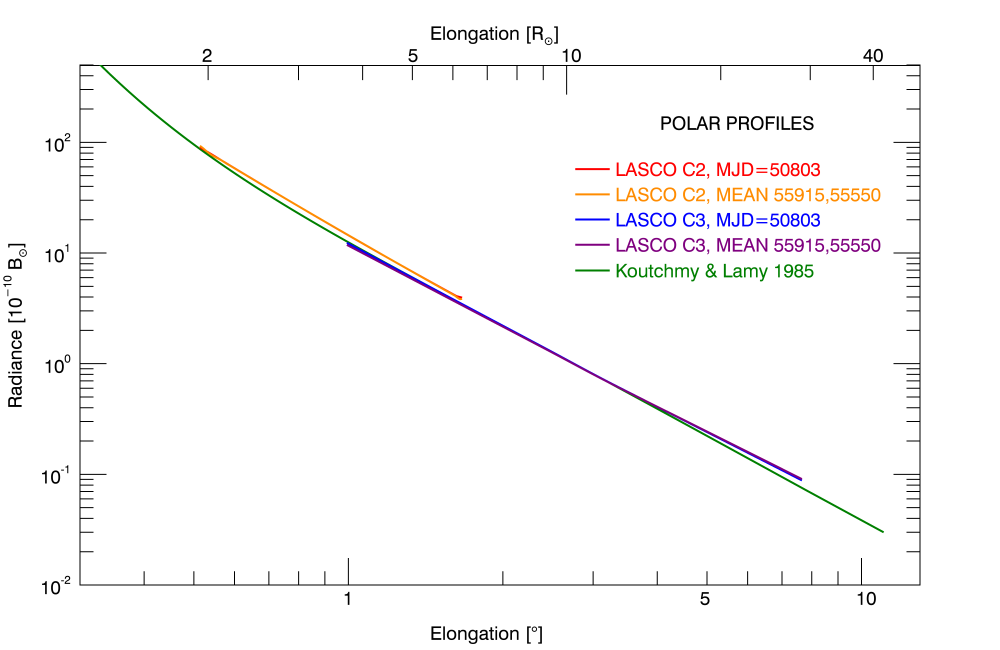}
\includegraphics[width=0.95\textwidth]{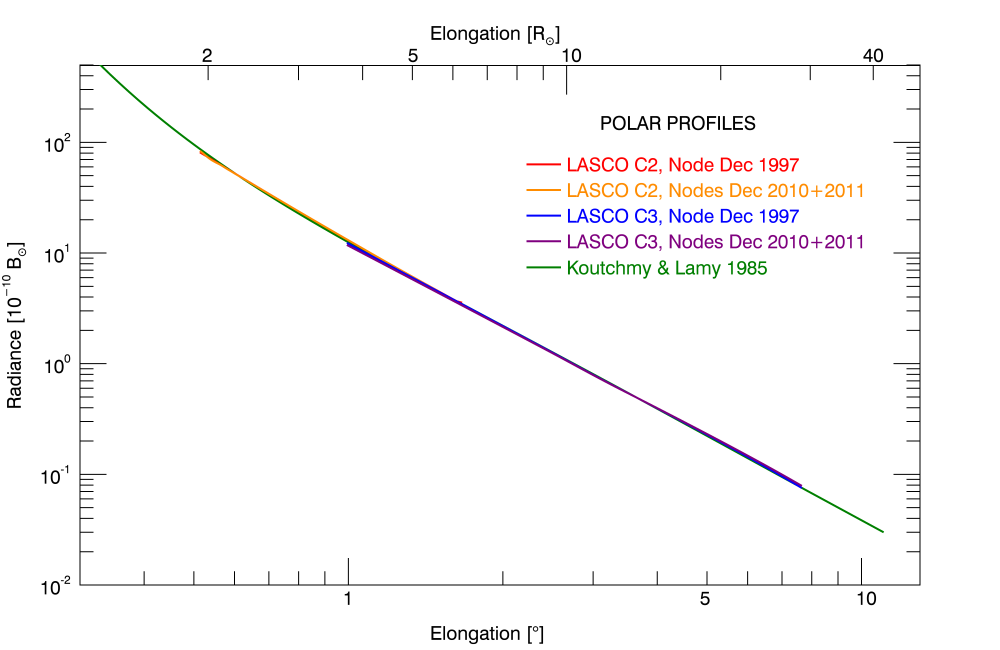}
\caption{Profiles of the radiance of the F-corona recorded by LASCO-C2 and C3 when located at 1\,AU in the plane of symmetry of the zodiacal cloud at the node of December 1997 and at the combined nodes of December 2010 and 2011 together with the Koutchmy--Lamy model (green curves).
The 2010+2011 profiles were systematically down-scaled by a factor of 0.92 to facilitate the comparison.
The upper and middle panels display the equatorial and polar profiles, respectively. 
In the lower panel, slight corrections similar to those used in the lower panel of Figure~\ref{fig:Prof_C2C3_50803} were applied to the C2 and C3 profiles as described in the text.
Although all curves were plotted, they are sometime indistinguishable when the agreement is nearly perfect.}
\label{fig:Prof_C2C3_55550_55915}
\end{figure*}

The photometric profiles coming from the images combining the nodes of December 2010 and 2011 exhibit similar properties with furthermore and as expected, a systematic enhanced radiance of 8\,\% with respect to the node of December 1997, in agreement with the temporal variation illustrated in Figure~\ref{fig:C2C3ringphot}. 
To simplify the comparison between these two cases, we considered the equatorial (average of east and west) and polar (average of north and south) profiles applying a global factor of 0.92 to the 2010+2011 profiles (Figure~\ref{fig:Prof_C2C3_55550_55915}).
The agreement is nearly perfect and only required the same slight adjustment of the polar 2010+2011 profiles as applied to the polar 1997 profiles, namely multiplying the C2 profile by a factor of 0.9 and subtracting a constant background of $2.1 \times 10^{-12}$\,\Bsun from the C3 profile.
These corrected profiles will therefore be adopted from now on.


\subsection{Comparison with Past Data of the F-corona}
\label{sec:comp}
Figure~1 of \cite{Kimura1998} quoted above and which compiles many past measurements is interesting by showing a global agreement on the general trend of the profile, but also systematic discrepancies by factors of 2 to 3.
The accumulation of data renders the comparison barely legible and inappropriate when looking at an accuracy at the level of 10 to 20\,\%.
We used a different approach and restricted our selection to profiles resulting themselves from scrutinized synthesis of selected data sets of presumed superior quality. 
\begin{itemize}
	\item 
	The Koutchmy--Lamy model already introduced in the above sub-section.
	\item
	The profiles given in the compendium entitled ``The 1997 reference of diffuse night sky brightness'' by \cite{Leinert1998} which ``takes the recent measurements into account as well as the fact that the scattering properties change due to the increasing diffraction peak at small scattering angles''.
	The two profiles are specified by power laws with exponents of -2.5 for the equatorial one and -2.8 for the polar one and by absolute radiance values at 4\Rsun\ (their Table 23).	
	\item
	The Cox's version of Allen's Astrophysical Quantities \citep{Cox2000} includes two complementary data sets, one given in the Section ``Corona'' (Table 14.19) covers the range 1.1 to 20\Rsun\ and the other in the Section ``Zodiacal Light'' (Table 13.8) covers the range 1\deg\ to 10\deg\ ($\approx$\,4 to 40\Rsun).
	There are a few slight discrepancies between the two data sets, but they are unimportant for our present purpose.
\end{itemize}
We made an exception by introducing the quasi space observations performed by \cite{Blackwell1955} at the eclipse of June 1954 from an open aircraft at an altitude of 30\,000 feet in excellent sky conditions.
The resulting equatorial and polar profiles are tabulated in his Table III. 

\begin{figure*}[htpb!]
\centering
\includegraphics[width=0.95\textwidth]{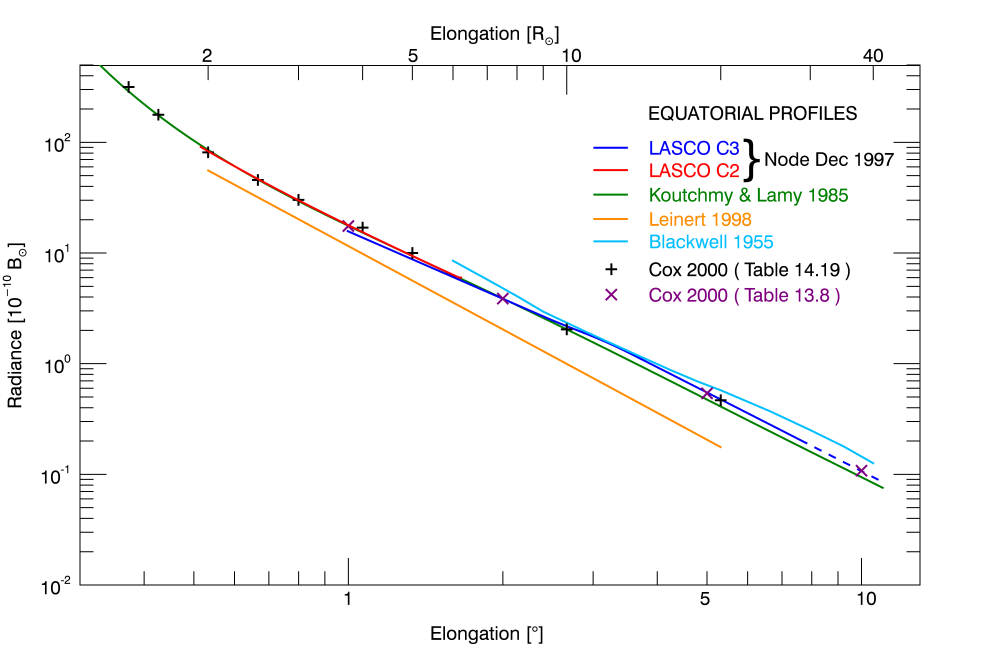}
\includegraphics[width=0.95\textwidth]{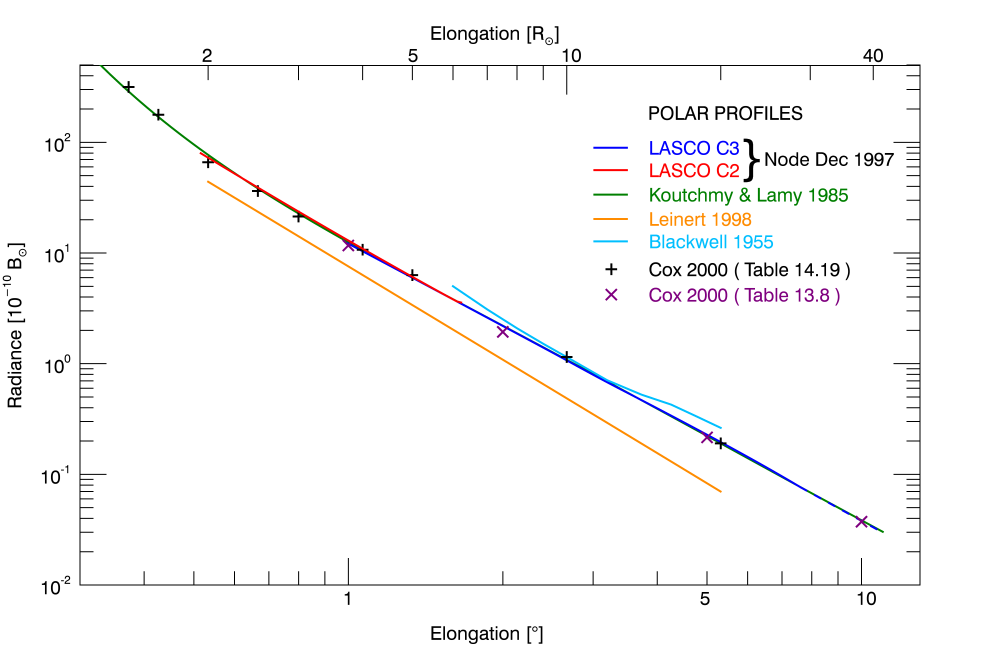}
\caption{Profiles of the radiance of the F-corona recorded by LASCO-C2 and C3 when located at 1\,AU in the plane of symmetry of the zodiacal cloud at the node of December 1997 together with the model of \cite{KoutchmyLamy1985}, the photometric profiles of \cite{Blackwell1955}, of \cite{Leinert1998}, and the data of \cite{Cox2000}.
The upper panel displays the equatorial profiles and the lower panel displays the polar ones.
In this latter panel, the slight corrections introduced in Section~\ref{sec:C2C3} were applied to the C2 and C3 profiles.
Although all curves were plotted, they are sometime indistinguishable when the agreement is nearly perfect.}
\label{fig:Compare_F}
\end{figure*}

We limit the comparison to the case of the node of December 1997 and Figure~\ref{fig:Compare_F} reveals the excellent agreement between the equatorial and polar profiles of LASCO, of the Koutchmy--Lamy model, and the data of \cite{Cox2000} data.
The profiles of \cite{Blackwell1955} are quite close with radiances at elongations of 2.5\deg\ to 3\deg\ in agreement with C3 and the data of \cite{Cox2000}.
However, they diverge at smaller and larger elongations. 
In the first case, this most likely arises from the fact that \cite{Blackwell1955} did not subtract the underlying contribution of the K-corona, although small, but non-negligible, at these small elongations.
In the second case, a residual contribution of the sky may be suspected. 
The profiles given by \cite{Leinert1998} are clearly off the main trend, both in radiance and gradient and consequently, are not further commented.
As already noted, the C3 equatorial profile exhibits a shoulder starting at $\approx$\,10\Rsun\ unlike the K--L model, but consistent with the \cite{Cox2000} data.
This supports its reality and therefore, the enhanced radiance beyond $\approx$\,10\Rsun\ compared with a model having a constant slope.  
Turning to the polar case, the C2 profile, the C3 profile once corrected as described in the above sub-section, and the \cite{Cox2000} data are in remarkable agreement.

\subsection{Comparison with the Zodiacal Light}
\label{sec:comp_ZL}
We adopted the same approach as in the above section, considering only synthetic tabulations of measurements of the zodiacal light radiance, and limiting the comparison to the case of the node of December 1997.
The all-sky ``Tenerife'' data set of \cite{Dumont1975, Dumont1976} at 502\,nm remains the most comprehensive and accurate.
It has been critically discussed, checked against space data and slightly improved for better smoothness by \cite{Levasseur1980}.
A further update was introduced by \cite{Leinert1998}: the data set was extended inward to an elongation of 15\deg\ from an original elongation of 30\deg\ in the ``Tenerife'' data set and the innermost values were slightly increased (presumably to best connect to the newly added values) together with those at higher latitudes.
Their Table 16 is reproduced in \cite{Cox2000} as Table 13.7 and the data are valid at 500\,nm for an observer at 1\,AU in the plane of symmetry of the zodiacal cloud.
The uncertainty is quoted at 10\,\% in the bright regions and at 20\,\% in the faint regions.
Note that the effect of the slight difference in wavelength -- 502 nm for the data of \cite{Levasseur1980} -- was neglected. 
We checked that the equatorial and polar radiance values as originally reported by \cite{Dumont1975, Dumont1976} and used by \cite{Lamy1986} for their inversion and those tabulated by \cite{Leinert1998} are quasi identical, thus ensuring that the volume scattering function derived by \cite{Lamy1986} remains valid.
The conversion of \SZLnosp, the traditional units for the zodiacal light, to \Bsun at 500\,nm is given by \cite{Leinert1998}: 1\,\SZL = 4.5 $\times$ $10^{-16}$\,\Bsunnosp.

A very valuable complement to the above data set is offered by the observations performed by the navigation cameras onboard the \textit{Clementine} spacecraft using the Moon to occult the Sun.
These two cameras were equipped with broad band filters whose rectangular equivalent bandpass extends from 560 to 710\,nm with a central wavelength of 635\,nm \citep{Hahn2002}.
We made use of the four profiles plotted in Figure~8 of this latter article and averaged the east--west and north--south profiles to generate the equatorial and polar profiles over their common range of elongation.
As a result, the former profile extends from $\approx$\,6\deg\ to $\approx$\,20\deg\ and the latter from $\approx$\,5\deg\ to $\approx$\,15\deg.
The original, individual profiles exhibit a double asymmetry barely visible at the inner limit of the \fov and progressively increasing with increasing elongations.
The north--south one was correctly interpreted by \cite{Hahn2002} as resulting from the inclination of the symmetry plane of the zodiacal cloud (PSZC) with respect to the ecliptic.
This is consistent with the observations having been taken in March and April, that is in quadrature with the nodes of the PSZC when the effect is maximized with north brighter than south, in agreement with the LASCO observations.
The east--west asymmetry was tentatively attributed to secular gravitational perturbations by the giant planets and this aspect will be addressed later in the general discussion.

The SECCHI-HI heliospheric imagers onboard the twin STEREO A and B spacecraft have rectangular \fovs\ to the east side of the Sun extending from 4\deg\ to 24\deg\ for HI-1 and from 19\deg\ to 89\deg\ for HI-2.
Photometric results from the STEREO-A/SECCHI-HI-1 instrument reported by \cite{Stenborg2018} are unfortunately very limited: only a single profile along the east direction between 5\deg\ and 24\deg\ obtained on 21 January 2008 (their Figure~4).
From their linear fit on a log--log scale, we derived for the sake of comparison the following expression for the radiance expressed in units of $10^{-10}$\Bsun: ${\rm B}_{\rm F}=24.546 \times \epsilon^{-2.337}$ where $\epsilon$ is the elongation in degree.
We note that \cite{Stenborg2018} plotted the different slopes (that is -$\nu_{\scriptscriptstyle\rm P}$) for the 16 roll sets that enabled the complete reconstruction of the coronal images.
In the equatorial case, there is little dispersion in the data and a very slight difference between the east and west sides with respective averages of $\nu_{\scriptscriptstyle\rm P}$\,=\,2.34 and 2.355, hence a difference of only 0.64\,\% and a global average value of 2.35 close to that determined above from their Figure~4.
In contrast, the polar values are highly dispersed, but with a net asymmetry between the north and south directions; we estimated a global average of 2.49. 

The Wide Field Imager for Solar Probe (WISPR) comprises two telescopes with rectangular \fovs\ to the west side of the Sun extending from 13.5\deg\ to 53.6\deg\ for WISPR-I and from 50\deg\ to 108\deg\ for WISPR-O.
The data obtained during the orbit inbound to the first perihelion at five heliocentric distances ranging from 0.336 to 0.166\,AU led \cite{Howard2019} to show that all photometric profiles along the axis of symmetry follow the same power law with an exponent of -2.31 down to an elongation of $\approx$\,25\deg; at shorter elongations ($\leq$\,25\deg), the radiance decreases with decreasing PSP heliocentric distance. 
This is essentially confirmed by the more recent analysis of \cite{Stenborg2021} using data acquired during the first five solar encounters (excluding encounter 3) with WISPR-I. 
From measurements along the axis of symmetry (Figure~\ref{fig:F_examples}), they determined an average slope of -2.295 (which we safely rounded to -2.30 since their quoted uncertainty of 0.006 probably reflects only the quality of the fits) valid down to an elongation of $\approx$\,25\deg. 
Both articles used ``equivalent'' elongations in units of \Rsun\ performed by dividing the real elongation (in degree) by the half the angular size of the Sun at the respective PSP distances where the data were acquired.
On the one hand, this allows superposing the profiles obtained at different heliocentric distances, but on the other hand, renders the profiles unphysical, especially since they are further all normalized to a common value at a given elongation.
As a consequence, this conceals the variation of the radiance with the heliocentric distance of the observer as already pointed out in Section~\ref{sec:dependence}.
In order to include the WISPR result in our comparison, we scaled the power law to closely match the other results with the following expression for the radiance in units of $10^{-10}$\Bsun: ${\rm B}_{\rm F}=19.53 \times \epsilon^{-2.3}$ where $\epsilon$ is the elongation in degree.

\begin{figure*}[htpb!]
\centering
\includegraphics[width=0.95\textwidth]{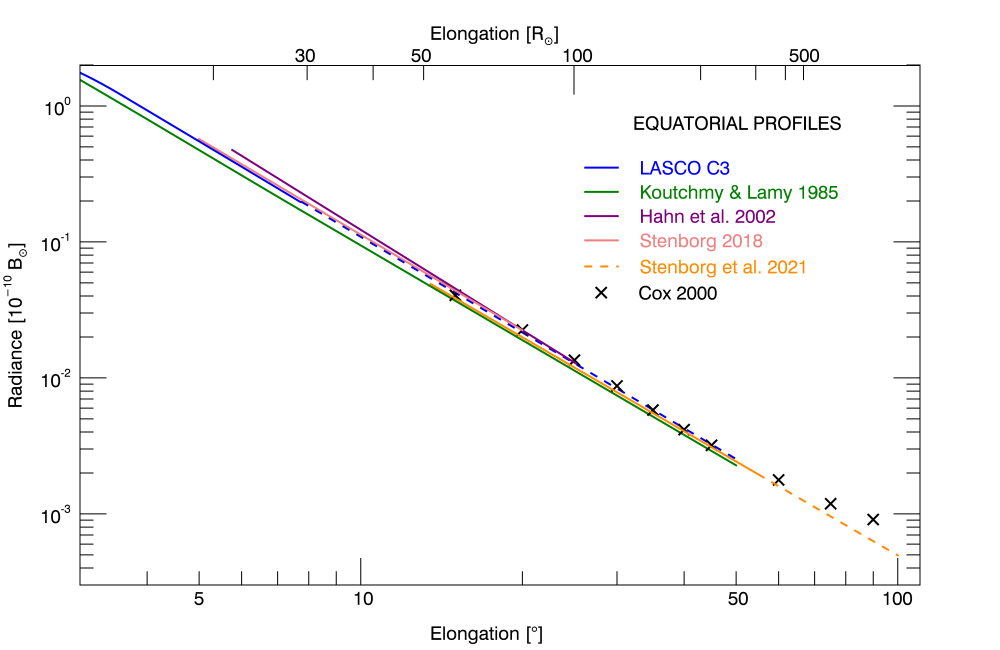}
\includegraphics[width=0.95\textwidth]{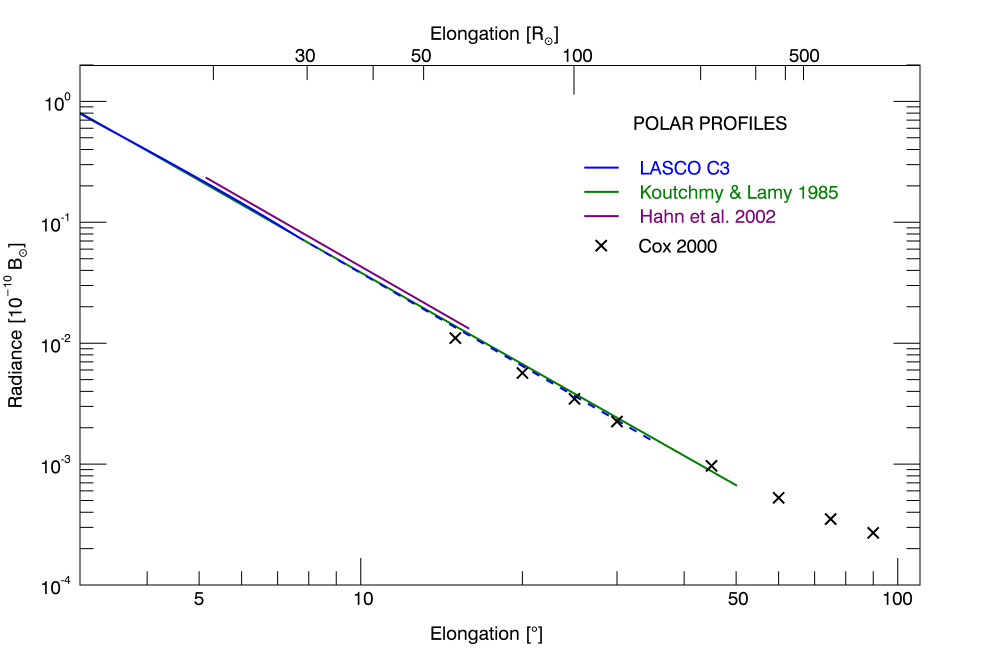}
\caption{Profiles of the radiance of the F-corona recorded by LASCO-C3 when located at 1\,AU in the plane of symmetry of the zodiacal cloud at the node of December 1997 together with the model of \cite{KoutchmyLamy1985}, the photometric profiles of \cite{Hahn2002}, \cite{Stenborg2018}, \cite{Stenborg2021}, and the data of \cite{Cox2000}.
The upper panel displays the equatorial profiles and the lower panel display the polar ones .
The dashed lines represent straightforward extrapolations of the original data assuming constant slopes.}
\label{fig:Compare_F_ZL}
\end{figure*}

Figure~\ref{fig:Compare_F_ZL} displays the LASCO-C3 profile further extended beyond 30\Rsun (the outer limit of the C3 \fovnosp) using a linear extrapolation on the log--log scale, the above four data sets of the zodiacal light, and the Koutchmy--Lamy model as our standard reference.
The connection between the extended C3 equatorial profile and the zodiacal light data of \cite{Cox2000} is excellent, implying that a power exponent of -2.33 holds to an elongation of $\approx$\,50\deg\ beyond which the slope becomes shallower.
In the case of the polar profiles, the connection is less satisfactory as it appears that the first two radiance values of the zodiacal light at elongations of 15\deg\ and 20\deg\ of \cite{Cox2000} are slightly too low so that an inward extrapolation would seriously diverge from the F-corona profile.
These two values would benefit from an upward revision, thus reinforcing the increase already introduced by \cite{Leinert1998}.
Consequently, the power exponent of -2.55 given by the C3 polar profile appears very appropriate and holds up to an elongation of $\approx$\,35\deg.
The results of \cite{Hahn2002} warrant two comments: i) their equatorial profile, whereas being in agreement with the \cite{Cox2000} data, has a slope steeper than those given by the LASCO data and the K--L model (a power exponent of -2.45 compared with -2.33) so that its inner extension clearly diverges, and ii) their polar profile, although it has a slope similar to that of the C3 profile, is however conspicuously brighter by $\approx$15\,\% and does not match the zodiacal light data. 
\cite{Hahn2002} quoted an uncertainty of 8\,\% in their calibration, perhaps it is too optimistic. 
The SECCHI-HI-1 equatorial profile of \cite{Stenborg2018} (their Figure~4), although being in agreement with the zodiacal light data, appears slightly too steep and if extrapolated inward, would tend to diverge from the C3 profiles and the K--L model.
In contrast, the WISPR profile exhibits a slope in excellent agreement with the extrapolated C3 profile and the K--L model. 
As noted above, the absolute scaling of its radiance does not come from \cite{Stenborg2021}, but was chosen by us so as to closely match the other curves for comparison purpose.
Table~\ref{tab:slope} summarizes the various determinations of the power exponent $\nu_{\scriptscriptstyle\rm P}$ of the equatorial and polar photometric profiles discussed above.
This leads to the robust determination of the following two ranges: 2.30 $\leq$ $\nu_{\scriptscriptstyle\rm P}$ $\leq$ 2.33 for the equatorial profile and 2.52 $\leq$ $\nu_{\scriptscriptstyle\rm P}$ $\leq$ 2.55 for the polar profile.
 
\begin{table}
\centering
\caption{Comparison of various determinations of the power exponent $\nu_{\scriptscriptstyle\rm P}$ of the equatorial and polar photometric profiles.}
\begin{tabular}{ccccc}
\noalign{\smallskip}\hline\noalign{\smallskip}
Source	& Range		& $\nu_{\scriptscriptstyle\rm P}$ Equ		& $\nu_{\scriptscriptstyle\rm P}$ Pol 	& Reference	\\
\noalign{\smallskip}\hline\noalign{\smallskip}
K--L model		& 2\deg -- 50\deg			& 2.31	& 2.52	& \cite{KoutchmyLamy1985}  \\                                      
Clementine 		&	5\deg -- 20\deg			& 2.45  & 2.55  & \cite{Hahn2002} \\
SECCHI-HI-1		& 5\deg -- 24\deg			& 2.35	& 2.49 	& \cite{Stenborg2018} \\
WISPR-I				& 13.5\deg -- 54\deg	& 2.31	& - 		& \cite{Howard2019} \\
WISPR-I				& 13.5\deg -- 54\deg	& 2.30	& - 		& \cite{Stenborg2021} \\
LASCO-C3		  &	3\deg -- 8\deg			& 2.33  & 2.55  & This work \\
\noalign{\smallskip}\hline\noalign{\smallskip}
\end{tabular}
\label{tab:slope}
\end{table}

\subsection{Color of the F-corona}
\label{sec:color}
\begin{figure*}[htpb!]
\centering
\includegraphics[width=0.99\textwidth]{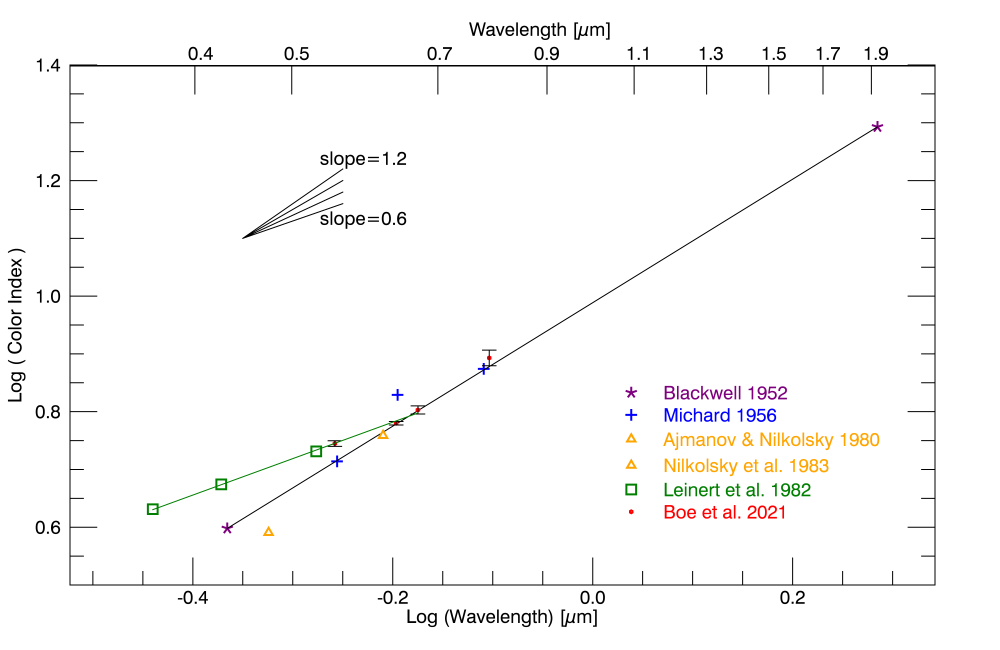}
\caption{The logarithm of the color indices  of the F-corona reported in the literature are displayed using a logarithmic scale for the wavelength.
The black line connects the two data points of \cite{Blackwell1952}.
The green line represents the linear regression (on the log--log scale) to the \cite{Leinert1982} zodiacal light data and to the \cite{Boe2021} F-corona data, excluding the value at 788.4\,nm.}
\label{fig:Color}
\end{figure*}

In the above section, the conversion of radiance units from \SZL to \Bsun assumed that the color of the F-corona is similar to that of the solar photosphere.
It is however appropriate to reconsider this assumption since several past works found that its spectrum exhibits a reddening at visible wavelengths.
Color data of the F-corona are extremely scarce and come exclusively from ground-based observations during solar eclipses.
Early attempts were briefly addressed by \cite{Blackwell1952} in his report of his own observations at the eclipse of 1952 of a considerable excess of infra-red radiation at 2.5\Rsun\ compared with that at 1.5\Rsun\ obtained by comparing the ratios of the radiances measured at 0.43 and 1.9\,$\mu$m at these two elongations.
The color index itself, that is the ratio of the F-corona radiances at the two wavelengths normalized to the photospheric ratio (${\rm B}_{\rm F}$/${\overline{{\rm B}_\odot}}$) at 2.5\Rsun\, was later derived by \cite{Michard1956} using photometric models of the K- and F-coronae and included in his Figure~8 together with his own measurements performed at three wavelengths, 556, 640, and 785\,nm and at $\approx$\,2\Rsun\ during the eclipse of 1955.
This figure displays the logarithm of the color index, arbitrarily normalized at a wavelength of 640\,nm, as a function of the inverse of the wavelength 1/$\lambda$. 
Ultimately, \cite{Michard1956} obtained a good agreement between his determinations and that derived from \cite{Blackwell1952} as described above. 
He further included the results of \cite{Allen1946} (improperly quoted as Allen 1940, this year being in fact that of the eclipse) as reported by \cite{Blackwell1952}, after applying the same procedure, but this raises concerns. 
Indeed, the original data set of \cite{Allen1946} (his Figure~7) displays considerable scatter from which it is difficult to extract a meaningful trend.
\cite{Allen1946} himself concluded in favor of a solar color whereas \cite{Michard1956} derived a red color. 
This confused situation explains why the \cite{Allen1946} result was excluded from the compilation performed by \cite{KoutchmyLamy1985}. 
This compilation includes more recent results, namely those of \cite{Ajmanov1980} and \cite{Nikolsky1983} obtained at the eclipses of 1972 and 1973, respectively and both at an elongation of $\approx$\,4\Rsun.
They are synthesized in their Figure~5 where the logarithm of the color indices are displayed as a function of wavelength $\lambda$, but using a linear variation in 1/$\lambda$. 
This corresponds to a flip of the wavelength axis of \cite{Michard1956}, but this offers the advantage of rendering the classical perception of the reddening as an increase of the color index with increasing wavelength.
Our own compilation (Figure~\ref{fig:Color}) further includes the recent results of \cite{Boe2021} obtained at the eclipse of 2019 and at four wavelengths between 529.5 and 788.4\,nm.
The F-corona was extracted from their images using a new inversion method and the reported color indices are averages over their whole \fov, that is up to an elongation of 3\Rsun. 
This is perfectly acceptable as detecting a putative color variation with elongation is realistically beyond the capability of the presently available data.
Figure~\ref{fig:Color} further includes the color index of the zodiacal light as summarized by \cite{Leinert1982} from measurements performed by the two \textit{Helios} spacecraft at 363, 425, and 529\,nm. 
These authors give the two corresponding color indices as slightly decreasing linear functions of the elongation and we used the constant values at zero elongation.
The uncertainties on the color index are only given by \cite{Boe2021} and by \cite{Leinert1982} for the zodiacal light ($\approx$3\,\%).
We suspect the uncertainties to be quite large in the case of the older photographic observations, or when the determinations were indirect such as the case of those of \cite{Blackwell1952}. 
The different data sets were all rescaled to an arbitrary common value at a wavelength of 640\,nm using interpolations or extrapolations when the required values were not available.
In the case of the data of \cite{Michard1956}, the interpolation used the two extreme data points as this led to a better global consistency with the other data sets.
Figure~\ref{fig:Color} displays the color index versus wavelength on a log-log scale suited to the determination of power laws.
We see that the straight line connecting the two data points of \cite{Blackwell1952} gives a good fit to the whole set of coronal values with a slope clearly steeper than that of the zodiacal light values.
This fit corresponds to a power law of the color index $\propto\,\lambda^{1.07}$ with only three outliers.
In summary, both F-corona and zodiacal light have colors redder than the Sun, the reddening being more pronounced for the F-corona.
However, an alternative trend emerges if we consider only the most recent data of \cite{Leinert1982} and \cite{Boe2021} as they can be reliably fitted by a shallower power law with an exponent of 0.63.
The extreme data point at $\lambda$ = 788.4\,nm of \cite{Boe2021} does depart from this law, but one may invoke a steepening of the reddening beyond approximately 650\,nm also supported by the value of \cite{Michard1956} at 785\,nm and that of \cite{Blackwell1952} at 1.9\,$\mu$m.
Of course, it may be argued that the \textit{Helios} data cannot be extrapolated to zero elongation but the consistency with the results of \cite{Boe2021} is striking.

Whatever the case, let us consider the impact of a coronal reddening on the photometric comparison between the radiance of the F-corona at 585\,nm as obtained by LASCO and that of the zodiacal light at 500\,nm. 
If we consider the first solution of the color index with a power exponent of 1.07, then a multiplicative factor of $(500/585)^{1.07}=0.85$ must be applied to the C2 and C3 data to scale them to 500 nm. 
In the case of the second solution with a power exponent of 0.63, the multiplicative factor becomes $(500/585)^{0.63}=0.91$.
This represents a 10\,\% correction, a value similar to the accuracy of the zodiacal light data in the bright regions \citep{Leinert1998} and of the LASCO F-corona data as discussed in Section~\ref{sec:comp}.
The first solution leads to a 15\,\% correction which is still modest.
Owing to the present large uncertainty affecting the color of the F-corona and to its rather limited impact, we choose to continue with the assumption of a solar color and the resulting conversion factor adopted in Section~\ref{sec:comp_ZL}.
We note that \cite{KoutchmyLamy1985} were on the same conservative line when they specified a spectral domain of 400 to 600\,nm for the validity of their model and in fact, its excellent photometric agreement with the LASCO data excludes a strong reddening in this spectral domain.

\section{Standard Model of the F-corona at 1\,AU}
\label{sec:model}
Having ascertained the reliability and the coherence of the photometry of the LASCO C2 and C3 images, we are now in a position to construct a composite giving, for the first time, a calibrated map of the F-corona from 2 to 30\Rsun.
It is based on the images obtained at the node of December 1997, corrected and scaled as described in the above sections, and therefore valid for an observer at a heliocentric distance of 1\,AU in the plane of symmetry of the zodiacal cloud.
Before proceeding, it was necessary to implement the corrections determined from the analysis of the photometric profiles (Section~\ref{sec:C2C3}). 
We subtracted the faint constant background of $1.3 \times 10^{-12}$\,\Bsun from the C3 images as it corrects the polar profile without affecting the equatorial one.
The case of C2 required a more elaborate treatment to progressively mitigate the correction of the polar profile, that is a down-scaling by a multiplicative factor of 0.9.
We constructed an image in polar coordinates [$\rho$,\,$\theta$] with $\theta$ ranging from 0\deg\ (polar direction) to 90\deg\ (equatorial direction) with a step of 1\deg.
Each radius had a constant value given by a linear function of $\theta$ varying from 0.9 at $\theta$\,=\,0\deg\ to 1 at $\theta$\,=\,90\deg.
This sector was replicated to cover the 360\deg\ range, transformed to Cartesian coordinates, and applied to the C2 image by simple multiplication.

\begin{figure}[htpb!]
\begin{center}
\includegraphics[width=0.9\textwidth]{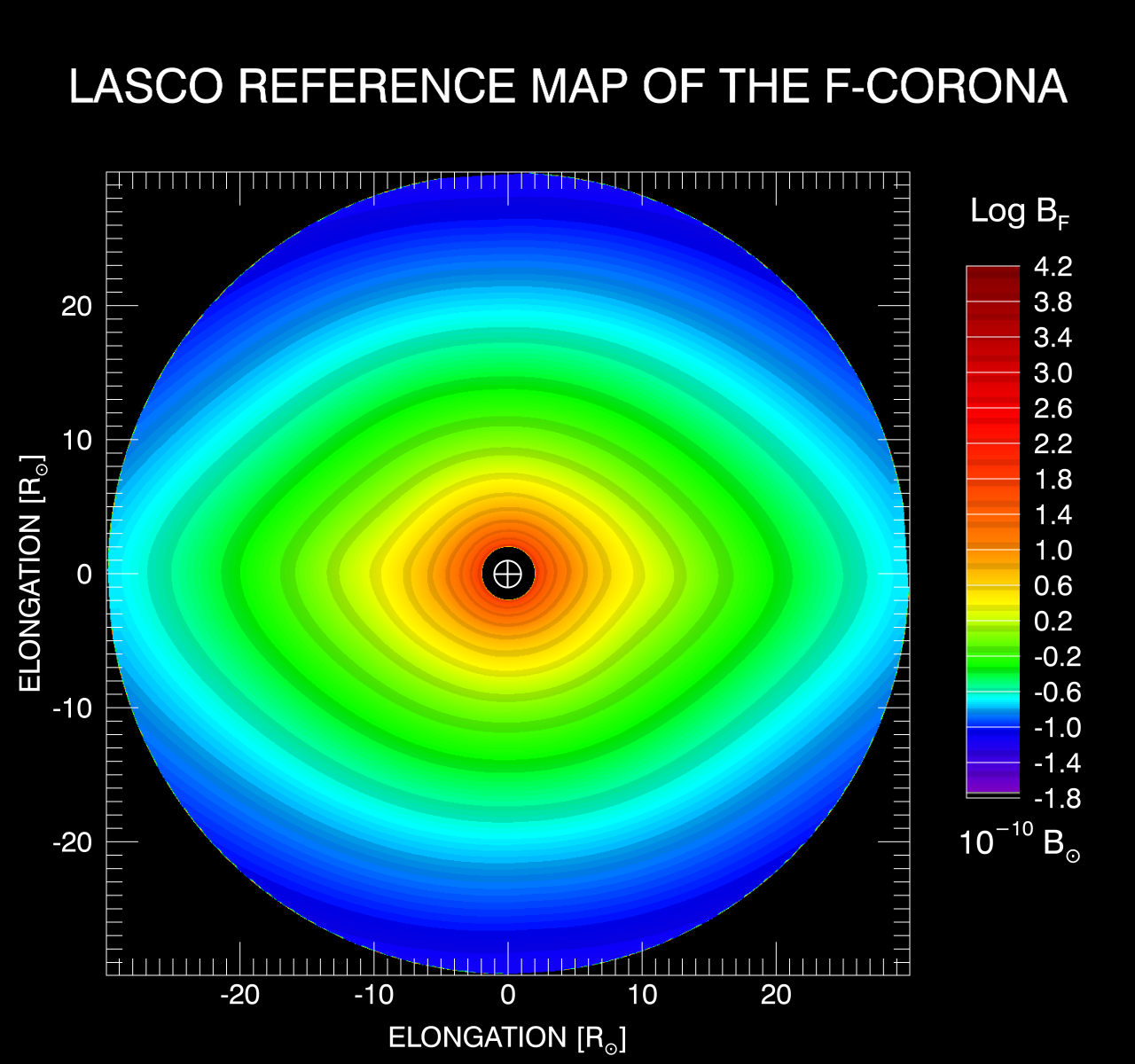}
\includegraphics[width=0.9\textwidth]{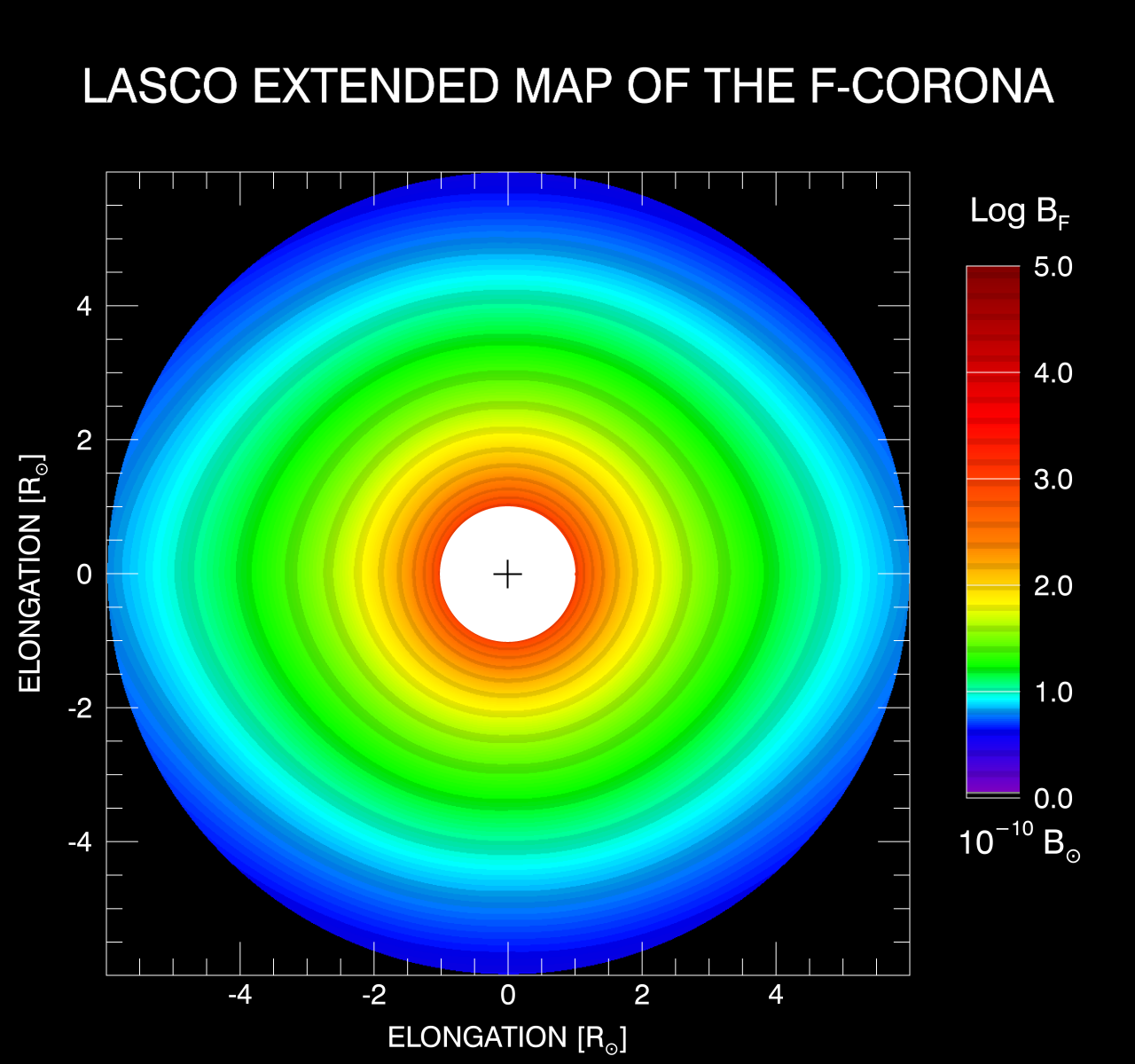}
\caption{Upper panel: LASCO reference map of the F-corona produced by a composite of the corrected C2 and C3 images obtained at the node of December 1997.
The black disk and the white circle centered on the Sun have a radius of 2 and 1\Rsun, respectively.
Lower panel: LASCO extended map of the F-corona produced by a composite of the Koutchmy--Lamy model and the above C2 image.
The white disk with a black cross at its center represents the solar disk.
The logarithm of the radiance is expressed in units of $10^{-10}$\,\Bsun and is coded according to the color bars.}
\label{fig:Fmap}
\end{center} 
\end{figure}

The composite was constructed in polar coordinates by introducing a frame of 14,250 $\times$ 720 pixels offering a radial scale of 500 pixels per solar radius and an angular scale of 2 pixels per degree.
The logarithm of the C2 and C3 images were transformed in polar coordinates and inserted in the frame, the juxtaposition taking place at 5.5\Rsun; the angular offset between the orientations of the C2 and C3 CCD detectors was taken into account.
A light smoothing was implemented over a transition region centered at 5.5\Rsun\ and 200 pixels wide by imposing a linear variation on each radial profile, a procedure that we have already used in the past \citep{Lamy2020}. 
The transformation to Cartesian coordinates was performed by imposing a scale of 17.103 pixels per solar radius similar to that of the C3 images, a \fov of 60 $\times$ 60\Rsun\ thus corresponding to a frame of 1026 $\times$ 1026 pixels, and an orientation such that the major axis of the ``elliptically'' shaped F-corona is horizontal.
This produced the LASCO reference map of the F-corona extending from 2 to 30\Rsun\ and displayed in Figure~\ref{fig:Fmap}.

We went one step further to fill the innermost region from 1 to 2\Rsun\ taking advantage of two favorable circumstances: i) the circularity of the isophotes at elongations $\lesssim$\,1.5\,\Rsun\ (see more detail in Section~\ref{sec:flattening} below), and ii) the excellent agreement between the C2 and the Koutchmy--Lamy radial profiles (Figure~\ref{fig:Prof_C2C3_55550_55915}). 
The composite was constructed in polar coordinates by introducing a frame of 2500\,$\times$\,720 pixels offering a radial scale of 500 pixels per solar radius and an angular scale of 2 pixels per degree.
The logarithm of the identical equatorial and polar profiles of the Koutchmy--Lamy model was replicated and inserted in the frame in the region extending from 1 to 1.5\Rsun.
The logarithm of the C2 image was transformed in polar coordinates and inserted in the frame in the region extending from 2 to 6\Rsun. 
The intermediate region from 1.5 to 2\Rsun\ was thus left blank at this stage to allow for some flexibility in the transition of the shape of the F-corona from circular to ``elliptical''. 
A linear interpolation on a log-log scale was applied on each radial profile to then bridge the two parts.
The transformation to Cartesian coordinates was performed by imposing a scale of 80.637 pixels per solar radius similar to that of the C2 images, a \fov of 12\,$\times$\,12\Rsun\ thus corresponding to a frame of 968\,$\times$\,968 pixels, and an orientation such that the major axis of the ``elliptically'' shaped F-corona is horizontal.
This produced the LASCO extended map of the F-corona from 1 to 6\Rsun\ displayed in Figure~\ref{fig:Fmap}.

These LASCO reference and extended maps of the F-corona are presented in tabular form in Appendix D; they will hopefully contribute to completing Tables~13.7 and 13.8 of \cite{Cox2000}.
Similar to the images, the two tables are subject to the following restrictions.
\begin{itemize}
	\item 
	They are valid for an observer at a heliocentric distance of 1\,AU in the plane of symmetry of the zodiacal cloud.	
	\item
	They are valid for approximately the first ten years of LASCO operations; beyond there are evidences of a slow variation of the general level of the radiance of the F-corona that will be discussed in Section~\ref{sec:stability}.
	\item
	They hold in the spectral range 450--600\,nm.
	\item
	The solar radius \Rsun\ must be considered as an absolute unit equal to 959.63 arcsec, that is 0.267\deg.
\end{itemize}

\section{Geometric Properties of the F-corona}
\label{sec:geometry}

\subsection{Center of the F-corona}
\label{sec:center}
The question of the center of the F-corona, of the zodiacal light and in turn, of the zodiacal cloud has not been a concern for a long time.
Regarding the F-corona, we are not aware of any observation or study in the relevant literature mentioning or even suggesting a deviation from the center of the Sun.
The same is true for the zodiacal light and this point was not even considered for instance in the series of articles by Dumont and co-workers (\eg \cite{Dumont1978}) nor in the detailed review of \cite{Leinert1990}.
The question of the off-centering of the zodiacal cloud, and consequently its asymmetries, arose from infrared surveys and notably from the first one performed by the IRAS satellite.
The discovery of a ``trailing/leading'' asymmetry was successfully interpreted in terms of an Earth's resonant ring (\cite{Dermott1984}; \cite{Dermott1985}). 
Briefly, asteroidal dust particles trapped in resonances with the Earth's mean motion form a toroidal circumsolar ring that contains the Earth embedded in a cavity and a local cloud or blob that trails the Earth in its orbit.
This was later confirmed and refined by the surveys performed by the follow-on infrared observatories COBE (\eg \cite{Dermott1996}; \cite{Kelsall1998}) and ARIKI (\eg \cite{Pyo2010}; \cite{Kondo2016}).
However, after removing the contribution of the ring, an asymmetry persisted in the survey maps and was interpreted in terms of an offset of the center of symmetry of the zodiacal cloud with respect to the center of the Sun due to the common forced eccentricities of the dust particle orbits \citep{Dermott1985}, very much like the forced inclinations result in an inclination of the plane of symmetry of the cloud with respect to the ecliptic (Section~\ref{sec:plane}). 
From the above surveys, an offset of 0.013 to 0.017\,AU was determined in the ecliptic plane with a much smaller value ($\approx$\,0.002\,AU) in the perpendicular direction.
A word of caution is warranted at this stage as infrared observations were performed at large elongations ($\approx$\,90\deg\ for IRAS and ARIKI, between 60\deg\ and 120\deg\ for COBE) so that they probe the outer zodiacal cloud beyond 1\,AU.
In addition, the different wavelength channels of the infrared observatories from 1.25 to 240\,$\mu$m probe different populations and different regions since the temperature of the dust grains depends upon their size and heliocentric distance.
Any geometric parameters of the cloud such as center and plane of symmetry may therefore change when derived from observations at different wavelengths.
Whatever the case, an average offset of 0.015\,AU -- that is 3.2\Rsun\ -- from the above results is so large that it is clearly inconceivable for the F-corona. 
This most likely implies that the zodiacal cloud at large does not have a unique center or, more precisely, has a progressively varying center when viewed from different directions probing different regions of the cloud. 
The immediate analogy is with the symmetry plane of the cloud known to be warped in the sense that there does not exist a single plane, but a surface experiencing a progressive deformation with increasing distance from the Sun (Section~\ref{sec:plane}).
  
Looking closer in, both \textit{Clementine} and SECCHI-HI have imaged the inner zodiacal light at elongations ranging from 5\deg\ to 25\deg.
As already mentioned in Section~\ref{sec:comp_ZL}, the \textit{Clementine} profiles exhibit an east--west asymmetry that could suggest an offset, but this was not quantified by \cite{Hahn2002}.
A similar east--west asymmetry appeared in the SECCHI HI-1 observations interpreted as an offset of $\approx$\,0.5\Rsun\ \citep{Stenborg2017} or $\approx$\,0.4\Rsun\ \citep{Stenborg2018}.
In contrast, such an asymmetry is absent in the LASCO equatorial profiles up to an elongation of 8\deg as conspicuously shown in the upper panel of Figure~\ref{fig:Prof_C2C3_50803}.
Furthermore, quasi perfect symmetry was also verified at other nodes, notably at those of December 2010 and 2011 used in our photometric analysis, thus excluding an offset of nearly half a solar radius that would result in a huge unbalance between the opposite profiles.
However, the display of the profiles on a logarithm scale over more than four decades may conceal minute asymmetries between the east and west sides and we used a more sensitive approach for a deeper look at this question.

This new approach relies on the monitoring of the coronal radiance integrated in two half-rings, east and west, an extension of the method presented in Section~\ref{sec:method}. 
A complete ring was simply separated in two parts by the vertical line (the column direction on the CCD detectors) passing through the center of the Sun.
Using half-rings allows mitigating the effect of the waddling of the F-corona as invoked in the case of the elongated windows in Section~\ref{sec:method}.
The rings themselves are located in the outer parts of the C2 and C3 \fovs\ where any offset was expected to be maximized; their inner and outer limits are 4.7 and 5.9\Rsun\ for C2 and 9.7 and 28.4\Rsun\ for C3.
The integrated radiances were normalized with respect to their respective areas, \ie the number of pixels in rings or half-rings.
The integrals in the rings were further used to normalize those in the half-rings to remove the periodic effect of the varying Sun-SoHO distance. 
A median filter of five pixels (\ie five days) was finally applied to smooth the temporal variations and facilitate the detection of any offsets on the difference curve ``east minus west''.

\begin{figure*}[htpb!]
\centering
\includegraphics[width=0.95\textwidth]{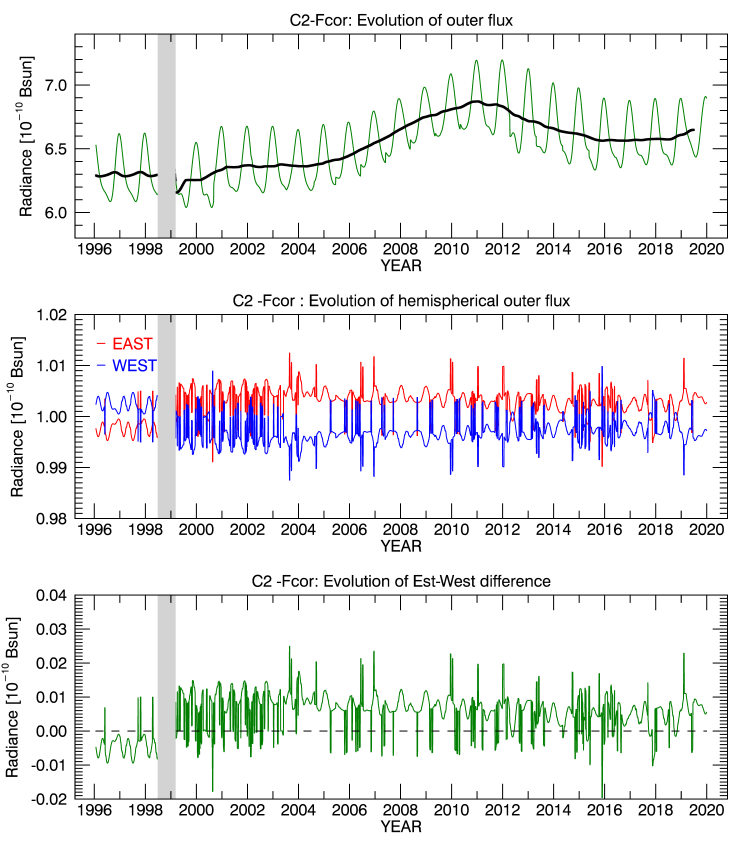}
\caption{Upper panel: temporal variation of the radiance of the C2 images of the F-corona integrated in the outer ring extending from 4.7 to 5.9\Rsun\ (green curve) and its running average over an interval of three years (black curve).
Middle panel: same as the upper panel for the east and west half-rings.
Lower panel: temporal  variation of the difference between the radiances integrated in the east and west half-rings.}
\label{fig:C2_half_rings}
\end{figure*}

\begin{figure*}[htpb!]
\centering
\includegraphics[width=0.95\textwidth]{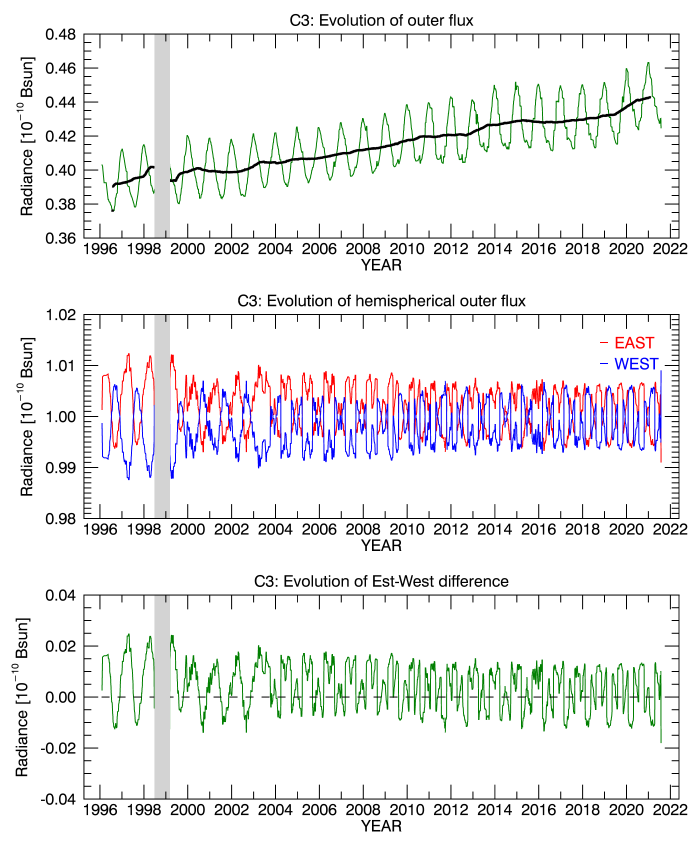}
\caption{Upper panel: temporal variation of the radiance of the C3 images of the F-corona integrated in the outer ring extending from 9.7 to 28.4\Rsun\ (green curve) and its running average over an interval of three years (black curve)
Middle panel: same as the upper panel for the east and west half-rings.
Lower panel: temporal  variation of the difference between the radiances integrated in the east and west half-rings.}
\label{fig:C3_half_rings}
\end{figure*}

Our results are presented in Figures~\ref{fig:C2_half_rings} and \ref{fig:C3_half_rings} for C2 and C3, respectively. 
The oscillations present in the temporal evolutions of the half-rings result from the fact that the fixed vertical division between the east and west sides does not rigorously track the waddling of the F-corona. 
Their amplitudes are typically $\pm$\,0.5\,\% for C3 and even less, $\pm$\,0.3\,\%, for C2 where the F-corona is more circular, thus reducing the influence of the waddling.
In the case of C3, the offset between the east and west half-rings is negligibly small and even null beyond 2010 as confirmed by the temporal evolution of the difference; this trend most likely results from the change of the orientation of SoHO from solar north to ecliptic north.
The mean value of the difference amounts to an incredibly low value of 4 $\times$ $10^{-13}$\Bsun.
In the case of C2, the offset between the east and west half-rings is systematic, but exhibits a reversal at the time of the loss of SoHO.
It is anyway extremely small with a mean value of the difference of 5 $\times$ $10^{-13}$\Bsun.
Translated in terms of a spatial offset between the centers of the Sun and the F-corona, this led to an upper limit of 0.03\Rsun.
However, we are convinced that the observed unbalance between the east and west half-rings are inherent to the procedure of centering the C2-Fcor images as discussed by \cite{Llebaria2021}, and that altogether C2 and C3 do not detect any unambiguous offset between the centers of the Sun and of the F-corona.
The C3 result in particular whose \fov partly overlaps that of SECCHI HI-1 clearly excludes the offset of 0.4 to 0.5\Rsun\ reported by \cite{Stenborg2017} and \citep{Stenborg2018}.

There are substantial differences between the LASCO on the one hand and the \textit{Clementine} and SECCHI HI-1 observations on the other hand that may explain the conflicting results.
Whereas LASCO-C2 and C3 secures full images of the corona, those of \textit{Clementine} and SECCHI HI-1 are mosaics of several fields (seven for the former, eight for the latter), a reconstruction which is prone to slight centering errors. 
Another aspect is that different stray light patterns in the different fields may also cause an artificial asymmetry. 
Stray light is indeed of major concern in this question and a relevant example is offered by the C3 observations: without the availability of images taken at opposite roll angles, it would have been impossible to detect the faint east--west stray light ramp and we would have incorrectly concluded on an east--west asymmetry in the C3 images of the F-corona.

\subsection{Flattening of the F-corona}
\label{sec:flattening}
Historically, the flattening of the corona was characterized by the Ludendorff index $\epsilon$ \citep{Saito1956} which considers altogether the K and F components.
Following the progress in the analysis of ground-based eclipse observations, in the separation of these two components, and the realization of the connection between the F-corona and the (flattened) zodiacal light, the picture emerged of a circular F-corona at small elongations, but becoming more and more flattened with increasing elongations, \eg \cite{Blackwell1955} and \cite{Saito1956}.
Later on and exploiting Skylab-ATM observations, \cite{Saito1977} derived two photometric profiles of the F-corona, equatorial and polar (their Figure~7), from which they derived the ``ellipticity'' defined by:
\begin{equation}
\epsilon_ {\scriptscriptstyle\rm S} = \frac{R_{\scriptscriptstyle\rm equ} - R_{\scriptscriptstyle\rm pol}}{R_{\scriptscriptstyle\rm equ}} = 1 - \frac{R_{\scriptscriptstyle\rm pol}}{R_{\scriptscriptstyle\rm equ}}
\label{eq:epsilon_saito}
\end{equation}
``where $R_{\scriptscriptstyle\rm equ}$ and $R_{\scriptscriptstyle\rm pol}$ are the radius vectors along the equator and pole corresponding to identical F-corona brightnesses''.
Note that we modified the original notation $\epsilon$ of \cite{Saito1977} to $\epsilon_{\scriptscriptstyle\rm S}$ to be more specific.
The ``ellipticity'' increases from $\approx$\,0 at 2\Rsun\ to $\approx$\,0.12 at 5.3\Rsun\ (their Figure~8).

\cite{KoutchmyLamy1985} introduced a new definition of the flattening index $\epsilon_{\scriptscriptstyle\rm F}$ via:
\begin{equation}
\epsilon_{\scriptscriptstyle\rm F} = \frac{R_{\scriptscriptstyle\rm equ}}{R_{\scriptscriptstyle\rm pol}}- 1
\label{eq:epsilon_kl}
\end{equation}
and compiled available flattening data until 1984 (their Figure~2) which exhibit considerable dispersion.
They pointed out the difficulties of correctly separating the K and F components and, in the case of ground-based measurements, of properly correcting for the sky background.
Eventually, they proposed a solution for the variation of $\epsilon_{\scriptscriptstyle\rm F}$ which, when plotted as a function of the logarithm of the elongation, is characterized by an approximate linear behaviour up to the connection with zodiacal light data.

In addition to the above difficulties and to the different definitions of the flattening, the quantities $R_{\scriptscriptstyle\rm equ}$ and $R_{\scriptscriptstyle\rm pol}$ are not consistently calculated by various authors.
Strictly speaking and as implemented by \cite{Saito1977}, they must be recorded along the equatorial and polar directions, but are sometime averaged over sectors of typically $\pm$\,20\deg centered on these directions (\eg \cite{Koutchmy1984}).
This has a mild effect on $R_{\scriptscriptstyle\rm pol}$ since the isophotes are quite flat over the poles, but dramatic on $R_{\scriptscriptstyle\rm equ}$ since the curvature of the isophotes is much more pronounced; the net impact is to systematically reduce $\epsilon_{\scriptscriptstyle\rm F}$ in comparison with its strict definition.

A final caveat is that the observer is not always located in the PSZC resulting in a prominently north-south asymmetry of the corona, but this distortion is generally alleviated by combining east--west and north--south profiles.
To make the situation even more problematic, \cite{Stauffer2018} introduced a pseudo-flattening index where in Equation~\ref{eq:epsilon_kl}, $R_{\scriptscriptstyle\rm pol}$ is replaced by $R$(25\deg) the radius measured at 25\deg\ from the equatorial direction which they tentatively related to the true flattening index.

The latest and most valuable contribution to-date come from SECCHI/HI-1A images taken between December 2007 and March 2014 and analyzed by \cite{Stenborg2018}.
For a set of isophotes extending over a range of elongation from 6\deg\ to 24\deg, they measured the lengths of their major and minor axes, calculated the flattening index according to Equation~\ref{eq:epsilon_kl} and found that it varies from 0.46 to 0.66 in the above range of elongation.

It is important to realize that the flattening gives only a very basic and limited characterization of the isophotes of the F-corona. 
As we will see later, the relevant information is contained in the shape of the isophotes which constrains the 3D spatial density of interplanetary dust.
However, we do consider it as the LASCO data fill a gap in the existing data and shed new light on its variation with elongation and the connection to the zodiacal light.

\begin{figure*}[htpb!]
\centering
\includegraphics[width=0.95\textwidth]{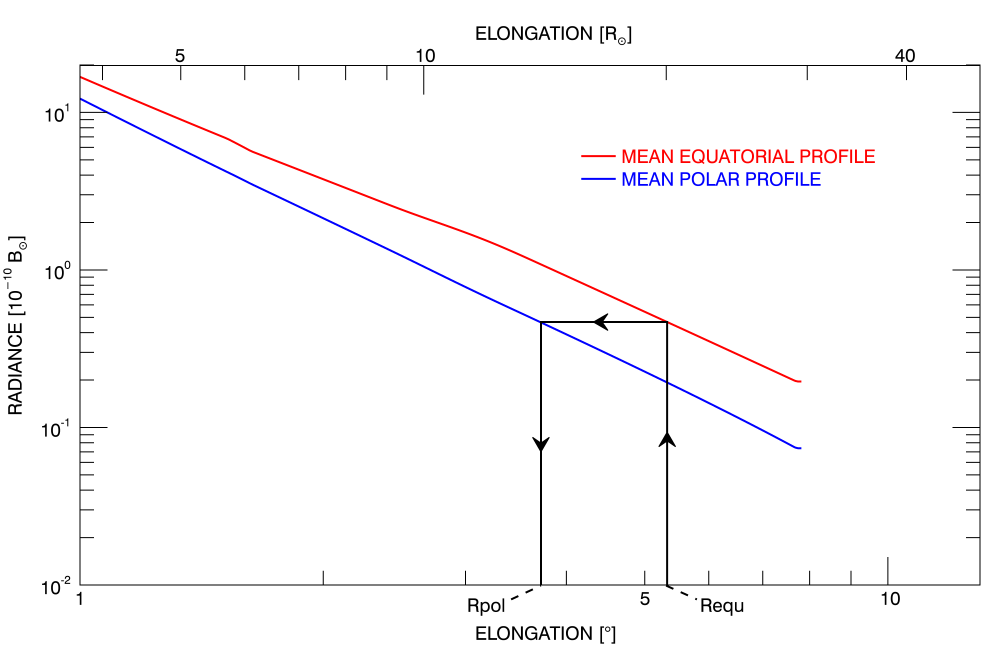}
\caption{Schematic of the procedure to calculate the flattening index from the equatorial and polar profiles.}
\label{fig:flat_scheme}
\end{figure*}

We implemented the strict definition given by \cite{KoutchmyLamy1985}, but following the suggestion of \cite{Stenborg2018}, replaced their notation by $f$ to avoid confusion with the elongation classically denoted by $\epsilon$.
The flattening index was calculated from the mean equatorial and polar photometric profiles constructed in Section~\ref{sec:C2C3} obtained from observations at the node of December 1997 and at the combined nodes of December 2010+2011 thus ensuring a symmetric configuration of the F-corona.
The method illustrated in Figure~\ref{fig:flat_scheme} is of utmost simplicity: a given $R_{\scriptscriptstyle\rm equ}$ specifies the radiance level on the equatorial profile which is translated to the polar profile specifying in turn $R_{\scriptscriptstyle\rm pol}$.

\begin{figure*}[htpb!]
\centering
\includegraphics[width=0.95\textwidth]{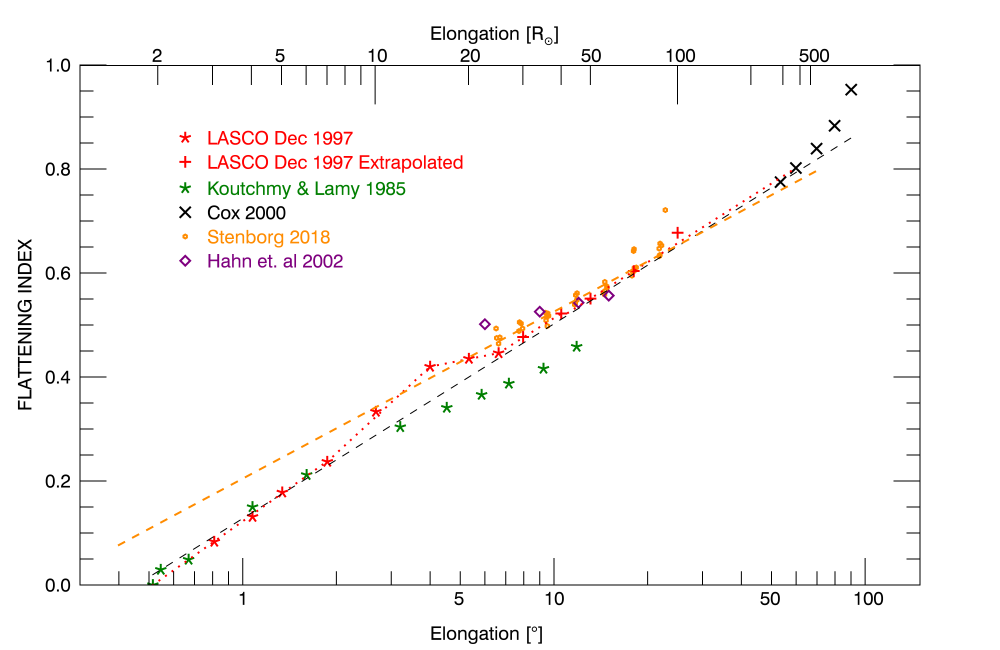}
\caption{Variation of the flattening index of the F-corona as a function of the logarithm of elongation. 
The LASCO results correspond to the node of December 1997 with additional values extrapolated beyond an elongation of 8\deg.
The dashed black line represents the linear fit to the LASCO data excluding the three data points between 2\deg\ and 6\deg.
The red dotted line corresponds to our optimal solution with four different linear fits.
The dashed orange line represents the linear fit to the \cite{Stenborg2018} data.}
\label{fig:flat}
\end{figure*}

The LASCO results for the December 1997 are displayed in Figure~\ref{fig:flat} on a log-lin diagram where the flattening index is plotted as a function of the logarithm of the elongation, further distinguishing the results derived from the profiles extrapolated beyond 30\Rsun.
For clarity, we omitted the results for the case of December 2010+2011 since they are strictly similar, the maximum difference between the two cases amounting to merely 0.3\,\%. 
Using the same method, we calculated the flattening index from the other profiles displayed in the same figure, namely those of \cite{KoutchmyLamy1985} (we did not use their Figure~2), and \cite{Hahn2002}. 
That of the zodiacal light was calculated from the radiance data along the plane of symmetry and along the perpendicular direction \citep{Cox2000}.
Finally, we directly incorporated the results of \cite{Stenborg2018} using their Figure~12.

The LASCO results are well fitted by a linear function of log($f$) except for three data points lying between 2\deg and 6\deg.
These three ``outliers'' are clearly associated with the shoulder in the equatorial profile highlighted in Section~\ref{sec:C2C3}. 
Splitting the linear fit in four sections (the red dotted line in Figure~\ref{fig:flat}) yields our optimal solution for the flattening index of the F-corona which further connects extremely well to that of the zodiacal light. 
The intercepts of the two fits at $f$=0 signaling perfect circularity of the isophotes are almost identical and correspond to an elongation of 0.5\deg\ $\pm$ 0.01\deg\ (1.9\Rsun), a value found by several authors, notably \cite{Saito1977} from Skylab-ATM observations.  

The results of \cite{Stenborg2018} are globally in good agreement with our solution, but there are clearly a substantial number of ``outliers''.
We calculated a linear fit to their data and found that its slope and its intercept at 0.23\deg\ (0.86\Rsun) markedly deviate from our solution.
In fact, \cite{Stenborg2018} showed that their data are best fitted by 2nd and 3rd order polynomials and that a realistic solution could only be obtained by a bootstrap linear fit.
This procedure considers that there are unknown errors in elongation and flattening index and it performs different linear fits on several hundred sub-samples of the data set.
Its extrapolation to $f$=0 yielded an intercept of 0.24\deg $\pm$ 0.01\deg\ (0.90\Rsun), close to the value we obtained above with the simple global linear fit, but altogether smaller than our result by a factor of two and implying that circularity of the isophotes is reached \textit{inside} the projected solar disk.

The variation of the flattening index derived from the profiles of the \cite{KoutchmyLamy1985} model is in excellent agreement with our result up to an elongation of 3\deg, but strongly deviates beyond.
This is readily explained by the absence of the shoulder in their equatorial profile so that, for a given radiance, their $R_{\scriptscriptstyle\rm equ}$ is systematically smaller than the LASCO $R_{\scriptscriptstyle\rm equ}$, henceforth a smaller index.
Altogether, the convergence of the extrapolated LASCO flattening to that of the zodiacal light and the agreement between the LASCO results and those of \cite{Stenborg2018} strongly support the reality of the shoulder in the equatorial profile. 
A couple of data points from the profiles of \cite{Hahn2002} closely agree with our solution, but the general trend is way off, a consequence of their equatorial profile being too steep as pointed out in Section~\ref{sec:comp_ZL}. 
 

\subsection{Shape of the Isophotes of the F-corona}
\label{sec:shape}
Up to now, we referred to the shape of the F-corona as ``elliptical'' as given by its visual perception, the ellipticity varying with elongation as characterized in the above section.
A strict ``elliptical'' model was developed by \cite{Fang1991} based on the two extreme profiles given by the Koutchmy--Lamy model.
Their purpose was not only to model the isophotes, but to construct realistic images of the F-corona, unavailable at that time, in preparation of the LASCO observations. 
For our present purpose of characterizing the shape of the isophotes, we limited ourselves to a set of five of them spanning the LASCO \fov and tested how different models compare with the LASCO reference map.
These five isophotes are defined by their equatorial elongations $R_{\scriptscriptstyle\rm equ}$ spanning the  5\,--\,25\Rsun\ interval with a step of 5\Rsun.

The case of the elliptical isophotes based on the K--L model is straightforward since a specified $R_{\scriptscriptstyle\rm equ}$ defines a corresponding $R_{\scriptscriptstyle\rm pol}$ following the scheme of Figure~\ref{fig:flat_scheme}, these two values defining the semi-major and semi-minor axes of the ellipse.

We next considered the two-dimensional model of Munro that resulted from space observations with Skylab ATM.
The publication itself, an HAO internal report dated 1985, is not easily accessible, but the formulation of the model can be found in \cite{Burkepile2017}.
The radiance in units of $10^{-10}$\Bsun is expressed as a function of the position angle $\theta$ reckoned from the polar direction and of the elongation from the center of the Sun $R$ in units of \Rsun\ by:
$${\rm B}_{\rm F}(R,\,\theta)=473.96 \times (1+0.23732 \times {\rm sin}^2(\theta)) \times R^{-2.72375}.$$
The equation of an isophote $R(\theta)$ for a given $R_{\scriptscriptstyle\rm equ}$ can be directly derived from this expression via:
$$ {\rm Log}(R) = {\rm Log}(R_{equ}) - 0.36714 \times {\rm Log}[1.23732/({1 + 0.23732 \times {\rm sin}^2(\theta)})]. $$
According to Munro, his model becomes increasingly less reliable below 2.3\Rsun, but this is not a concern as our first isophote starts at 5\Rsun.   

The analysis of the SECCHI/HI-1 images led \cite{Stenborg2018} to propose super-ellipses as better suited to representing the isophotes of the F-corona.
When centered and oriented with the (equatorial) major axis along the x-axis, the super-ellipses are defined in Cartesian coordinates (x,\,y) by the following equation:
$$ ({\rm x}/a)^n + ({\rm y}/b)^n = 1, $$ 
where $a$ and $b$ are the super-ellipse semi-major and semi-minor axes, respectively, and $n$ is an exponent that defines the overall shape of the super-ellipses.
\cite{Stenborg2018} showed that the exponent $n$ is linked to the flattening index and we used the green curve of their Figure~15 representing a third-degree polynomial of their proposed relationship between $n$ and the flattening index.
Then the procedure to construct the isophote specified by $R_{\scriptscriptstyle\rm equ}$ proceeded as follows.
We determined the corresponding $R_{\scriptscriptstyle\rm pol}$ using the LASCO photometric profiles of Figure~\ref{fig:Prof_C2C3_50803}, the scheme of Figure~\ref{fig:flat_scheme}, and the flattening index from Figure~\ref{fig:flat}.
We then read the exponent $n$ from Figure~15 of \cite{Stenborg2018} and setting $a=R_{\scriptscriptstyle\rm equ}$ and $b=R_{\scriptscriptstyle\rm pol}$ completed the set of required parameters to draw the super-ellipses.

\begin{figure*}[htpb!]
\centering
\includegraphics[width=\textwidth]{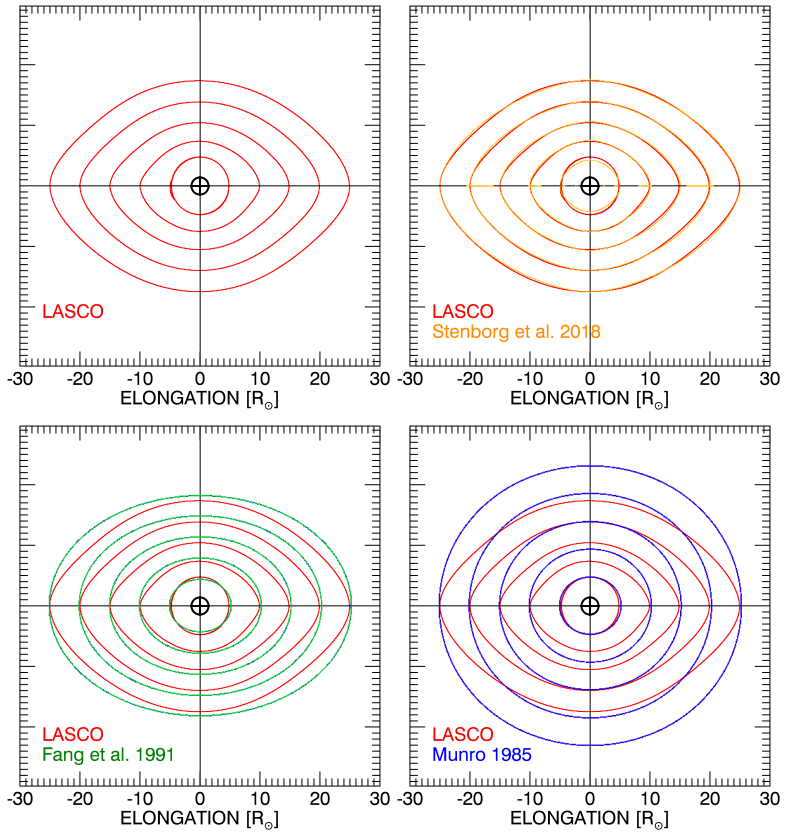}
\caption{Comparison of a set of isophotes extracted from the LASCO reference map of the F-corona with different models: elliptical model of \cite{Fang1991}, 2D model of Munro, and super-elliptical model of \cite{Stenborg2018}.}
\label{fig:C3_contours}
\end{figure*}

Figure~\ref{fig:C3_contours} displays the comparison of the selected isophotes of the LASCO reference map with those of the above three models. 
The contours produced by the model of Munro are by far the worst being much too circular.
The elliptical contours produced by the model of \cite{Fang1991} fare better, but are surpassed by the super-ellipse model of \cite{Stenborg2018}.

Whereas, this is a noteworthy achievement, the scientific usefulness of finding an analytic description of the isophotes remains questionable. 
It does not help at all in the derivation of the 3D distribution of the density of interplanetary dust which presently can only be tackled by forward modeling: starting from a priori distributions, tests and trials are conducted until the calculated radiance map matches the observed one (\cite{Giese1986}; \cite{Giese1989}).
However, a possible application would consist in generating maps of the F-corona as attempted by \cite{Fang1991}, that would be useful when only K+F observations are available in order to retrieve the K-corona by subtraction.
This is however quite challenging as we have seen that the very appearance of the F-corona depends upon the location of the observer, that is the time of the year for a ground-based observer. 
A specific difficulty with the super-ellipse model of \cite{Stenborg2018} arises from the derivation of the exponent $n$ from its relationship with the flattening index that does not appear very robust.
Indeed, \cite{Stenborg2018} conceded that their model is not fully accurate as deviations are found toward the north and south poles, the modeled isophotes appearing almost elliptical -- and not super-elliptical -- in those regions (their Figure~13).
Rather than modeling the shape of the isophote, a more useful approach would consist in deriving a reliable 3D distribution of the dust density, which in itself would be scientifically valuable, so as to then calculate radiance maps for any location in the solar system.

\section{Plane of Symmetry of the Inner Zodiacal Cloud}
\label{sec:plane}
The question of the plane of symmetry of the zodiacal cloud characterized by the longitude of its ascending node $\Omega_A$ and its inclination $i$ has long been a matter of debate between the observers of the zodiacal light as illustrated by the following two examples.
From their all-sky ``Tenerife'' survey and specifically from the radiance in quadrature with the Sun obtained in June and December 1966 (\ie when the Earth was close to the putative plane of symmetry), \cite{Dumont1968} found the best agreement with the invariable plane of the solar system.
Ten years later, from their observations at elongations of 32\deg\ to 50\deg\ at Mount Haleakala, \cite{Misconi1978} pointed out that the photometric axis of their isophotes was closer to the orbital plane of Venus than to the invariable plane. 
However, these observations took place in February, March, and April (their illustration is for 17 March 1966) when the north-south asymmetry is maximum and this may have (partly) biased their determination.
Whatever the case, they suggested that ``there may be a multiplicity of symmetry planes, that is the position of the symmetry plane may vary with the distance to the Sun'', an idea which was later developed by \cite{Misconi1980}. 

Concentrating now on the inner zodiacal cloud and F-corona, \cite{MacQueen1968} performed a balloon experiment on 9 January 1967 (hence, very close to the December node) that scanned the corona at 2.2\,$\mu$m between 5 and 8\Rsun. 
He found that the angular position of the interpolated peak of their isophotes coincided closely with the invariable plane.
A different conclusion was however reached by \cite{MacQueen1973} from their analysis of photographs made from the command module of Apollo 16 on 22--23 April 1972 and covering elongations 8 to 40\Rsun.
This time, they found that the peak of their isophotes were displaced north of the ecliptic by 3\deg\,$\pm$1\deg, at least in the region beyond 20\Rsun.
However, they recognized that the dates of the observations favored the north--south asymmetry and finally concluded that ``the situation is hardly clear''.

From their rocket observations performed on 2 July 1971 (hence, very close to the June node) at elongations of 15\deg, 21\deg, and 30\deg, \cite{Leinert1974} noticed that the line of maximum brightness was inclined with respect the ecliptic, with deviations of -3.7\deg\,$\pm$\,0.5\deg\ west and +2.5\deg\,$\pm$\,0.5\deg\ east of the Sun.
In a follow-on article, \cite{Leinert1976} built a 3D model of the zodiacal cloud and its brightness, and they found that their rocket observations and other observations as well were best fitted by a plane of symmetry with $\Omega_A$\,=\,66\deg\,$\pm$\,11\deg\ and $i$\,=\,3.7\deg\,$\pm$\,0.6\deg, ``implying a significant deviation from the invariable plane towards the planes of the inner planets and the solar equator''.

Next came the \textit{Helios} observations: although the two \SC\ were active during several years, only the data covering one orbit of each of \textit{Helios~1} (DOY 240/1975 to 70/1976) and \textit{Helios~2} (DOY 72 to 240/1976) chosen for minimum changes in spin axis orientation were analyzed for the determination of the PSZC.
Using suitable observing geometries, \cite{Leinert1980} reached their goal without recourse to model calculations, unlike the previous attempt with the rocket flight. 
This led to a substantial revision of the longitude of the ascending node to $\Omega_A$\,=\,87\deg\,$\pm$\,4\deg\ and a slight reduction of the inclination to $i$\,=\,3.0\deg\,$\pm$\,0.3\deg.

\cite{Stenborg2017} analyzed SECCHI/HI-1 images of the eastern side of the Sun taken between 2007 and 2012 to characterize the plane of symmetry of the inner zodiacal cloud in the elongation range 5\deg\ to 24\deg.
Both inclination and ascending node were found to be constant with $\Omega_A$\,$\approx$\,83\deg\ and $i$\,$\approx$\,3.7\deg\ in the outer half of the \fov, then evolving to reach $\Omega_A$\,$\approx$\,57\deg\ and $i$\,$\approx$\,6\deg\ at about 5\deg\ elongation confirming the intuition of \cite{Leinert1976}.

To complete the picture, we come back to the outer zodiacal cloud as observed in the infrared by the three satellites IRAS, COBE, and IRAKI whose surveys were extensively modeled by \cite{Rowan2013}, \cite{Kelsall1998}, and \cite{Kondo2016}, respectively.
Their solutions for the symmetry plane together with the above results for the inner zodiacal cloud are regrouped in Table~\ref{tab:inc_omega}.
It is interesting to note that, whereas the inclination of the outer cloud is indeed close to that of the invariable plane ($\approx$\,2\deg\ versus 1\deg), this is not the case of its longitude of the ascending node whose narrow range of 76\deg\ to 78\deg\ significantly deviates from that of the invariable plane (107\deg).
In fact, it is curiously much closer to that of Venus (76\deg) and even to the solar equator (75\deg).

\begin{table}
\centering
\caption{Comparison of the longitudes of the ascending node $\Omega_A$ and of the inclinations $i$ of the equatorial plane of the Sun, of the orbital planes of planets, and of the plane of symmetry of the zodiacal cloud according to different sources.}
\begin{tabular}{ccccc}

\noalign{\smallskip}\hline\noalign{\smallskip}
Source				& Range	(\deg)					& $\Omega_A$ (\deg)		& $i$ (\deg)					& Reference	\\

\noalign{\smallskip}\hline\noalign{\smallskip}
\multicolumn{5}{c}{SUN \& PLANETS} \\
Solar Equ.					& -								& 75									& 7.3									& - \\
Mercury							& -								& 48									& 7.0									& - \\
Venus								& -								& 76									& 3.4									& - \\
Mars								& -								& 50									& 1.85								& - \\
Jupiter							& -								& 100									& 1.3									& - \\
Inv. Plane					& -								& 107									& 1.6									& - \\
\noalign{\smallskip}\hline\noalign{\smallskip}

\multicolumn{5}{c}{VISIBLE AND NEAR-INFRARED OBSERVATIONS} \\
\noalign{\smallskip}\hline\noalign{\smallskip}
Balloon				& 1.3 -- 2.1		& -										& $\approx$\,1.6 			& \cite{MacQueen1968} \\
Apollo 				& 2 -- 10				& -										& 3\,$\pm$1						& \cite{MacQueen1973} \\
Rocket				& 15 -- 30			& 66\,$\pm$11					& 3.7\,$\pm$0.6 			& \cite{Leinert1976}  \\
HELIOS				& 16 -- 160			& 87\,$\pm$4					& 3.0\,$\pm$0.3 			& \cite{Leinert1980}  \\
SECCHI-HI-1		& 15 -- 24			& $\approx$\,83				& $\approx$\,3.7 			& \cite{Stenborg2017} \\
SECCHI-HI-1		& $\approx$\,5 			& $\approx$\,57		& $\approx$\,6.0 			& \cite{Stenborg2017} \\
LASCO		  		&	0.5 -- 8			& 87.6\,$\pm$0.15		  & 7.5 to 3.0  	   		& This work \\
\noalign{\smallskip}\hline\noalign{\smallskip}

\multicolumn{5}{c}{INFRARED OBSERVATIONS} \\
\noalign{\smallskip}\hline\noalign{\smallskip}
COBE					& 94										& 77.7\,$\pm$0.6			& 2.030\,$\pm$0.02 		& \cite{Kelsall1998} \\		
IRAS					& 60 -- 120							& 78.0								& 1.5 								& \cite{Rowan2013}   \\		
IRAKI					& 90										& 75.9\,$\pm$0.1			& 2.047\,$\pm$0.007 	& \cite{Kondo2016}   \\
\noalign{\smallskip}\hline\noalign{\smallskip}
\end{tabular}
\label{tab:inc_omega}
\end{table}

Our characterization of the plane of symmetry of the inner zodiacal cloud based on 25 years of LASCO images take advantage of favorable observational geometries, an approach successfully adopted by \cite{Leinert1980}.
To do so, we proceeded in two steps.
In a first step, we determined the times at which SoHO crosses this plane which in turn define its line of nodes and finally the longitude of its ascending node $\Omega_A$.
At these times, the F-corona appears symmetric and the axis of symmetry corresponds to the projection of the PSZC seen edge-on onto the sky plane.
It is also known as the photometric axis as it corresponds to the locus of the maximum radiances when considering photometric profiles perpendicular to it.
The aim of the second step was precisely to analyze these symmetric images so as to determine the position angle (PA) of the axis of symmetry which directly yields the inclination of the PSZC with respect to the ecliptic plane. 
As the PSZC is warped, the photometric axis is curved as conspicuously seen on the C2 and C3 images of the F-corona at the node of June 1997 (Figure~\ref{fig:Final_F}) and on the WISPR image of 5 November 2018 (Figure~\ref{fig:F_examples}).
For simplicity, it may be viewed as a set of concentric planar annuli whose inclination with respect to the ecliptic plane is progressively increasing with decreasing heliocentric distances. 

\subsection{The Ascending Node of the Plane of Symmetry}
\label{sec:node}
The determination of the times of SoHO crossing the PSZC  relies on the temporal monitoring of the coronal radiance integrated in the two small symmetric windows, ``north'' and ``south'', centered on the vertical passing through the center of the Sun as illustrated in Figure~\ref{fig:windows}.
Their centers are located at an elongation of 5.5\Rsun\ for C2 and  25\Rsun\ for C3.

\begin{figure}[!htpb]
\centering
\includegraphics[width=\textwidth]{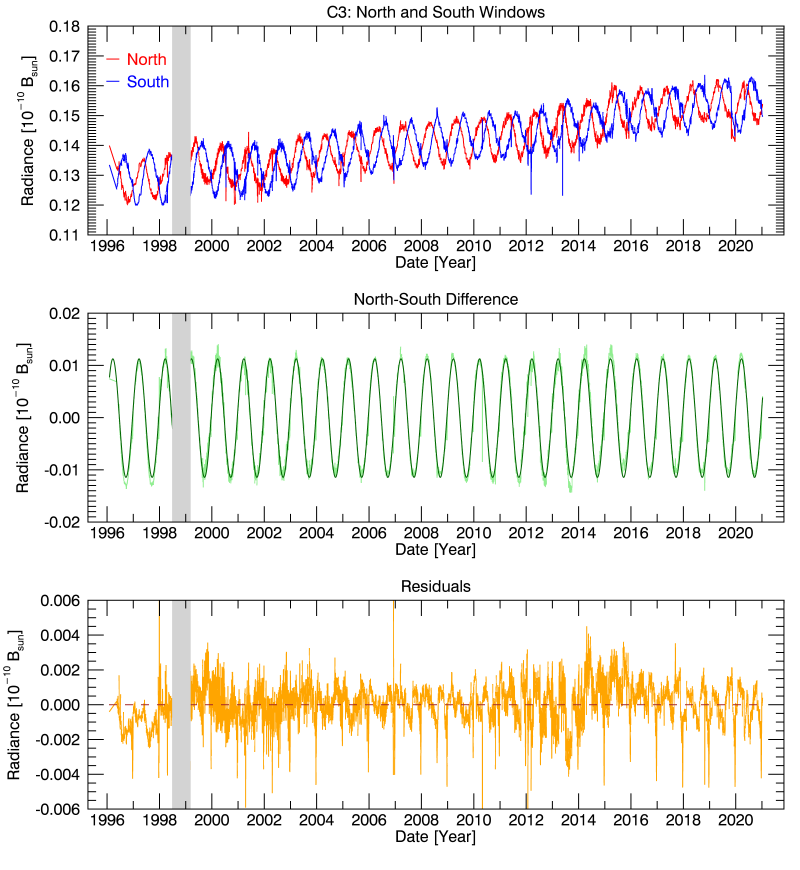}
\caption{Upper panel: temporal variation of the north and south radiances integrated in the corresponding windows in the C3 \fov illustrated in Figure~\ref{fig:windows}.
Second panel: temporal variation of the difference between the north and south radiances and its first harmonics.
Lower panel: residuals between the ``north minus south'' difference and the first harmonics.}
\label{fig:C3_NS_Profiles}
\end{figure}

Let us first consider the case of C3 (Figure~\ref{fig:C3_NS_Profiles}) as the leverage offered by its distant windows amplifies the temporal variations. 
Leaving aside for the moment the long-term trend, the two temporal variations exhibit a periodicity of one year, but with an offset of six months. 
This confirms the periodic oscillation of SoHO about the PSZC and the crossings of the two curves specify the dates of the nodes.
The relative variations of the radiances at the nodes, alternatively increasing and decreasing, impose that the ascending nodes take place in late December and the descending node in late June.

The accurate determination of the dates of the nodes was performed by considering the ``north minus south'' difference and fitting to its temporal variation a sinusoid with the mean period of a tropical or equinoxial year (365.2422 days) that in fact corresponds to the first harmonics of the difference. 
The dates of the nodes given by the zeros of the first harmonics do coincide with the points where the differences ``data minus first harmonics'' are close to zero.
These differences are displayed as residuals in Figure~\ref{fig:C3_NS_Profiles} and are extremely low, typically $\pm\,10^{-13}$\Bsun, but slightly larger during two time intervals clearly in phase with the two solar maxima.
This effect most likely results from the shot noise associated with the K-corona which is not eliminated in the process of separating the two coronal components and is even more pronounced in the case of C2 as shown below.

The resulting dates of the ascending node are listed in Table~\ref{tab:omega} and they lie between December 20 and 21. 
This table also displays the dates of the preceding September equinox which are not fixed but vary with the period of a tropical year.
Then the longitude of the ascending node $\Omega_A$ were simply calculated from the time difference between the date of a given December node and that of the preceding September equinox. 
The 25 determinations are given in Table~\ref{tab:omega} and range from 87.56\deg\ to 87.59\deg.

\begin{table}
\caption{Dates of the ascending node of the plane of symmetry of the zodiacal cloud and those of the preceding September equinox given in two formats, MJD and calendar.
The last column lists the resulting longitudes of the ascending node $\Omega_A$ of the plane of symmetry of the zodiacal cloud.} 
\label{tab:omega}
\centering
\begin{tabular}{ccccc}
\hline\noalign{\smallskip}
NODE & NODE & EQUINOX & EQUINOX & $\Omega_A$ \\
\noalign{\smallskip}
MJD & Calendar date & MJD & Calendar date & Deg \\
\noalign{\smallskip}\hline\noalign{\smallskip}
50437.61 & 1996-12-20.61 & 50348.75 & 1996-09-22.75 & 87.58 \\
50802.85 & 1997-12-20.85 & 50714.00 & 1997-09-23.00 & 87.57 \\
51168.09 & 1998-12-21.09 & 51079.23 & 1998-09-23.23 & 87.58 \\
51533.34 & 1999-12-21.34 & 51444.48 & 1999-09-23.48 & 87.58 \\
51898.58 & 2000-12-20.58 & 51809.73 & 2000-09-22.73 & 87.57 \\
52263.82 & 2001-12-20.82 & 52174.96 & 2001-09-22.96 & 87.58 \\
52629.06 & 2002-12-21.06 & 52540.21 & 2002-09-23.21 & 87.57 \\
52994.30 & 2003-12-21.30 & 52905.45 & 2003-09-23.45 & 87.57 \\
53359.55 & 2004-12-20.55 & 53270.69 & 2004-09-22.69 & 87.58 \\
53724.79 & 2005-12-20.79 & 53635.93 & 2005-09-22.93 & 87.58 \\
54090.03 & 2006-12-21.03 & 54001.17 & 2006-09-23.17 & 87.58 \\
54455.27 & 2007-12-21.27 & 54366.41 & 2007-09-23.41 & 87.58 \\
54820.52 & 2008-12-20.52 & 54731.66 & 2008-09-22.66 & 87.58 \\
55185.76 & 2009-12-20.76 & 55096.89 & 2009-09-22.89 & 87.59 \\
55551.00 & 2010-12-21.00 & 55462.13 & 2010-09-23.13 & 87.59 \\
55916.24 & 2011-12-21.24 & 55827.38 & 2011-09-23.38 & 87.58 \\
56281.48 & 2012-12-20.48 & 56192.62 & 2012-09-22.62 & 87.58 \\
56646.73 & 2013-12-20.73 & 56557.86 & 2013-09-22.86 & 87.59 \\
57011.97 & 2014-12-20.97 & 56923.10 & 2014-09-23.10 & 87.59 \\
57377.21 & 2015-12-21.21 & 57288.35 & 2015-09-23.35 & 87.58 \\
57742.45 & 2016-12-20.45 & 57653.60 & 2016-09-22.60 & 87.57 \\
58107.70 & 2017-12-20.70 & 58018.83 & 2017-09-22.83 & 87.59 \\
58472.94 & 2018-12-20.94 & 58384.08 & 2018-09-23.08 & 87.58 \\
58838.18 & 2019-12-21.18 & 58749.33 & 2019-09-23.33 & 87.57 \\
59203.42 & 2020-12-20.42 & 59114.58 & 2020-09-23.58 & 87.56 \\
\noalign{\smallskip}\hline\noalign{\smallskip}
\end{tabular}
\end{table}

\begin{figure}[!htpb]
\centering
\includegraphics[width=\textwidth]{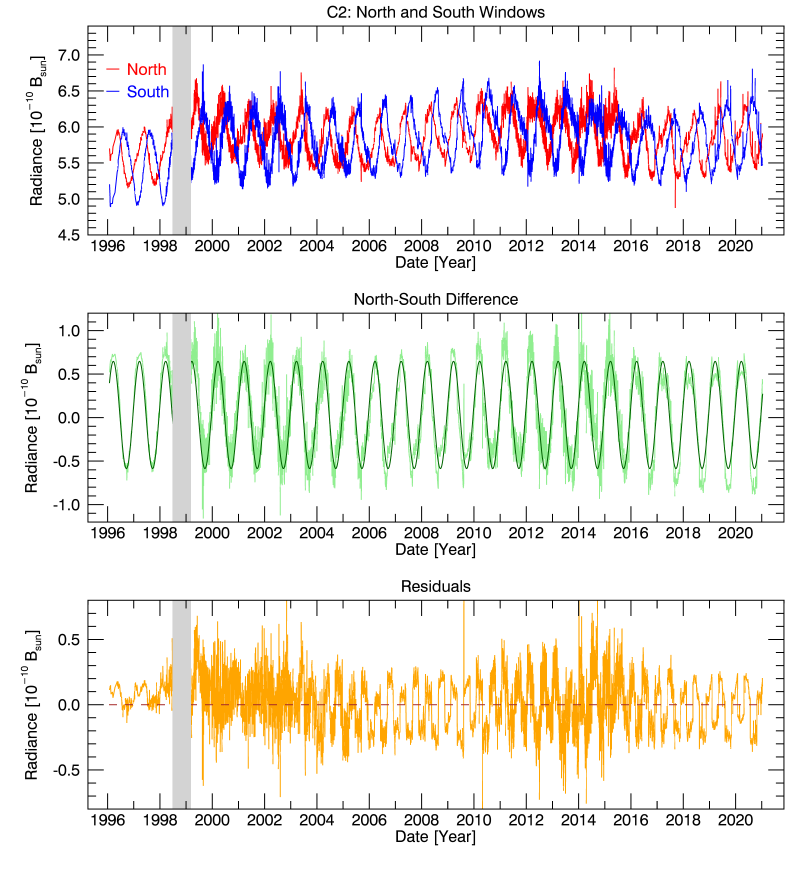}
\caption{Upper panel: temporal variation of the north and south radiances integrated in the corresponding windows in the C2 \fov illustrated in Figure~\ref{fig:windows}.
Second panel: temporal variation of the difference between the north and south radiances and its first harmonics.
Lower panel: residuals between the ``north minus south'' difference and the first harmonics.}
\label{fig:C2_NS_Profiles}
\end{figure}

We performed a similar analysis of the C2 images using two data sets, the ``F+SL'' and ``Fcor'' images as introduced in Section~\ref{sec:C2} and found that the former set yielded more accurate results than the latter one. 
The temporal variations of the radiances integrated in the four windows, of the ``north minus south'' difference, and of the residuals with respect to the first harmonics are displayed in Figure~\ref{fig:C2_NS_Profiles}.
As a consequence of the presence of the stray light and probably of remnants of the K-corona, the C2 residuals (lower panel of Figure~\ref{fig:C2_NS_Profiles}) are noisier that those of C3 (lower panel of Figure~\ref{fig:C3_NS_Profiles}) with an even stronger correlation of the fluctuations with the two solar cycles.
Nevertheless, the sinusoidal fit with a period of 365.2422 days led to the same dates of the nodes, and finally to the same determinations of $\Omega_A$ as found in the case of C3 (Table~\ref{tab:omega}), thus offering an independent validation.

Although we are not aiming at an accuracy of a few hundredth of a degree for $\Omega_A$, it is interesting to point out that very slight differences between the determinations do not come from the data but from the fact that we used a mean value of the tropical or equinoxial year (365.2422 days) whereas we used the exact dates of each September equinox.
Rounding at two decimal places may also artificially increase the differences.
The real uncertainties coming from the data themselves are difficult to propagate and we resorted to the bootstrap method as already mentioned in Section~\ref{sec:flattening} in the context of the analysis of the flattening index by \cite{Stenborg2018}.
We found standard deviations of 0.15\deg\ and 0.09\deg\ for the C2 and C3 determinations of $\Omega_A$, respectively, consistent with our expectation of the superiority of the C3 data for this investigation.

\subsection{The Inclination of the Plane of Symmetry}
\label{sec:inclination}
With the dates of the nodes at hand, we analyzed the corresponding symmetric images of the F-corona in order to determine the orientation of their photometric axis and in turn, the inclination of the PSZC.
There are different approaches to tackle this problem and, for instance, \cite{Stenborg2017} used the location of the ``nose'' of a set of iso-contours, the ``nose'' being defined by the maximum distance of each isophote to the center of the Sun.

We introduced a different method in Section~\ref{sec:C2C3} by rotating a set of isophotes until its configuration and the mirrored one visually coincided (Figure~\ref{fig:control_tilt}).
Whereas this was appropriate to extract the photometric profiles of the F-corona along the major and minor axes of the isophotes, we found it insufficiently accurate to determine the inclination of the PSZC with a precision of a fraction of a degree.
This became exceedingly challenging as the ellipticity of the isophotes decreases with elongation to the point of making them circular at $\approx$\,2\Rsun\ as illustrated in Figure~\ref{fig:flat}; in other terms, the problem becomes degenerate. 

\begin{figure}[!htpb]
\centering
\includegraphics[width=0.8\textwidth]{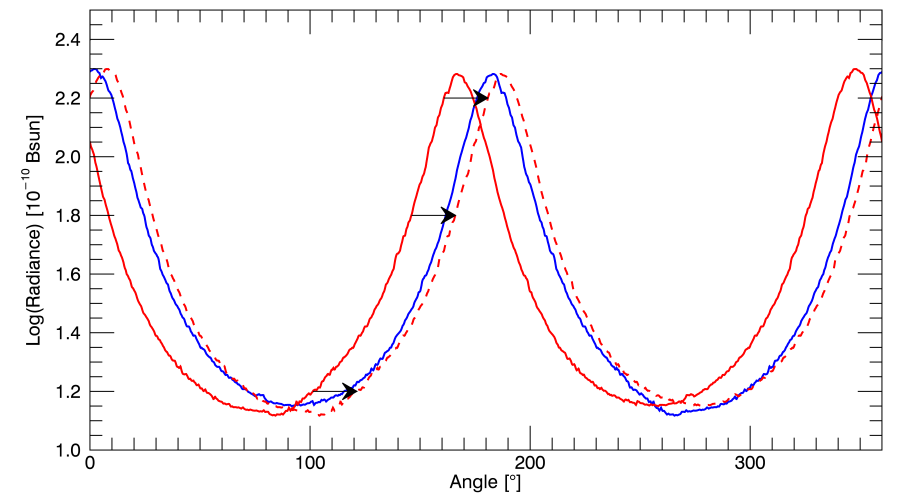}
\caption{Illustration of the method implemented to determine the position angle of a point located on the photometric axis of the F-corona.
The circular profile passing through this point (blue curve) and its reversed or flipped version (red curve) are overplotted.  
The red profile is shifted rightward until it coincides with the blue profile via a minimization process illustrated in (Figure~\ref{fig:minimization}). 
In this example, the red profile is slightly over-shifted to ease its visualization (red dashed line).}
\label{fig:2prof}
\end{figure}

\begin{figure}[!htpb]
\centering
\includegraphics[width=0.49\textwidth]{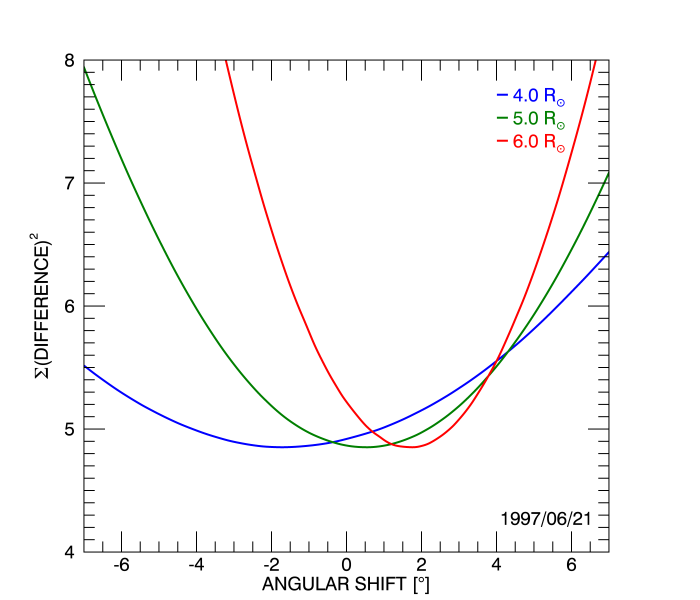}
\includegraphics[width=0.49\textwidth]{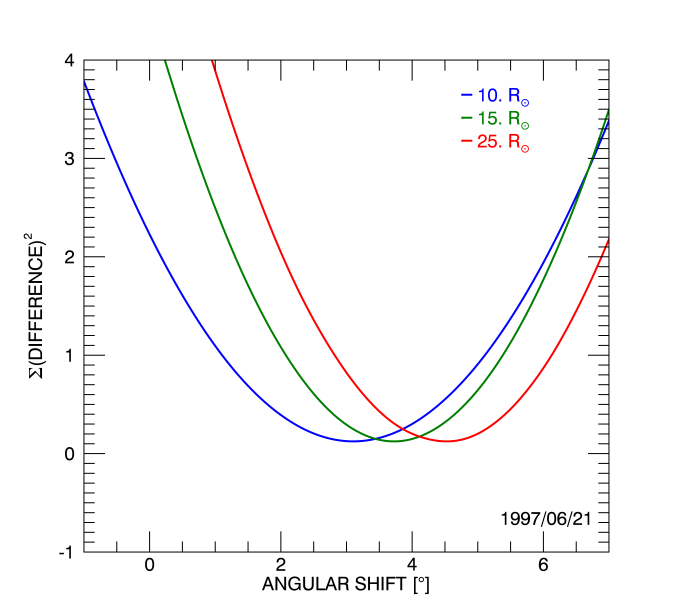}
\caption{Illustrations of the minimization process where the sum of the square of the differences between the two profiles of Figure~\ref{fig:2prof} are plotted versus the angular shift between them.
The minimum of each curve yields the value of the shift.
The left and right panels display the case of a C2 and a C3 image, respectively at different elongations.}
\label{fig:minimization}
\end{figure}

The new procedure exploits the global symmetry of the considered images of the F-corona and operates on the logarithm of the images to reduce their dynamics.  
We transformed the images from Cartesian to polar coordinates with its origin at the center of the Sun and with radial and angular sizes of 250 and 3600 pixels, respectively.
The radial sampling matches that of the Cartesian images of 512 $\times$ 512 pixels and the angular sampling of 0.1\deg\ per pixel ensures a sufficient accuracy.
At each radial distance, two profiles were extracted in opposite directions and were plotted as a function of the polar angle reckoned from the x-axis parallel to the rows of the CCD detector.
In the general case where the photometric axis is not aligned with this axis, the two profiles are offset as illustrated in Figure~\ref{fig:2prof}.
Optimal coincidence of these two profiles was achieved by minimizing the sum of the square of the differences between each point of these two profiles. 
The variation of this sum with angular shift exhibits a quasi parabolic shape ensuring an accurate determination of the optimal shift (Figure~\ref{fig:minimization}), and half its value yielded the polar angle $\xi$ of each point of the photometric axis reckoned from the x-axis of the detector.

The derivation of the inclination of the PSZC from the above angles was performed by considering the different sky-projected polar directions, solar north pole (SNP) and ecliptic north pole (ENP), with all angles reckoned from the y-axis of the CCD detector, \ie the upward direction of its columns, and measured counter-clockwise.
The inclination at each point of the photometric axis is thus given by the direction defined by ZNP = $\xi$ + 90\deg\ with respect to the ENP direction. 
Its determination involves the SoHO roll angle (\ie the position angle of the SoHO optical Z-axis reckoned from the SNP direction), the offset angles between this Z-axis and the y-axes of the C2 and C3 detectors,
and the angle between the SNP and ENP directions given by the ephemeris.

\begin{figure}[!htpb]
\centering
\includegraphics[width=0.49\textwidth]{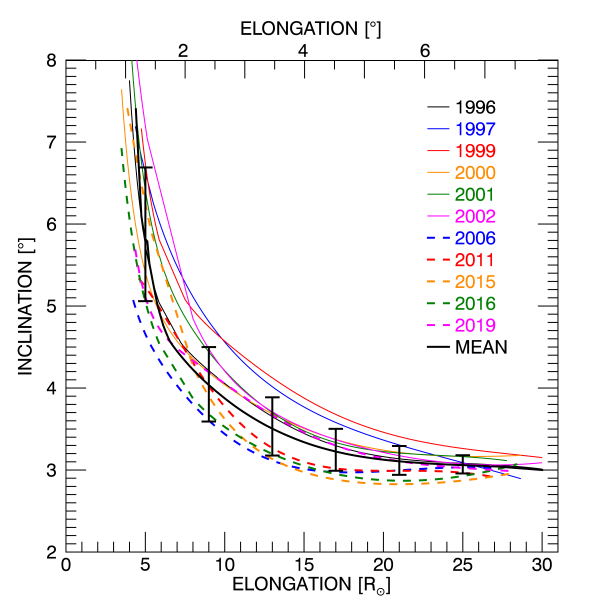}
\includegraphics[width=0.49\textwidth]{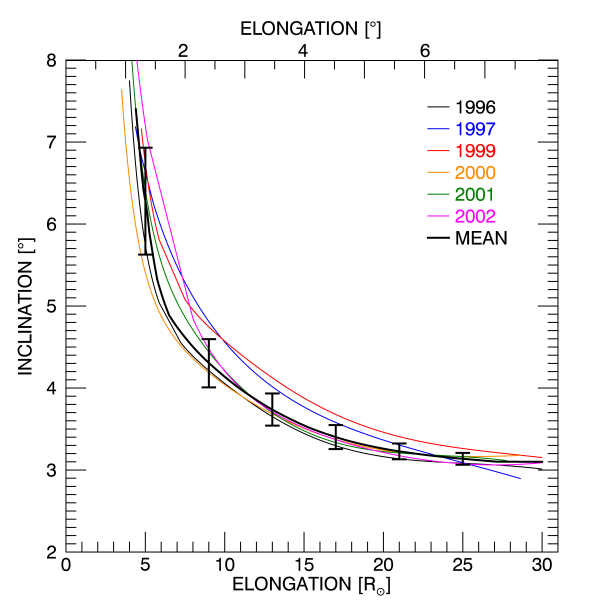}
\caption{Variation of the inclination of the plane of symmetry of the inner zodiacal cloud with solar elongation as determined from the LASCO-C2 and C3 observations of the F-corona.
The left panel displays the results for eleven years spanning the interval [1996\,--\,2019] and the mean variation (black curve).
In the right panel, the display is restricted to six years spanning the interval [1996\,--\,2002].}
\label{fig:inclination_lasco}
\end{figure}

Our analysis was performed for eleven years spanning the interval [1996\,--\,2019] and for each year, we averaged the determinations at the June and December nodes. 
The eleven variations are displayed in Figure~\ref{fig:inclination_lasco} together with the mean and the standard deviations at selected elongations.
The dispersion of the curves is rather large reaching $\approx$\,1\deg\ at mid elongations ($\epsilon$\,$\approx$\,4\deg) and even more at smaller elongations.
The latter effect is directly attributed to the increasing difficulty of determining the polar angle $\xi$ at these elongations where the extracted profiles become more and more circular, a degeneracy already mentioned above.
It appears that, with one exception (year 2019), the curves tend to cluster into two groups, before and after 2002, possibly pointing to the impact of the discontinuation of the SoHO definitive attitude files which provided highly accurate values of the roll angle until 2002.
Another difficulty may be invoked as the switch of the SoHO orientation from solar north to ecliptic north reduced the angular excursion of the tilt of the symmetry axis thus rendering the determination of the polar angle $\xi$ less accurate.

We therefore decided to propose an alternative, probably more reliable, solution restricted to the first six years spanning the interval [1996\,--\,2002] as displayed in the right panel of Figure~\ref{fig:inclination_lasco}. 
The continuous increase of the inclination from an ``asymptotic'' value of 3.1\deg at an elongation of 7.5\deg\ to a value of $\approx$\,7.5\deg\ at an elongation of $\approx$\,1.2\deg\ is thus better ascertained with standard deviations of 0.07\deg\ and 0.7\deg\ at the outer and inner limits of the LASCO \fovnosp, respectively.
This confirms early conjectures of a warped surface of symmetry of the zodiacal cloud by \eg \cite{Misconi1980} -- although his proposed deviation was limited to the inclination of Venus -- and by \cite{Leinert1976} who envisioned ``a significant deviation from the invariable plane towards the planes of the inner planets and the solar equator'', a trend also suggested by the results of \cite{Stenborg2017}.

\begin{figure}[!htpb]
\centering
\includegraphics[width=\textwidth]{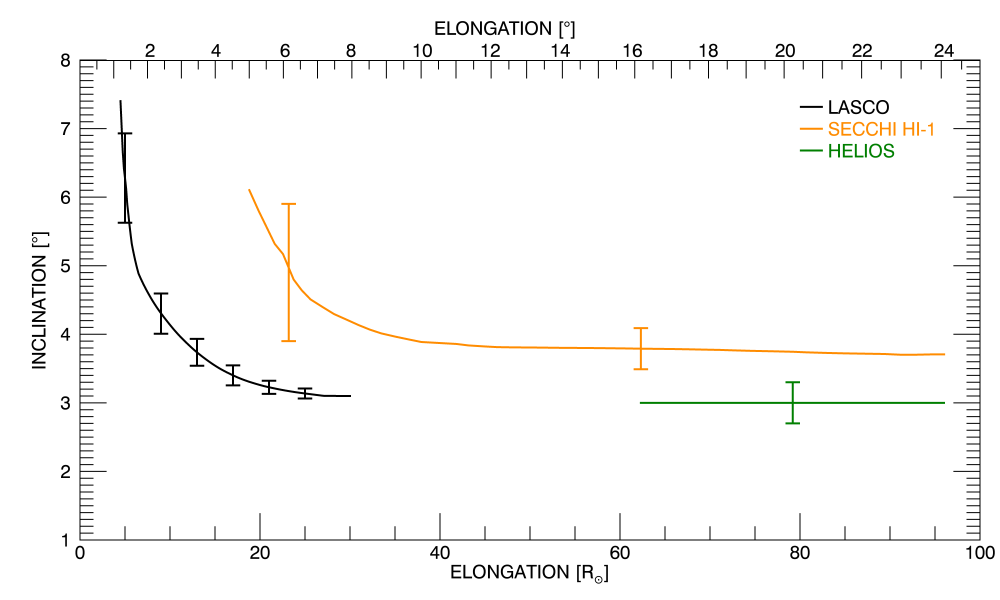}
\caption{Variation of the inclination of the plane of symmetry of the inner zodiacal cloud with solar elongation.
The mean profile of the six LASCO determinations reported in the right panel of Figure~\ref{fig:inclination_lasco} is compared with the \textit{Helios} result of \cite{Leinert1980} and the SECCHI HI-1 result of \cite{Stenborg2017}.}
\label{fig:inclination_all}
\end{figure}

Figure~\ref{fig:inclination_all} reveals the excellent consistency between the LASCO result and that of \textit{Helios} \citep{Leinert1980} as their inclination of 3.0\deg\,$\pm$0.3\deg\ determined down to an elongation of 16\deg\ smoothly connects to our ``asymptotic'' value at an elongation of 7.5\deg; this strongly suggests that the subsequent rapid increase of the inclination with decreasing elongation takes place at $\approx$\,7.5\deg.
Further considering the excellent agreement between their and our determination of the longitude of the ascending node, 87\deg\,$\pm$4\deg\ and 87.6\deg\,$\pm$0.15\deg, respectively,  considerably reinforces the robustness of the two solutions.

In order to incorporate the SECCHI HI-1 result presented in Figure~6 of \cite{Stenborg2017}, we combined their strongly filtered smooth black curve at large elongations with their mildly filtered light-blue curve for the rising branch to preserve the strong gradient of the original data, and we estimated two error bars from the dispersion of their data points. 
If the SECCHI and the LASCO results agree on increasing inclination with decreasing elongation, the profiles of this variation are in clear disagreement. 
Their constant, ``asymptotic'' value amounts to 3.7\deg -- also in disagreement with the \textit{Helios} result -- and the rising branch starts at about 10\deg, the inclination reaching 6\deg\ at about 5\deg\ elongation compared with our value of 3.3\deg.
But the most serious discrepancy concerns the longitude of the ascending node as we found it constant throughout the LASCO \fov whereas \cite{Stenborg2017} reported an abrupt decrease from $\approx$\,83\deg\ (incidentally close to our result) down to $\approx$\,57\deg\ over an elongation range of less than 10\deg.

\section{Stability of the Inner Zodiacal Cloud}
\label{sec:stability}
The stability of the zodiacal cloud and of its optical manifestation the zodiacal light, have been a matter of debate similar to that of the plane of symmetry.
Both are difficult measurements, but the latter case is even more difficult as it requires long-term observations under strictly similar conditions with rigorously calibrated instruments.
An historical summary of the variety of early investigations may be found in \cite{Leinert1975}, but significant progress have been achieved thereafter that led to the view of a quite stable zodiacal light.

Indeed, from 11 years of ground-based ``Tenerife'' data, \cite{Dumont1978} concluded that any minimum-to-maximum variation larger than 10\,\% was highly improbable.
This constraint was reduced by the \textit{Helios} observations over 11 years (from December 1974 to February 1986) which set limits of $\pm$\,2\,\% for secular variations (\cite{Leinert1982}; \cite{Leinert1989}). 
In practice, their analysis was limited to a small 1\deg $\times$ 5.6\deg\ patch of sky centered at ecliptic coordinates $\beta$\,=\,-16\deg, $\lambda - \lambda_\odot$\,=\,-62\deg.
Furthermore, 6150 out of the $\approx$\,7000 individual measurements ($\approx$\,88\,\%) were obtained during the first eight years of observation. 
In any case, both \cite{Dumont1978} and \cite{Leinert1989} additionally concluded on the total absence of correlation between the brightness of the zodiacal light with solar activity. 

The Solar Mass Ejection Imager (SMEI) survey performed by \cite{Buffington2016} extended over 8.5 years (February 2003 to September 2011) and nine ecliptic sky locations were monitored in order to detect temporal variations.
Relying on a long-term photometric stability of 0.25\,\% certified by an analysis of three stable sidereal objects, they established a 1-$\sigma$ value of 0.3\,\% for zodiacal light change over the above time interval.

The specific case of the stability of the F-corona was investigated by \cite{Morgan2007} by comparing LASCO-C2 images obtained in September and October 1996 (the minimum of Solar Cycle 23) on the one hand, and in December 2000 (its maximum) on the other hand. 
They found that the brightness remained unchanged, although the varying Sun\,--\,SoHO distance which was not considered, should have led to a small, detectable variation.
Further out and based on SECCHI/HI-1A images taken from December 2007 to March 2014 ($\approx$\,6.3 years), \cite{Stenborg2018} found that the brightness profile along the photometric axis of the eastern side of the F-corona between 5\deg\ and 24\deg\ elongation experienced a subtle secular variation of less than 0.1\,\%, although it is unclear how this relates to the accuracy of the measurements.
Finally, the WISPR observations during the first two years of PSP (first five orbits) and specifically, the brightness profiles along one side of the symmetry axis from 7.65 to almost 130\Rsun\ led \cite{Stenborg2021} to conclude on a stable F-corona; variations below 5\,\% were observed but attributed to either electron corona structures or to the Milky Way.

At various stages of our analysis of the LASCO-C2 and C3 images of the F-corona, we pointed out long-term evolutions, but deferred their detailed investigation to the present section.
The monitoring of the integrated radiances in the ring common to C2 and C3 in Section~\ref{sec:SOHOLASCO} and illustrated in Figure~\ref{fig:C2C3ringphot} revealed a common trend over 25 years: atop a progressive increase at a rate of 0.26\,\% per year (6.6\,\% over 25 years), a steeper increase took place starting in 2006, culminating in 2010--2011, and then returning to the nominal trend in 2016. 
The radiances integrated in half-rings used to study the center of the F-corona (Section~\ref{sec:center}) and in small windows to investigate the symmetry plane (Section~\ref{sec:plane}) exhibit also long-term evolutions.
For instance, in the case of C3 where the windows are centered at 25\Rsun\, the two integrated radiances north and south underwent a quasi monotonic increase at a rate of 0.94\,\% per year throughout the 25 years of observation (Figures~\ref{fig:C3_NS_Profiles}). 
An increase was also noticed in the radial profiles of the radiance (Section~\ref{sec:C2C3}) and the averaged 2010+2011 profiles had to be down scaled by 8\,\% to match those of 1997.
Altogether, this suggests complex long-term, so-called secular variations, but before we further proceed, let us insist on two important points.
\begin{itemize}
	\item 
		The temporal variations described above are totally uncorrelated with solar activity over two cycles.
	\item 
	The stability of the photometric responses of C2 and C3 is quite remarkable and for instance a steady decline of only 0.3\,\% per year was determined for C2.
	This results from the meticulous calibration of C2 based on thousands measurements of stars present in its \fov (\cite{Llebaria2006}; \cite{Gardes2013}; \cite{Colaninno2015}; \cite{Morgan2015}).
	This calibration is continuously monitored and annually updated ensuring a photometric accuracy better than 1\,\% (Figure~9 of \cite{Lamy2020}). 
	The C3 calibration is scaled to that of C2 as explained in \cite{Lamy2021} and in Section~\ref{sec:C2C3} so that its photometric performance is similar.
\end{itemize}

\begin{figure*}[!htpb]
\centering
\includegraphics[width=0.95\textwidth]{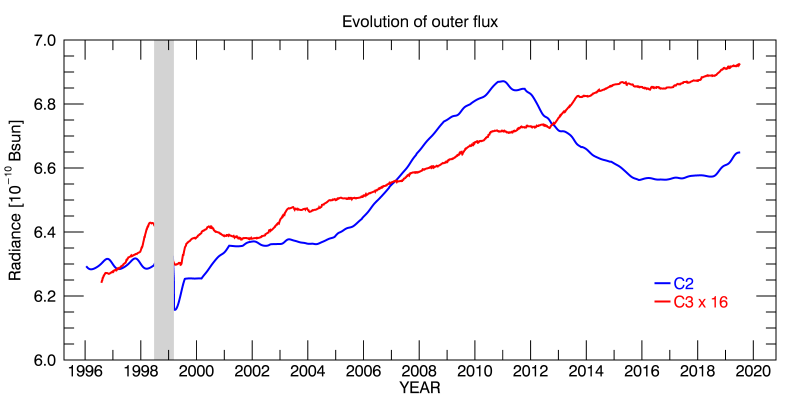}
\caption{Comparison of the long-term evolutions of the radiance of the F-corona recorded by C2 and C3 in broad outer rings of their \fovs.
The C3 radiance is up-scaled by a factor of 16 to match the C2 radiance.}
\label{fig:c2c3long_term}
\end{figure*}

To ascertain the variations detected by LASCO, we relied on the radiances integrated in the broad outer rings introduced in Section~\ref{sec:center}.
The selection of these rings ensured that they are representative of large parts of the C2 and C3 \fovs\ while avoiding  the inner parts where possible remnants of the K-corona and stray light might affect the radiances. 
The long-term evolutions displayed in the upper panels of Figures~\ref{fig:C2_half_rings} and \ref{fig:C3_half_rings} are now plotted altogether in Figures~\ref{fig:c2c3long_term} where the C3 radiance curve was up-scaled by a factor of 16 to compensate for the large difference between the C2 and C3 radiances.
The C3 radiance underwent a quasi monotonic increase and a linear fit to the variation yields a global variation of 11.5\,\% over 25 years translating to a rate of 0.46\,\% per year.
The trend is consistent with that found in the north and south windows, but in these highly localized areas, the increase was twice larger. 
The evolution of the C2 radiance is consistent with that found in the ring common to C2 and C3 (Section~\ref{sec:C2C3}): atop a baseline progressive increase at a rate of 0.2\,\% per year (5\,\% over 25 years, slightly smaller than the radiance in the common ring), a steeper increase took place starting in 2006, culminating in 2010\,--\,2011, and then returning to the baseline in 2016.
The culmination itself represents a 5.6\,\% increase with respect to the baseline.

\begin{figure*}[!htpb]
\centering
\includegraphics[width=0.95\textwidth]{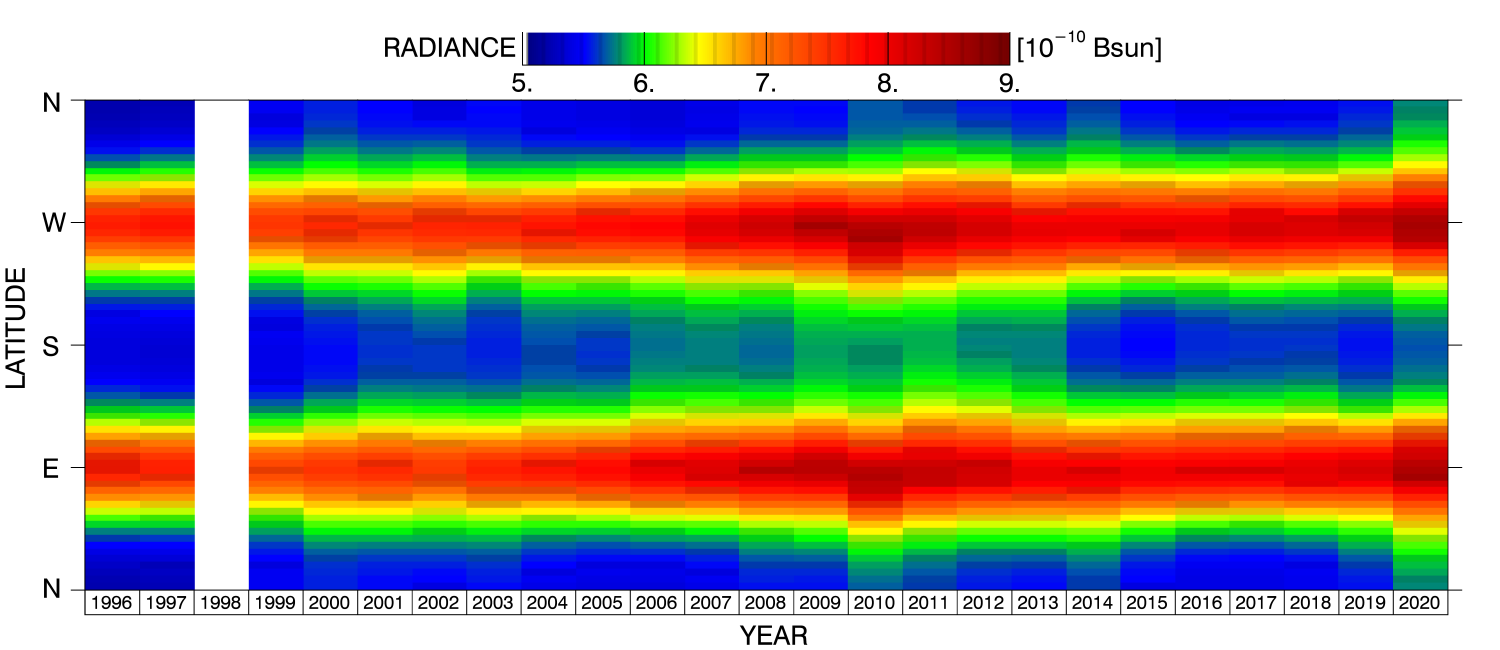}
\includegraphics[width=0.95\textwidth]{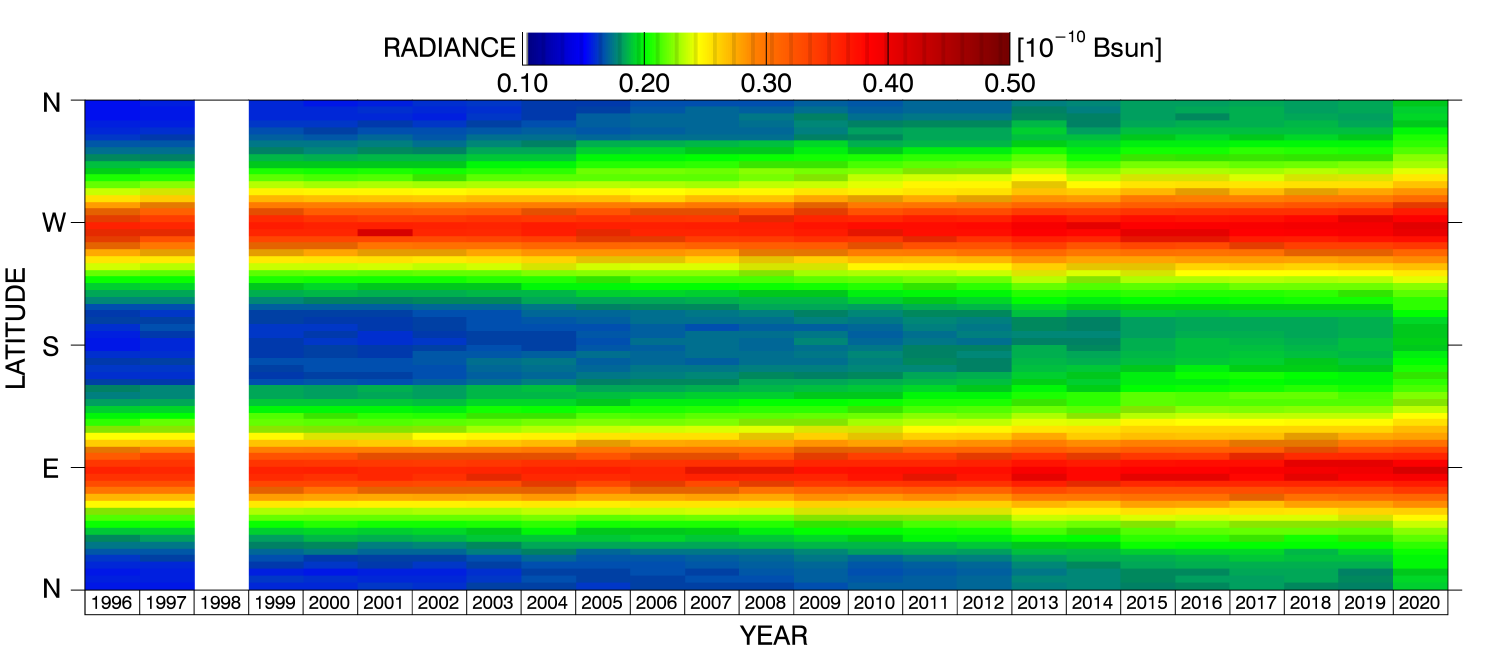} 
\caption{Stackmaps constructed from C2 (upper panel) and C3 (lower panel) images of the F-corona at the December nodes using the broad outer rings defined in the text.
Each column corresponding to a given node is replicated over each year for better legibility.}
\label{fig:stackmap_C2C3}
\end{figure*}

We finally turn to the stackmaps introduced in Section~\ref{sec:method} in an attempt to decipher the spatio-temporal variations of the radiance of the F-corona. 
As a reminder, they were generated at each node of June and December to benefit from the simplicity of a symmetric corona using the same broad outer rings.
However, the results are similar for these two nodes so that we only display the case of the December nodes in Figures~\ref{fig:stackmap_C2C3}.
A global view of these two stackmaps confirms that the general evolutions are fully consistent with that recorded in the rings: for C2 a bulge over the years 2005 to 2016 superimposed on a regular increase and for C3, a continuous regular increase throughout the 25 years of observation.
A closer look reveals that these variations tend to globally affect the F-corona.
This is particularly the case of the C3 stackmap as the radiance at all latitudes evolves in the same balanced way.
The situation is slightly more contrasted in the case of C2 as minute yearly fluctuations are perceptible, but likely imputable to the difficulties of accurately reconstructing the Fcor images in comparison with C3.
It however appears that the uprising in the years 2006 to 2016 was slightly stronger at high southern latitudes than at northern latitudes. 
 
In conclusion, we have strong evidences of complex secular variations affecting the radiance of the F-corona totally uncorrelated with solar activity and probably related to the intrinsic evolution of the interplanetary dust cloud and its supplies.
The outer \fov appears to experience a global regular increase whereas the situation is more contrasted in the inner \fov with a pronounced uprising over ten years and some evidence of different behaviours in the northern and southern regions.

\section{Discussion and Implications for the Properties of Interplanetary Dust}
\label{sec:discussion}
In the course of our analysis, we compared our LASCO results with those of SECCHI/HI-1 reported by \cite{Stenborg2017} and \cite{Stenborg2018} because of the overlap between the respective \fovs.
Whereas we found points of convergence such as the suitability of the super-ellipse model of the isophotes and the increase of inclination of the PSZC with decreasing elongation, there are areas of serious disagreement.
We first address these problems before discussing the general implications of the LASCO results on our understanding of the zodiacal cloud.

A major concern -- already pointed out in Section~\ref{sec:photometry} -- with the two aforementioned articles, as well as that of \cite{Stenborg2021}, arises from the systematic confusion between the two power exponents $\nu_{\scriptscriptstyle\rm ZL}$ which characterizes the variation of the radiance with the heliocentric distance of the observer at constant elongation $B_{\rm F}(d)$ and $\nu_{\scriptscriptstyle\rm P}$ which characterizes the variation of the radiance with elongation at constant heliocentric distance of the observer $B_{\rm F}(\epsilon)$. 
These two exponents are equal only in the case of the profiles along the long axis of symmetry and further under restrictive conditions, in particular the fact that the properties of the IDPs must be the same everywhere (Section~\ref{sec:photometry}).
However, we have compelling evidences that this is not true as, for instance, the albedo of the IDPs varies with heliocentric distance.
As a consequence, we do not have yet a robust determination of $\nu_{\scriptscriptstyle\rm ZL}$ other than that obtained by \cite{Leinert1981} on the basis of the \textit{Helios} observations.

We regret the quasi absence of photometric results in the article of \cite{Stenborg2018}, for instance radial profiles along characteristic directions (\eg equatorial and polar) and calibrated images.
There are further difficulties when these authors associate an elongation with a specific location in space, \ie heliocentric distance.
An elongation defines a \los and the corresponding radiance results from the integration all along this \los so that IDPs along a large range of heliocentric distance are involved as illustrated by Figure~\ref{fig:los}.

These authors reported in their two articles that the symmetry center of the inner zodiacal cloud is displaced from the Sun center by about 0.5\Rsun\ in the direction of the average location of Jupiter during the epoch studied (2007\,--\,2012).
On the one hand, the LASCO results refute this offset and on the other hand, the underlying idea that Jupiter deforms the zodiacal cloud by direct tidal attraction is incorrect.
The cloud does not behave like an ocean and gravitational forces exerted by Jupiter and other planets translate in secular perturbations of the orbital elements of the IDPs. 
This general problem was treated by \cite{Wyatt1999} with direct application to the offset and warp of the zodiacal cloud and they found that the offset is in the (fixed) direction of the apocenter of the perturbing planet, not in the direction of the planet itself.
The IDPs responsible for the inner zodiacal light/F-corona are located within 1\,AU, a region where their orbital evolution is far more complex. 
It was investigated by \cite{Gustafson1986} based on direct numerical integration, considering IDPs of 30 $\mu$m radius released from slightly beyond 1\,AU and perturbed by radiation forces, Jupiter, Mars, Earth, and Venus.
They established the dominating influence of the inner planets leading to marked evolutions of their eccentricity, inclination, and longitude of ascending node.
In fact, an experiment in which \cite{Gustafson1986} removed these inner planets led to a completely different behavior with no evolution of eccentricity and longitude of ascending node as illustrated by comparing their Figures~2 and 4.
There is an ultimate contradiction to the influence of Jupiter on the zodiacal cloud observed in the outer half of the \fov of SECCHI HI (15\deg\,--\,24\deg) as \cite{Stenborg2017} found a plane of symmetry very close to the orbital plane of Venus and then, close to the equatorial plane of the Sun in the inner half of the \fov (5\deg\,--\,15\deg) for which they invoked Lorentz forces.


\subsection{Photometric Properties of the F-corona and Inner Zodiacal Light}
\label{D:photometry}
One of the main achievements of our analysis of the LASCO observations over 25 years consists in the highly accurate photometry of the F-corona with an uncertainty of 5\,\% in a \fov ranging from 2 to 30\Rsun\ (0.5\deg\ to 8\deg).
The resulting equatorial and polar radiance profiles are found to be in excellent agreement with the Koutchmy-Lamy model, demonstrating that ground-based eclipse data carefully scrutinized and synthesized remain extremely valuable. 
There is however an exception as this procedure washed out the shoulder conspicuous on the C3 equatorial profile starting at $\approx$\,2.5\deg\ and extending to $\approx$\,3.5\deg. 
We showed that this shoulder has indirect justifications allowing a smooth connection between the equatorial profiles of the F-corona and of the zodiacal light on the one hand and also of their flattening on the other hand. 
Intermediate radial profiles from the C3 images in between the equatorial and polar directions show that the shoulder progressively subsides with increasing latitudes, becoming undetectable above $\approx$\,45\deg.

A careful inspection of Figure~1 of \cite{Kimura1998} which compiles many past measurements along the direction of the solar equator does reveal a systematic shoulder similar to that found on the C3 equatorial profile. 
The presence of this feature from the mid-fifties until today strongly suggests that it is permanent, at least on a time scale of several decades.
The term ``shoulder'' remains purely descriptive and a more appropriate physical interpretation is possibly in terms of a deficit in the inner radiance profile starting at elongations of 3.5\deg\ to 5\deg.
This deficit can be quantified by comparing the inward extrapolation of the outer profile with the inner one since they are quasi parallel and it amounts to $\approx$20\,\% on the basis of the C3 profile.  
In the frame work of the evolution of the inner zodiacal cloud, the question arises which process is at work to explain this subsidence of the radiance over such a narrow range of elongation. 
However and because of the integration over lines-of-sight, the region where this putative process acts on the interplanetary dust particles (IDPs) is fairly large as $\approx$75\,\% of the radiance comes from within $\approx$\,60\Rsun\, that is $\approx$\,0.3\,AU (Figure~\ref{fig:los}).
Grains composed of silicate minerals are known to sublimate at much smaller heliocentric distances, typically within 10\,--\,12\Rsun\ \citep{Lamy1974a}, even if they are fluffy aggregates \citep{Kimura2002} so they are not likely to explain the observed deficit. 
This is not the case of organic grains that are prominently ejected from comets and which have a very different thermodynamical behaviour.
Considering the case of tholin, \cite{Lamy1988} found that grains with radius larger than 5\,$\mu$m reach a temperature of 400\,K at 60\Rsun\ (and even higher for smaller grains), sufficient to initiate their thermal dissociation or degradation. 
A similar conclusion was reached by \cite{Cottin2004} in the case of grains composed of the polymer of formaldehyde (polyoxymethylene or POM), even at lower temperatures.

We therefore conjecture that this process is responsible for the disappearance of organic matter either as grains or as mantles or as inclusions in a silicate matrix at heliocentric distances up to $\approx$\,0.3\,AU and in turn, possibly explains the observed subsidence of the radiance. 
This conjecture is consistent with the interpretation by \cite{Lasue2007} of the decrease of the local polarization at 90\deg\ scattering angle ${\rm P}_{90^\circ}$ between 1.5 and 0.5\,AU  as resulting from the ``progressive disappearance of the solid carbonaceous compounds towards the Sun, probably linked to the presence of an extended zone of thermal degradation''.
It is further consistent with the concurrent increase with decreasing heliocentric distance of the geometric albedo at 90\deg\ scattering angle of IDPs starting at $\approx$\,0.3\,AU as shown in Figure~1 of \cite{Mann1998}. 
More generally, the albedo of IDPs has been shown to follow this trend between 0.5 and 1.5\,AU \citep{Dumont1988} and power laws of the heliocentric distance have been determined with exponents of -0.3\,$\pm$\,0.3 \citep{Reach1991} and -0.32\,$\pm$\,0.05 \citep{Renard1995}.
All these trends are indeed consistent with the progressive disappearance of absorbing, dark organic matter, thus reinforcing the contribution of the dielectric, somewhat brighter silicate materials.
However, they have been prominently observed near the symmetry plane of the zodiacal cloud and it is unclear whether they persist outside this plane at increasing latitudes. 
We noted that the subsidence of the radiance progressively disappear with increasing latitudes becoming  undetectable above $\approx$\,45\deg\ and this may raise a difficulty for our conjecture.
Indeed, according to \cite{Hahn2002}, at ecliptic latitudes larger than 45\deg, more than 90\,\% of the cross section is contributed by dust grains that are in comet-like orbits, for instance Halley-type and Oort cloud comets which are rich in organic matter.
Clearly, we are still far from understanding all aspects of the zodiacal cloud in detail.

As a final remark, it is interesting to point out that low values of the local polarization ${\rm P}_{90^\circ}$ mentioned above are also required to explain the unexpectedly very low polarization of the F-corona up to an elongation of at least 10\Rsun\ \citep{Lamy2021}.  
This is because a large fraction of the observed radiance comes from dust close to the Sun that scatters solar light at angles around 90\deg.

\subsection{Geometry of the Zodiacal Cloud}
\label{D:geometry}
The LASCO results on the plane of symmetry of the inner zodiacal cloud and, to some extent those of SECCHI/HI-1, demonstrate for the first time the early conjecture of a deviation towards the planes of the inner planets and of the solar equator.
In detail, they do differ on i) the profile of the variation of the inclination of the PSZC with elongation (Figure~\ref{fig:inclination_all}) and ii) on the longitude of nodes which we found constant and equal to the \textit{Helios} value whereas \cite{Stenborg2017} found a rapid decrease with decreasing elongation.
The behaviour of the inclination leads to the view that the symmetry surface tracks the local Laplace plane which is itself distorted by the presence of each individual planet, hence the observed progression to values close to the orbital inclination of Venus (3.4\deg) and Mercury (7.0\deg), and ultimately to the equatorial plane of the Sun (7.3\deg). 
The situation is less clear regarding the longitude of the nodes, but the robustness and convergence of the \textit{Helios} and LASCO results support our value of 87.6\deg.
Recalling the results from the infrared observations (COBE, IRAS, IRAKI), then the longitude of nodes appears to be limited to a narrow range of 76\deg\ to 88\deg\ throughout the zodiacal cloud.

There are unfortunately very limited theoretical studies that could help with the interpretation of these results.
We already mentioned the numerical simulation performed by \cite{Gustafson1986} which has the merit of including gravitational perturbations by the planets Venus, Earth, Mars, and Jupiter, and to highlight the role of the inner ones, but is however limited to 200 trajectories. 
The sample test trajectory given in their Figure~2 does show that, once within the orbit of Venus, the particle can reach inclinations up to 5.6\deg. 
In their investigation of the cometary origin of the zodiacal cloud, \cite{Nesvorny2010} performed similar numerical simulations (further including Neptune), but for several thousands of particles; unfortunately, they did not report on their orbital evolution.
Beyond these studies, it remains to be investigated whether Mercury and the Sun play a role in shaping the innermost part of the symmetry surface of the zodiacal cloud. 
In the case of the Sun and considering the J coefficients of the spherical harmonics describing its gravitational potential, its oblateness and consequently the J2-term are known to be small.
Furthermore, the J2-term mainly affects the longitude of nodes and the argument of perigee; higher order terms such as J4 would have to be considered.

Likewise the symmetry surface, the zodiacal cloud does not have the Sun as a unique center of symmetry and infrared surveys at large elongations and the present LASCO results suggest  a varying center when viewing the cloud in different directions.
At 90\deg\ elongation, thus viewing the outer cloud, the offset of its center of symmetry with respect to the Sun amounts to typically 3.2\Rsun.
It vanishes to zero at $\approx$\,8\deg\ elongation when viewing the inner cloud according to our results which however disagree with that of SECCHI/HI-1.
Thanks to the work of \cite{Wyatt1999}, we have some theoretical background to explain this offset.
They showed how the gravitational influence of a planet with an eccentric orbit causes a brightness asymmetry in a circum-stellar disk by imposing a forced eccentricity on the orbits of the dust particles, thus shifting the center of symmetry of the disk away from the star in the direction of the forced pericenter of the perturbing planet.
It remains to be investigated in detail how the interplay of the different planets with their different eccentricities can explain the observed varying offset in the solar system.

\subsection{Shape of the Isophotes and Implications for the Spatial Density of Interplanetary Dust}
\label{D:density}
As already mentioned in Section~\ref{sec:flattening}, the shape of the isophotes -- and not the flattening -- effectively constrains the three-dimensional spatial density of interplanetary dust. 
Although it is beyond the scope of the present article to fully investigate this question, we outline the general trend given by the shape of the F-corona resulting from the LASCO observations as a guideline for future research.
The method required that, starting from realistic analytic expressions of the density, we calculate radiance maps as integrals of the light scattered by the IDPs along the \los defined by each pixel of the maps.
We considered the simple case of an axi-symmetric cloud centered at the Sun and an observer located at 1\,AU in the PSZC, and we naturally used the formalism involving the VSF (Appendix A) so that the radiance along a given \los is expressed by a single integral:

\begin{equation}
\label{eq:F_formula}
B_{\rm F}(\beta, \epsilon) \propto \int_\epsilon^\pi \Psi_0(\theta) \ V(r, \beta_\pi) \ d\theta \\
\end{equation}

\noindent where $r$ and $\epsilon$ are the heliocentric distance and the solar elongation of the elementary scattering volume (ESV), respectively, $\beta$ is the angle between the observer\,--\,EVS direction and the PSZC, $\beta_\pi$ is the angle between the Sun\,--\,EVS direction and this plane, and $V(r, \beta_\pi)$ describes the spatial density of the cloud.
The geometry of the scattering is illustrated in Figure~1 of \cite{Lamy1986}.
$V(r, \beta_\pi)$ is classically expressed as the product of two independent functions $R$ and $G$ of the variables $r$ and $\beta_\pi$, respectively: $V(r, \beta_\pi)$ = $R(r)G(\beta_\pi)$.

For $R(r)$, we adopted the power law $r^{-1}$ as derived from the inversions carried out by \cite{Dumont1975} and by \cite{Lamy1986}. 
This choice is further justified by dynamical arguments as it results from the action of the Poynting-Robertson effect on dust particles injected at large heliocentric distances \citep{Leinert1990}.
Of direct interest to the present study, various functions $G(\beta_\pi)$ were listed and analyzed by \cite{Giese1986}, prominently the so-called ellipsoid, fan, modified fan, and cosine models. 
We rejected layer models where the variation of the density is a function of the distance from the PSZC as these authors found that they do not reproduce the observed flattening of the isophotes at elongations $\lesssim$\,15\deg.

\begin{figure}[!htpb]
\centering
\includegraphics[width=0.75\textwidth]{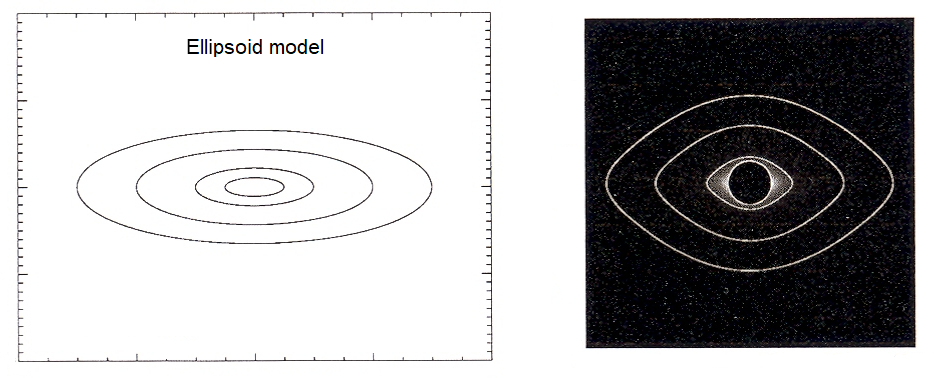}
\includegraphics[width=0.75\textwidth]{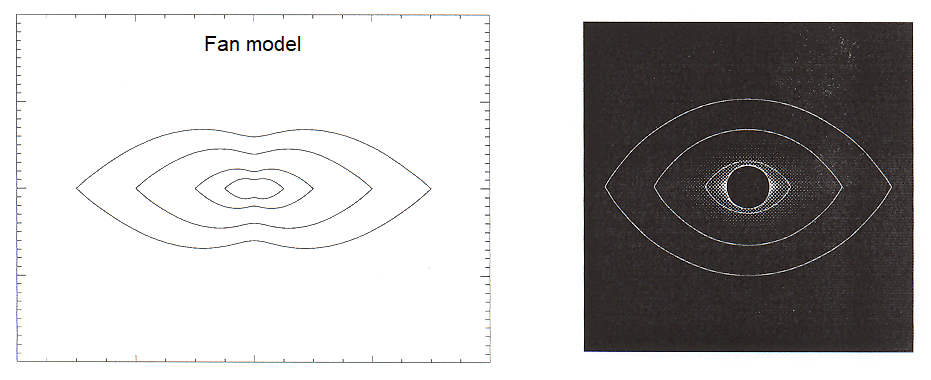}
\includegraphics[width=0.75\textwidth]{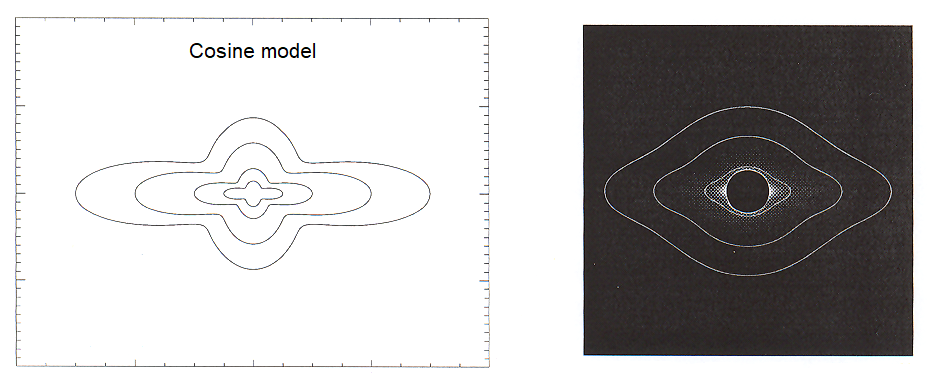}
\caption{Illustration of different models of the three-dimensional spatial density of interplanetary dust (left column) and the resulting isophotes of the radiance (right column).
In the left panels, the lines represent the intersections of different surfaces of equal number density with the helioecliptic meridian plane.
The right panels display three typical isophotes extracted from the radiance maps.
The rows correspond to the ellipsoid (upper panels), flattened fan (middle panels), and cosine (lower panels) models of the dust density.}
\label{fig:isodensity}
\end{figure}

The cosine models are described by the generalized analytical expression:
$$ G(\beta_\pi) = k + (1-k)\,(\cos\beta_\pi)^{-\nu_{\scriptscriptstyle\rm C}} $$
\noindent where $k$ and $\nu_{\scriptscriptstyle\rm C}$ are two parameters.
According to \cite{Giese1986}, these simple models yield isodensity surfaces practically identical to more complicated versions of the fan model and to the sombrero models of \cite{Dumont1976}: they are all characterized by a bulge of the isodensity surfaces above the solar poles.
The typical example displayed in Figure~\ref{fig:isodensity} using parameters $k$\,=\,0.2 and $\nu_{\scriptscriptstyle\rm C}$\,=\,44 shows that this bulge strongly distorts the isophotes making them totally unrealistic when compared with the observations. 

The fan models are described by the expression:
$$ G(\beta_\pi) = {\rm exp}(-\gamma_{\scriptscriptstyle\rm F}\!\left| \sin\beta_\pi \right|) $$
\noindent which requires a single parameter $\gamma_{\scriptscriptstyle\rm F}$.
In contrast with the cosine models, the fan models create a depression of the isodensity surfaces above the solar poles.
But this depression does not necessarily appear in the isophotes as illustrated in Figure~\ref{fig:isodensity} using $\gamma_{\scriptscriptstyle\rm F}$\,=\,2.6. 
On the flip side, these models produce a pinch in the ``nose'' of the isophotes translating in a severe discontinuity in the PSZC which are not observed. 
These weaknesses led to the rejection of these fan models.
The modified fan model proposed by \cite{Lumme1985} which introduces an extra term of $(\sin\beta_\pi)^2$ in the exponential presents the same major defect and was consequently not considered.

The ellipsoid models are given by the analytic expression:
$$ G(\beta_\pi) = \left[1 + (\gamma_{\scriptscriptstyle\rm E} \ \sin\beta_\pi)^2 \right]^{\nu_{\scriptscriptstyle\rm E}} $$
\noindent where $\gamma_{\scriptscriptstyle\rm E}$ and $\nu_{\scriptscriptstyle\rm E}$ are two parameters.
We found that $\gamma_{\scriptscriptstyle\rm E}$\,=\,4.5 and $\nu_{\scriptscriptstyle\rm E}$\,=\,-0.65 led to isophotes quite close to the observed ones in the C3 \fovnosp, in particular they render well the rounded shape of the ``nose'' of the isophotes (Figure~\ref{fig:isodensity}).
Consequently, we concluded that the ellipsoid models offer by far the most promising representation of the spatial density of interplanetary dust.
However, they must be improved to allow the shape of the isophotes to evolve with elongation, in particular their progressive circularization with decreasing elongations.
This probably can be achieved by allowing the parameters to vary with either $r$ or $\beta_\pi$ or both.

As a final remark, we point out the serious disagreement between our conclusion and the nearly all-sky (elongation $>$20\,\deg) empirical model of the zodiacal brightness proposed by \cite{Buffington2016} based on Solar Mass Ejection Imager data. 
They found that their model most resemble the modified fan model of \cite{Lumme1985} and indeed their Figure~2, although limited to a quarter of the sky, indicates that the isophotes have pinched ``noses'', thus exhibiting a discontinuity along the longitude axis.
This is definitively contradicted by the images presented in Figure~\ref{fig:F_examples} -- one of which extends to an elongation of 50\deg\ -- and by numerous ground-based photographs of the zodiacal light. 
Out to the two other maps presented by \cite{Buffington2016} in their Figure~1 based on data from \cite{Leinert1998} and \cite{Kwon2004}, only the latter one has isophotes with a rounded ``nose'', although they exhibit a suspect distortion at mid-latitudes.

\subsection{Stability of the Inner Zodiacal Cloud}
\label{D:stability}
In Section~\ref{sec:stability}, we reviewed past evidences of the stability of the zodiacal cloud based on reliable space observations, prominently \textit{Helios} and SMEI, leading to upper limits on secular variations of $\pm$\,2\,\% and 0.3\,\%, respectively.
However, we pointed out that the measurements were performed in small patches of the sky, typically a few deg$^2$ and over limited time spans, typically eight years.
The LASCO observations offer for the first time a global view of the temporal evolution of the F-corona over a record time interval of 25 years.
We found that, in the C3 \fovnosp, the global radiance experienced a continuous, quasi monotonic increase of 11.5\,\%.
The situation is more complex in the C2 \fovnosp: atop the same type of continuous increase, but limited to 5\,\% over 25 years, a steep increase occurred in 2006, reached a maximum of 5.6\,\% with respect to the baseline in 2010\,--\,2011, to be followed by a decrease so that the evolution returned to the baseline in 2016.

We may have a relevant theoretical explanation of the secular increase of the radiance, at least that observed in the C3 \fovnosp.
In their review of interplanetary dust, \cite{Leinert1990} pointed out that the lifetime of meteoroids with mass $\geq\,10^{-8}$ kg (a radius of approximately 100\,$\mu$m) is dominated by collisions, the destroyed mass re-appearing in the form of small fragments, a scenario confirmed by the zodiacal cloud model of \cite{Nesvorny2010}.
\cite{Leinert1990} calculated that collisions are adding 9 $\times$ 10$^{-32}$ kg m$^3$ s$^{-1}$ of dust particles with radii $<$ 100 $\mu$m at the expense of large meteoroids whereas only 4 $\times$ 10$^{-33}$ kg m$^3$ s$^{-1}$ are needed to compensate the Poynting--Robertson losses. 
They figured out that the spatial density of IDP would consequently increase by 10\,\% in 3000 years at 1\,AU, and in only 30 years at 0.1\,AU.
This is consistent with the elaborate treatment of the collisional evolution of the inner zodiacal cloud by \cite{Szalay2021} who found that the majority of collisional erosion occurs in the range 10 to 20\Rsun. 
The above heliocentric distances lie within the \fov of C3 and it is remarkable that the increase of 10\,\% in 30 years given by \cite{Leinert1990} is comparable to our result of 11.5\,\% in 25 years, granting that the collisional model depends upon many assumptions. 
LASCO may well have found direct evidence of the role of collisions in the dynamical evolution of the zodiacal cloud by observing its relevant inner part and over a long enough time interval.
Conversely, observations of the outer part and on shorter time intervals are not sufficient to detect a temporal evolution.

The case of the surge observed in the C2 \fov between 2006 and 2016 (and also in the innermost part of the C3 \fovnosp) is not amenable to a simple explanation and it is not clear which process or processes can be invoked to explain such a sudden event.

\subsection{Small Scale Structures}
\label{D:structures}
The zodiacal cloud is known to be pervaded by local structures which manifest themselves as spatial features, prominently in the infrared but also in the visible, superimposed on the uniform zodiacal light.
Well-established examples are asteroidal dust bands, cometary dust trails, and rings associated with Earth, Venus, and Mercury.
In addition, meteoroid streams including $\beta$-streams produced by collisions between meteoroid streams and zodiacal dust are detected via their impacts \citep{Szalay2021}.
Much more disputed are several features whose presence in the F-corona has been occasionally reported, notably from near-infrared and infrared observations, and connected to the hypothetical existence of circum-solar ring(s), the most popular one being located at approximately 4\Rsun.
The ups and downs of this saga and related theoretical analysis have given birth to a wealth of articles and an extensive synthesis can be found in \cite{Kimura1998}. 
Their Figure~3 reveals that, curiously, these features were no longer seen after 1985 when infrared cameras started to be used; then, interest in this question vanished.
The important point is when ``detected'' usually at solar eclipses, these features manifest themselves as sharp and bright peaks whose radiance exceeds the underlying F-corona by factors up to two.  
Needless to say that i) such huge enhancements should have been seen in images of the corona taken by professional and amateur astronomers present at those eclipses and ii) they require a substantial event to provide the required amount of dust, event which can hardly have gone unnoticed. 
Twenty five years of LASCO observations demonstrate the total absence of such localized density enhancements, a conclusion also supported by six years of SECCHI/HI-1 observations \citep{Stenborg2018}.

It has been conjectured that dust deposited by sun-grazing comets could explain the occasional appearance of brightness enhancements in the F-corona (\eg \cite{Hodapp1992}; \cite{Macqueen1994}).
Strong arguments against this scenario were presented by \cite{Kimura1998}.
We wish to emphasize the very short time-scale of these events illustrated by the thousands of sun-grazing comets discovered by LASCO as the released dust as well as their usually tiny (typically a few meters) nucleus completely disappear in a few days with no trace left behind.
This is even the case of spectacular comets such as C/2011 W3 Lovejoy \citep{Raymond2018} and C/2012 S1 ISON \citep{Druckmuller2014} -- both having a nucleus size of a few hundred meters -- for which however the dissipation took a little longer, typically a few weeks.
It is therefore difficult, not to say impossible, to invoke the comet scenario to explain any permanent features such as asymmetries in the F-corona. 

In summary, we fully concur with the past assertion by \cite{Mann2004} that we highlighted in our introduction, namely ``that under present conditions no prominent dust ring exists near the Sun''

\subsection{Dust-free Zone}
\label{sec:structures}
We complete our discussion by addressing the question of the dust-free zone (hereafter abbreviated to DFZ) and start with a historical note.
It is repeatedly stated in the relevant literature (\eg \cite{Kimura1998}; \cite{Mann2004}; \cite{Howard2019}; \cite{Stenborg2021}; \cite{Stenborg2022}) that \cite{Russell1929} predicted a dust-free zone around the Sun.
In reality, the work of \cite{Russell1929} was motivated by the question of whether wide, shallow, and diffuse absorption bands observed in stellar spectra could be produced by meteorites falling into or passing  near a star. 
He was therefore concerned, not by the disappearance of dust particles, but by the possibility of a solid body large enough ``a foot or more in diameter'' to survive until colliding with the Sun so as to sublimate enough gas down to the photosphere.
To give the eager reader the answer to this burning question, \cite{Russell1929} concluded that, in the case of the Sun, ``meteoritic matter near the Sun cannot exert enough effective absorption in the spectrum to produce the equivalent of a single narrow Fraunhofer line''.
A more appropriate and, as a matter of fact, older reference is the work of \cite{Anderson1924} who tackled the question of the survival of carbon particles near the Sun and concluded that they cannot -- and other materials as well -- exist within 1\Rsun\ of the photosphere (2\Rsun\ from the center of the Sun).  

Following the work of \cite{Over1958} who studied the temperature and sublimation of quartz particles near the Sun, more realistic materials were considered (\eg \cite{Lamy1974a}; \cite{Mukai1974}; \cite{Kimura2002}; \cite{Shestakova2018}).
They all converge to a DFZ whose radius depends upon the composition of the dust, thus ranging from approximately 2.5 to 25\Rsun.
For a given material, the sublimation is very fast once the dust particle has reached a critical temperature and this implies a well-defined DFZ.
In reality, this is not the case since IDPs are composed of various minerals and their temperature further depends upon their size, shape, porosity and orbital evolution: as their size shrinks, the ratio of the gravitational attraction over the radiation pressure force changes causing oscillations of their orbit and possibly escape as $\beta$-meteoroids (\eg \cite{Shestakova2018}).
However, this orbital evolution prominently applies to sub-micronic IDPs whose contribution to the radiance along a \los is probably very small.
The bottom line is that there most likely exists a fuzzy region extending over several solar radii where the spatial density of dust decreases progressively with decreasing heliocentric distance up to the point where the most refractory compounds disappear, thus defining the DFZ.  

As noted in the introduction, \cite{VdH1947} already realized that the DFZ is basically unconstrained by observations at 1\,AU.
It was not even detected at 0.3\,AU as the \textit{Helios} observations only allowed to establish that it must reside within 0.09 AU, that is $\approx$\,19\Rsun\ \citep{Leinert1978}.
It was only during the fourth and fifth encounters of the \textit{Parker Solar Probe} at a perihelion distance of 28\Rsun\ that \cite{Stenborg2021} unambiguously observed a decrease of the radial gradient of the brightness profile along the axis of symmetry, starting at $\approx$\,19\Rsun\ down to the shortest elongation of $\approx$\,7.65\Rsun\ allowed by the WISPR \fovnosp.
They interpreted this ``dust depletion zone'' using a simple model where the spatial density of dust stays constant between 19 and 10\Rsun\ then decreases exponentially to zero at 3\Rsun.
We note that the break of the density profile at 19\Rsun\ precisely corresponds to the upper limit given by \cite{Leinert1978}.
Forthcoming PSP observations at closer heliocentric distances down to ultimately 9.86\Rsun\ will undoubtedly allow a much more detailed characterization of the depletion and dust-free zones.

\section{Summary}
\label{sec:summary}
The review of space observations of the F-corona, the extensive analysis of 25 years of LASCO C2+C3 observations, and the connection to the zodiacal light led us to the following main results. 

\begin{enumerate}
	\item 
LASCO was able to detect the variation of the radiance of the F-corona with the heliocentric distance $d$ of SoHO despite its very small excursion (typically 0.973 to
1.009\,AU).
The resulting power laws have exponents $\nu_{\scriptscriptstyle\rm ZL}$ equals to 2.22\,$\pm$\,0.04 and 2.45\,$\pm$\,0.13 in the C2 and C3 \fovs, respectively. 
The \textit{Helios} result of 2.35 valid at elongations $\epsilon\leq$\,50\deg\ was deduced from its scanning photometers mounted at fixed angles of $\pm$16\deg and $\pm$31\deg to the probes orbital plane -- hence reaching a minimum elongation of 16\deg -- from widely distant vantage points at 0.3 and 1\,AU.
Despite the very different observational conditions, the LASCO and \textit{Helios} results appear consistent, but the lower C2 value suggests an evolution of the spatial density of dust in the inner zodiacal cloud, possible linked to the presence of the depletion and the dust-free zones.
	\item
Considering 1997 as representative of the first few years of LASCO observations and the December node to ensure the north-south symmetry of the images, the radiance profiles along the major and minor axes of the ``elliptically'' shaped F-corona (assimilated to the equatorial and polar directions, respectively) extracted from the C2 and C3 images and scaled to 1\,AU are in quasi perfect agreement, only requiring two minor corrections.
The opposite profiles, east-west and north-south, were further found in agreement thus confirming the symmetry of the F-corona when observed at the nodes. 
	\item
The above profiles were further found in excellent agreement with the Koutchmy--Lamy model \citep{KoutchmyLamy1985} with however one significant difference: the LASCO equatorial profile exhibits a shoulder starting at $\approx$\,10\Rsun\ so that it exceeds the K--L profile by $\approx$17\,\% while they remain parallel beyond $\approx$\,12\Rsun. 
The LASCO equatorial and polar profiles connect extremely well to the corresponding standard profiles of the zodiacal light. 
In the elongation range where these profiles are characterized by constant power exponents, that is typically 5\deg\, to 50\deg, we found that the exponents are robustly constrained to 2.30 $\leq$ $\nu_{\scriptscriptstyle\rm P}$ $\leq$ 2.33 for the equatorial and 2.52 $\leq$ $\nu_{\scriptscriptstyle\rm P}$ $\leq$ 2.55 for the polar profiles.
	\item
The above shoulder is best interpreted as a decrease of the radiance within an elongation of $\approx$\,10\Rsun\ and affecting the F-corona up to mid-latitudes.
We conjectured that it may be explained by the disappearance of organic material by thermal dissociation or degradation that takes place within $\approx$\,0.3\,AU.
On the one hand, this scenario is plausible since i) it is established that the bulk of IDPs are of cometary origin with consequently large fractions of organic constituents, and ii) there exists other evidences of this process.
On the other hand, it faces the difficulty of the absence of the radiance deficit at high latitudes. 
	\item
The radiance profiles at the nodes of December 2010 and 2011 exhibit similar properties except for a systematic enhanced radiance of 8\,\% with respect to those recorded at the node of December 1997.
This is part of a general temporal increase of the radiance of the F-corona highlighted by the 25 years of continuous observations by LASCO whereas previous, much shorter records concluded on its stability.	
This general trend is best illustrated by the quasi monotonic increase at a rate of 0.46\,\% per year of the integrated radiance in the C3 \fovnosp.
The evolution of that of C2 is more complex: atop a more modest general increase at a rate of 0.2\,\% per year, a sharp increase occurred in 2006, reached a maximum  in 2010\,--\,2011, and returned to the baseline evolution in 2016.
LASCO may well have confirmed the predictions of theoretical models of the evolution of the inner dust cloud showing that collisions are adding more mass of dust particles with radii $<$\,100\,$\mu$m (at the expense of large meteoroids) that can be removed by the Poynting--Robertson effect.
It is quite remarkable that the increase rate of 0.33\,\% per year at 0.1\,AU calculated by \cite{Leinert1990} is comparable to the rate of 0.46\,\% per year measured in the C3 \fovnosp.
	\item
The question of the color of the F-corona is not yet clearly settled.
Whereas it is agreed that both F-corona and zodiacal light have colors redder than the Sun, the reddening being more pronounced for the F-corona, the available data do not allow a well-characterized variation of the color index with wavelength. 
We found two possible power laws for this variation with exponents of 1.07 and 0.63 with, in the latter case, a steepening beyond approximately 650\,nm.
This however results in a small correction (10 to 15\,\%) if one wants to convert the LASCO data at 585\,nm to a reference wavelength of 500\,nm.	
	\item
Based on the images obtained at the node of December 1997 and introducing minor corrections, we built a composite of the C2 and C3 images that led to the LASCO reference map of the F-corona from 2 to 30\Rsun\ complemented by the LASCO extended map from 1 to 6\Rsun.
This at last fills the missing inner region in the all-sky standard tables of the radiance of the zodiacal light.
Both maps are valid for an observer at a heliocentric distance of 1\,AU in the plane of symmetry of the zodiacal cloud and in the spectral range 450--600\,nm.
We noted that ``The 1997 reference of diffuse night sky brightness'' of \cite{Leinert1998} gives an incorrect model of the F-corona.
	\item
We set an upper limit of 0.03\,R$_\odot$ for the offset between the center of the Sun and that of the F-corona based on C2-Fcor images and we suspected that it prominently results from the inherent difficulties of correctly centering those images. 
The C3 images support the absence of offset to an elongation of 30\Rsun, thus excluding the large offset of of 0.4 to 0.5\Rsun\ reported by \cite{Stenborg2017} and \cite{Stenborg2018}.
	\item
The fattening index of the F-corona follows a linear increase with the logarithm of elongation that connects very well to that of the zodiacal light.
The shoulder present in the LASCO equatorial profile produces a slight hump above the variation, centered at approximately 4\deg. 
The index reaches zero (\ie circularity of the isophotes) at an elongation of 0.5\deg\ $\pm$ 0.01\deg\ (1.9\Rsun).
Beyond this limit, the shape of the isophotes is best described by super-ellipses	as proposed by \cite{Stenborg2018} with an exponent linked to the flattening index.
	\item
Among the different classical models of the spatial density of interplanetary dust, ellipsoid, fan, and cosine, only the first one leads to isophotes whose shape correctly matches that of the observed isophotes. 
A future task would consist in allowing the two parameters involved in the ellipsoid model to vary with heliocentric distance in order to account for the increasing ellipticity of the isophotes with elongation although this may turn out to be insufficient implying the need for more complex models.	
	\item
The plane of symmetry of the zodiacal cloud is known to be warped so that for simplicity, it may be viewed as a set of concentric planar annuli whose inclination	and possibly longitude of the ascending node vary with heliocentric distance.
Our analysis of the LASCO data and, to some extent those of SECCHI/HI-1, demonstrate for the first time the early conjecture of a deviation from the invariable plane towards the planes of the inner planets and of the solar equator. 
They do differ on the longitude of the ascending node as the LASCO data led to a constant value of 87.6\deg\ equal to that of \textit{Helios} whereas \cite{Stenborg2017} found a rapid decrease with decreasing elongation.
Presently, there are unfortunately very limited studies of the orbital evolution of the inner dust cloud that could contribute to the interpretation of these results.
	\item
During its 25 years of observation of the F-corona, LASCO did not detect any small scale structures such as putative rings occasionally reported during solar eclipses, in particular that located at approximately 4\Rsun. 
Dust deposited by sun-grazing comets disappear in a matter of a few days and at most, a few weeks for the brightest ones.
	\item
The dust-free zone remains difficult to characterize due to the complexity and the variety of the processes at work on dust particles very near the Sun.
There is probably a fuzzy region extending over several solar radii where the spatial density of dust progressively decreases with decreasing heliocentric distance up to a distance where the most refractory materials disappear. 
Its outer border was constrained to $\approx$\,19\Rsun\ by the \textit{Helios} observations \citep{Leinert1978} a value confirmed by \cite{Stenborg2021} on the basis of WISPR observations during the fourth and fifth encounters of the \textit{Parker Solar Probe} at a perihelion distance of 28\Rsun.
They did found a decrease of the radial gradient of the brightness profile along the axis of symmetry, starting at $\approx$\,19\Rsun\ down to the shortest elongation of $\approx$\,7.65\Rsun\ allowed by the WISPR \fovnosp.
\end{enumerate}

Our in-depth review of past results complemented by a detailed presentation of the LASCO observation benefiting from its remarkable photometric stability and unprecedented time coverage (25 years) led to an up-to-date characterization of the F-corona and unveiled new aspects that should trigger new theoretical investigations.
It further provides the framework for the analysis and interpretation of the new views from different vantage points in both heliocentric distance and inclination offered by PSP and SoLO and by other forthcoming space missions as well.

\bigskip

\maketitle
\noindent {\bf Note added in proof}\\
Upon completion of this review, \cite{Stenborg2022} published an extension of their previous analysis \citep{Stenborg2021} based on new WISPR observations of the dust density depletion near the Sun performed during the five most recent encounters (6 to 10) of the \textit{Parker Solar Probe}, the perihelion distance having then decreased from 20.35 to 13.28\Rsun.
Both analysis were limited to the brightness profile along the axis of symmetry and interpreted using a simple model where the spatial density of dust is allowed to progressively decrease with decreasing heliocentric distance to ultimately drop to zero, thus defining the border of the DFZ.
\cite{Stenborg2022} confirmed their earlier result of the onset of the ``dust depletion zone'' (DDZ) at 19\Rsun\ and were able to constrain its inner boundary (\ie the extent of the DDZ) to a most likely value of $\approx$\,5\Rsun. 
In addition, their recent measurements did not reveal any signature of a circumsolar ring or enhanced dust density.
Altogether, these new results fully support our analysis (Section~\ref{sec:structures}) and our conclusions (Section~\ref{sec:summary}).
The extent of the DDZ at $\approx$\,5\Rsun\ allows us to characterize the most refractory IDPs that survive down to this distance.
On the basis of the thermophysical and dynamical studies performed by \cite{Lamy1974a}, \cite{Lamy1974b}, and \cite{Shestakova2018}, a silicate with optical properties intermediate between obsidian and andesite and in fact, close to basalt would be relevant.
This implies that the most refractory IDPs that survive down to the above heliocentric distance are likely composed of moderately absorbing silicates with an extinction coefficient (the imaginary part of the complex index of refraction) at visible wavelengths of $\approx$\,0.0008 \citep{Pollack1973}.


\begin{acknowledgements}
We thank P. Rocher for providing the ephemeris data, O. Floyd for Figure~\ref{fig:los}, and V. Pag\'e for Figure~\ref{fig:isodensity}.
We very much appreciated the congratulations of the Reviewer on our work as well as valuable comments.
The LASCO-C2 project at the Laboratoire Atmosph\`eres, Milieux et Observations Spatiales is funded by the Centre National d'Etudes Spatiales (CNES).
LASCO was built by a consortium of the Naval Research Laboratory, USA, the Laboratoire d'Astrophysique de Marseille (formerly Laboratoire d'Astronomie Spatiale), France, the Max-Planck-Institut f\"ur Sonnensystemforschung (formerly Max Planck Institute f\"ur Aeronomie), Germany, and the School of Physics and Astronomy, University of Birmingham, UK.
SoHO is a project of international cooperation between ESA and NASA.
\end{acknowledgements}

\clearpage
\section*{Appendix A: Volume Scattering Function}
\label{AppendixVSF}
\begin{figure}[htpb!]
\begin{center}
\includegraphics[width=\textwidth]{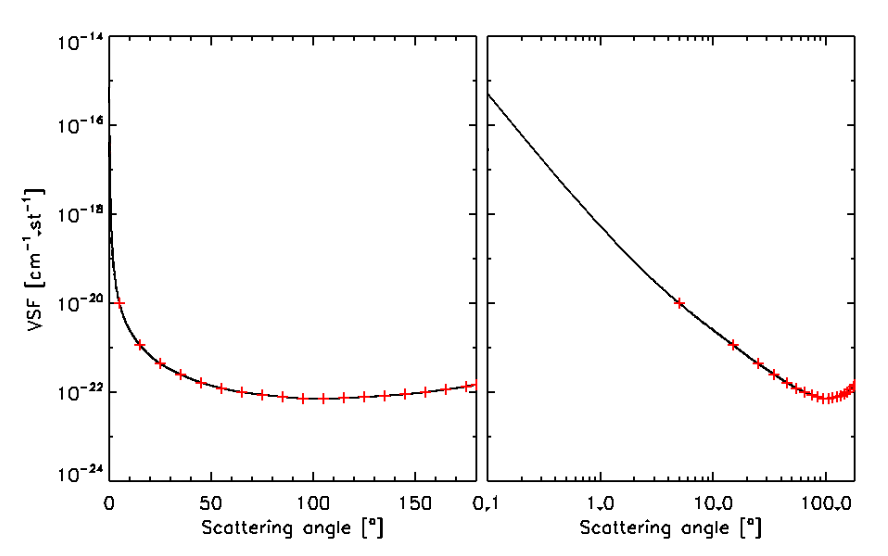}
\caption{Volume scattering function $\Psi_0$ at 1~AU displayed on lin-log (left panel) and log-log (right panel) scales. 
The red symbols represent the original data from \cite{Lamy1986}.}
\label{fig:vsf}
\end{center} 
\end{figure}

The inversion of photopolarimetric data of the F-corona and zodiacal light and the retrieval of the volume scattering function $\Psi_0$ at 1~AU performed by \cite{Lamy1986} was limited to a scattering angle of 5\deg. 
The present investigation and particularly the elaboration of Figure~\ref{fig:los} required an extension of this work to smaller angles. 
We present in Table~\ref{tab:vsf} the complete data set (old and new results) in tabular form which was lacking in the article of \cite{Lamy1986}.
In addition, two figures displays the variation of the VSF with scattering angle $\Psi_0(\chi)$ using lin-log and log-log scales, see Figure~\ref{fig:vsf}.

\begin{table}[!h]
\begin{center}
\caption{Tabulated values of the volume scattering function $\Psi_0$ (${\rm cm}^{-1}\,{\rm st}^{-1}$) at 1\,AU as a function of scattering angle $\chi$ (degree).}
\label{tab:vsf}
\begin{tabular}{c|c||c|c}
$\chi$ & $\Psi_0$  & $\chi$ & $\Psi_0$\\
\hline
0   	& 5.0E-16  	& 75  & 8.78E-23 \\ 
0.3		& 1.73E-17	& 85  & 7.85E-23 \\
0.5		& 3.84E-18	& 95  & 7.2E-23  \\
1			& 5.15E-19	& 105 & 7.15E-23 \\
3			& 3.09E-20	& 115 & 7.4E-23  \\
5   	& 1.0E-20		& 125 & 7.8E-23  \\
15  	& 1.15E-21 	& 135 & 8.3E-23  \\
25		& 4.38E-22 	& 145 & 9.0E-23  \\
35  	& 2.5E-22  	& 155 & 1.0E-22  \\
45  	& 1.65E-22 	& 165 & 1.15E-22 \\
55  	& 1.23E-22 	& 175 & 1.34E-22 \\
65  	& 1.0E-22		& 180 & 1.5E-22
\end{tabular}
\end{center}
\end{table}

We remind the interested readers that using this VSF $\Psi_0$ for the calculation of the radiance of the F-corona/zodiacal light at any location and in any direction in interplanetary space requires its scaling according to the $d^{-0.3}$ power law, $d$ being the heliocentric distance of the observer. 
This is inherent to the formalism developed by \cite{Lamy1986} that led to a single VSF compatible with all observational data.

\clearpage
\section*{Appendix B: Determination of the C3 Secondary Stray Light Ramp}
\label{AppendixRamp}
The C3 images suffer from a non-radially symmetric pattern of stray light, so-called diagonal ramp, and \cite{Morrill2006} presented a solution for its determination in the case of routine unpolarized images of 1024$\times$1024 pixels obtained with the broadband ``clear'' filter.
Later on, \cite{Lamy2021} realized that their method had a flaw and presented a revised procedure as well as an extension to the ``orange'' unpolarized and polarized images of 512$\times$512 pixels.

\begin{figure}[htpb!]
\begin{center}
\includegraphics[width=\textwidth]{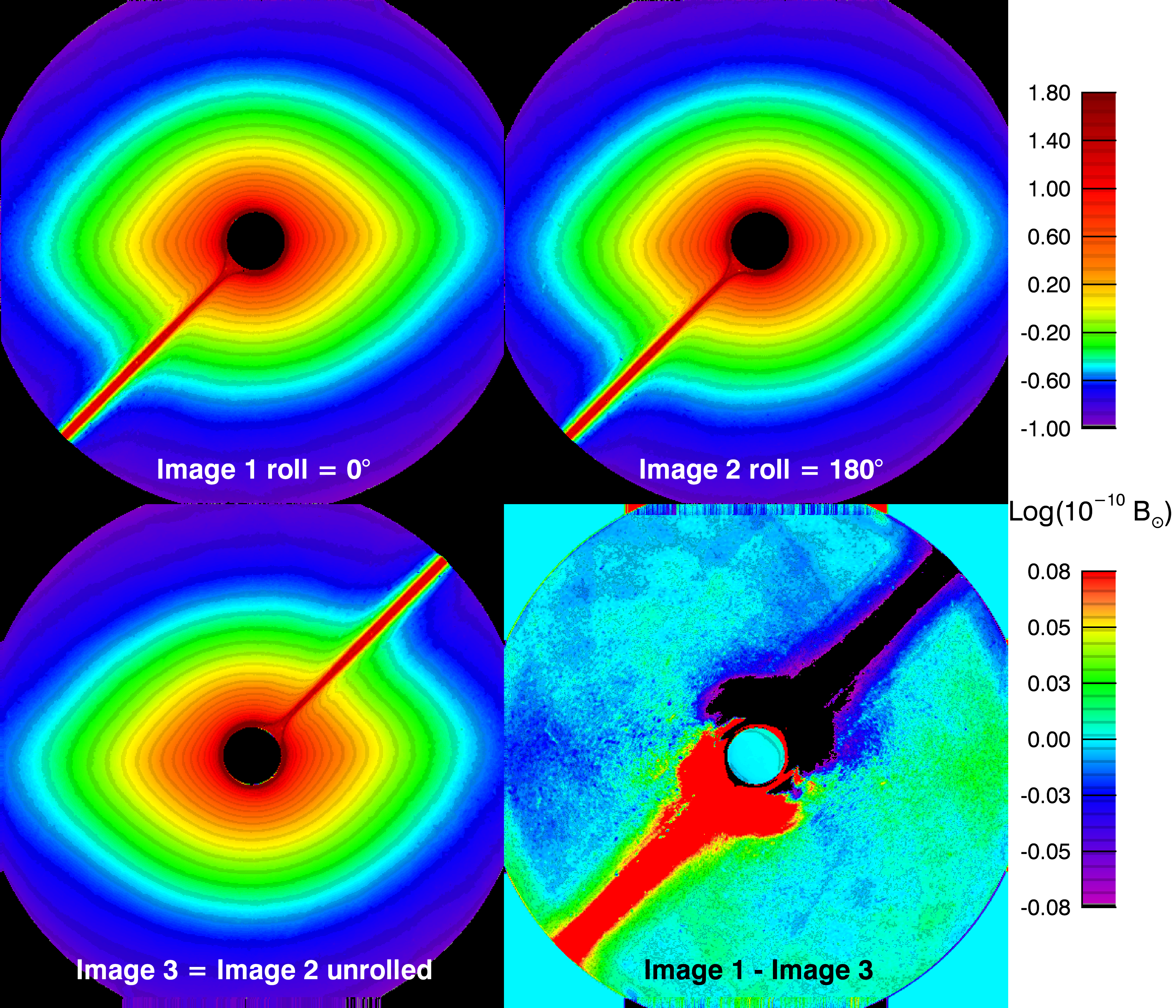}
\caption{These four images illustrate the presence of a secondary east--west stray light ramp.
The upper left image displays a C3 ``F+SL'' image taken at a SoHO roll angle of 0\deg, just before a roll operation.
The upper right image displays the image taken at a SoHO roll angle of 180\deg, just after the roll operation.
The lower left image displays the latter image counter-rotated by 180\deg.
The logarithm of the radiance of these three images are shown according to the upper color bar at right.
The lower right image displays the difference between the third and first images on a linear scale according to the lower color bar at right.}
\label{fig:test55408}
\end{center} 
\end{figure}

\begin{figure}[htpb!]
\begin{center}
\includegraphics[width=\textwidth]{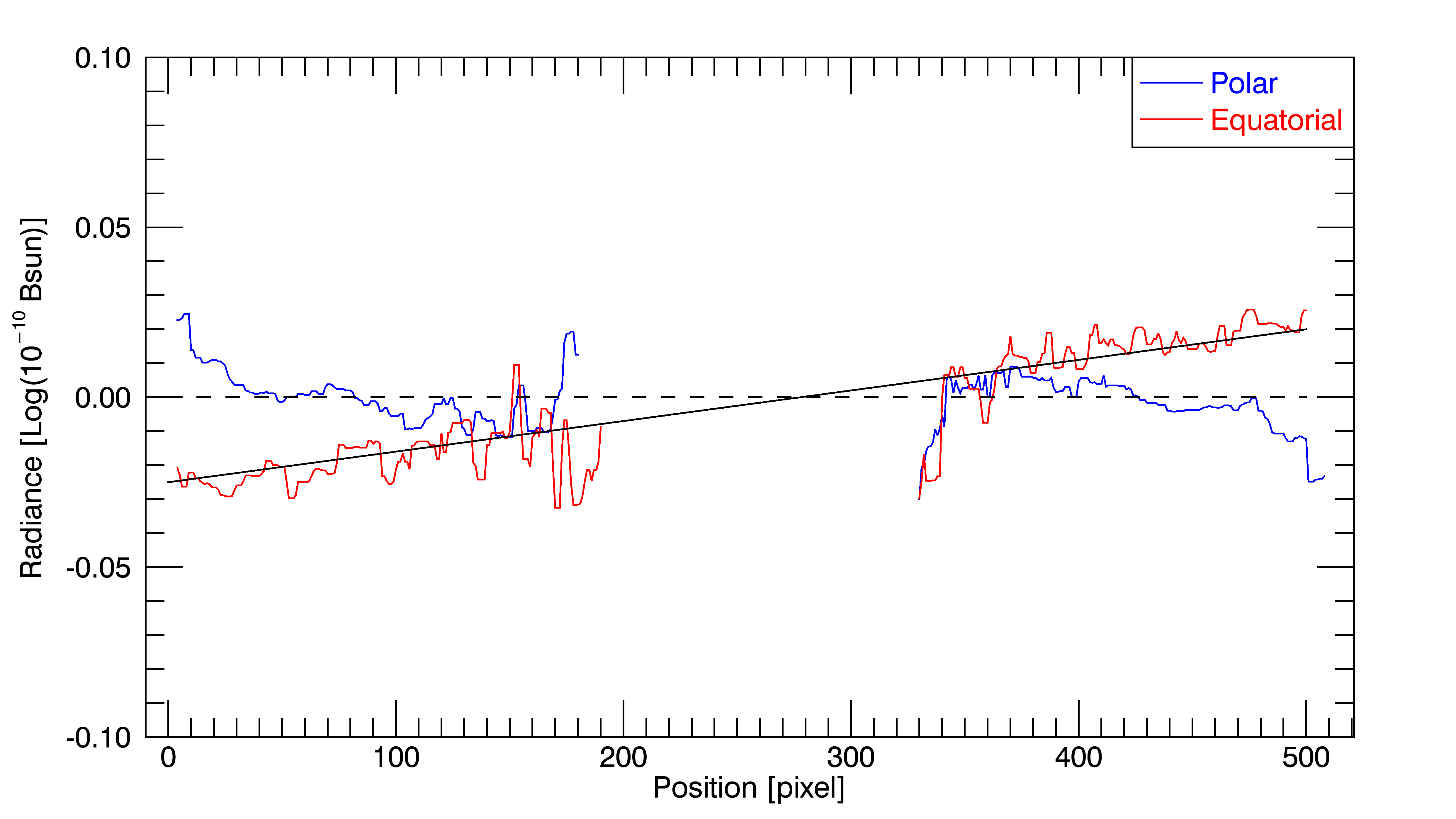}
\caption{Horizontal (red lines) and vertical (blue lines) diametral profiles extracted from the difference image shown in the lower right corner of Figure~\ref{fig:test55408}.
Whereas the vertical profile is nearly flat with a radiance of $\approx$\,0, the horizontal one exhibits a monotonic increase best represented by an inclined straight line (black line).}
\label{fig:diff55408}
\end{center} 
\end{figure}

In the course of the present work, we discovered an additional stray light component in the C3 ``F+SL'' images revealed by the comparison of images obtained just before and just after several rolls of SoHO of 180\deg.
The conspicuous artificial asymmetry suggested the presence of an east--west (EW) ramp as illustrated in Figures~\ref{fig:test55408} and \ref{fig:diff55408}.
We independently checked the routine, ``clear'' (unpolarized) images and found that this effect is absent so that it only affects the ``F+SL'' images which make use of the polarized images.
We conjectured that it may result from using the same \emph{diagonal} ramp for the three polarized images, an assumption imposed by the limited signal-over-noise (S/N) ratio of these images and consequently, the impossibility to determine three separate, reliable ramps.

\begin{figure}[htpb!]
\begin{center}
\includegraphics[width=\textwidth]{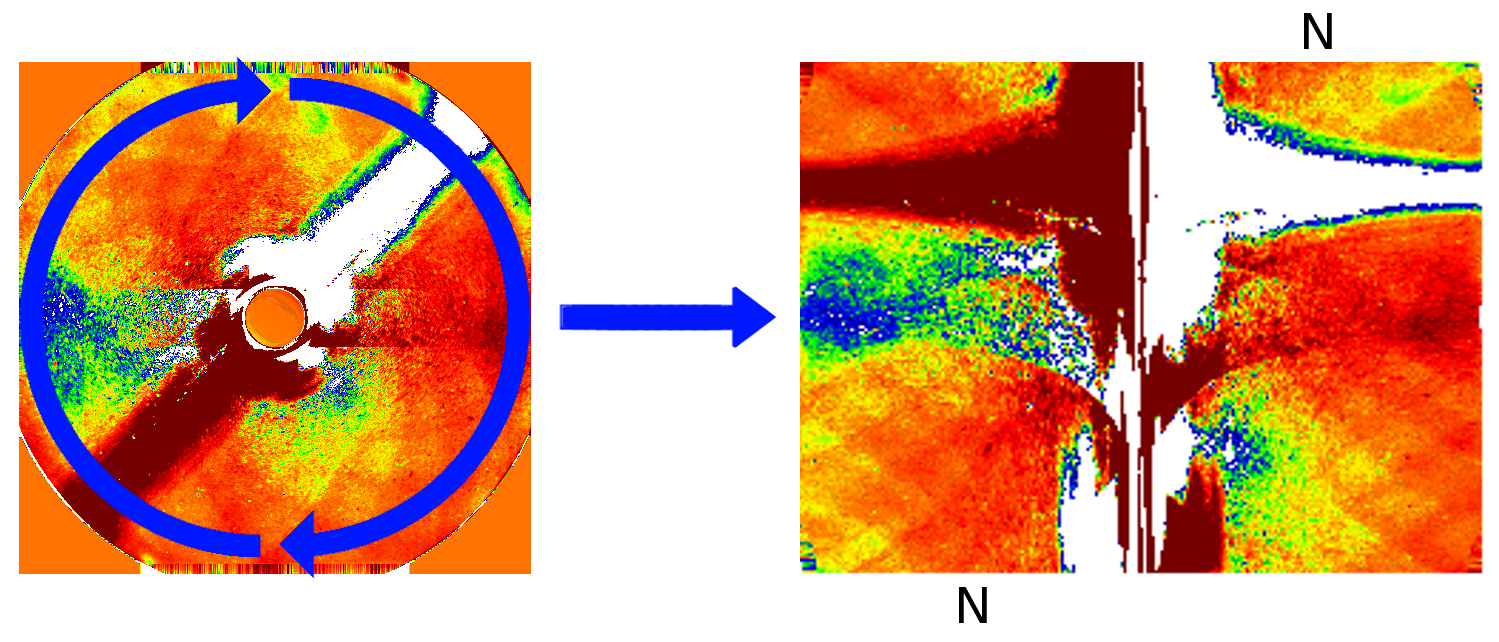}
\caption{Illustration of the two polar transformations applied to the difference image of Figure~\ref{fig:test55408} (lower right panel) to generate the concatenated image at right.
The pylon then appears as the aligned brown and white horizontal streaks whereas the EW ramp is the horizontal structure just below indicated by the blue arrow.}
\label{fig:ramp_polar}
\end{center} 
\end{figure}

\begin{figure}[htpb!]
\begin{center}
\includegraphics[width=\textwidth]{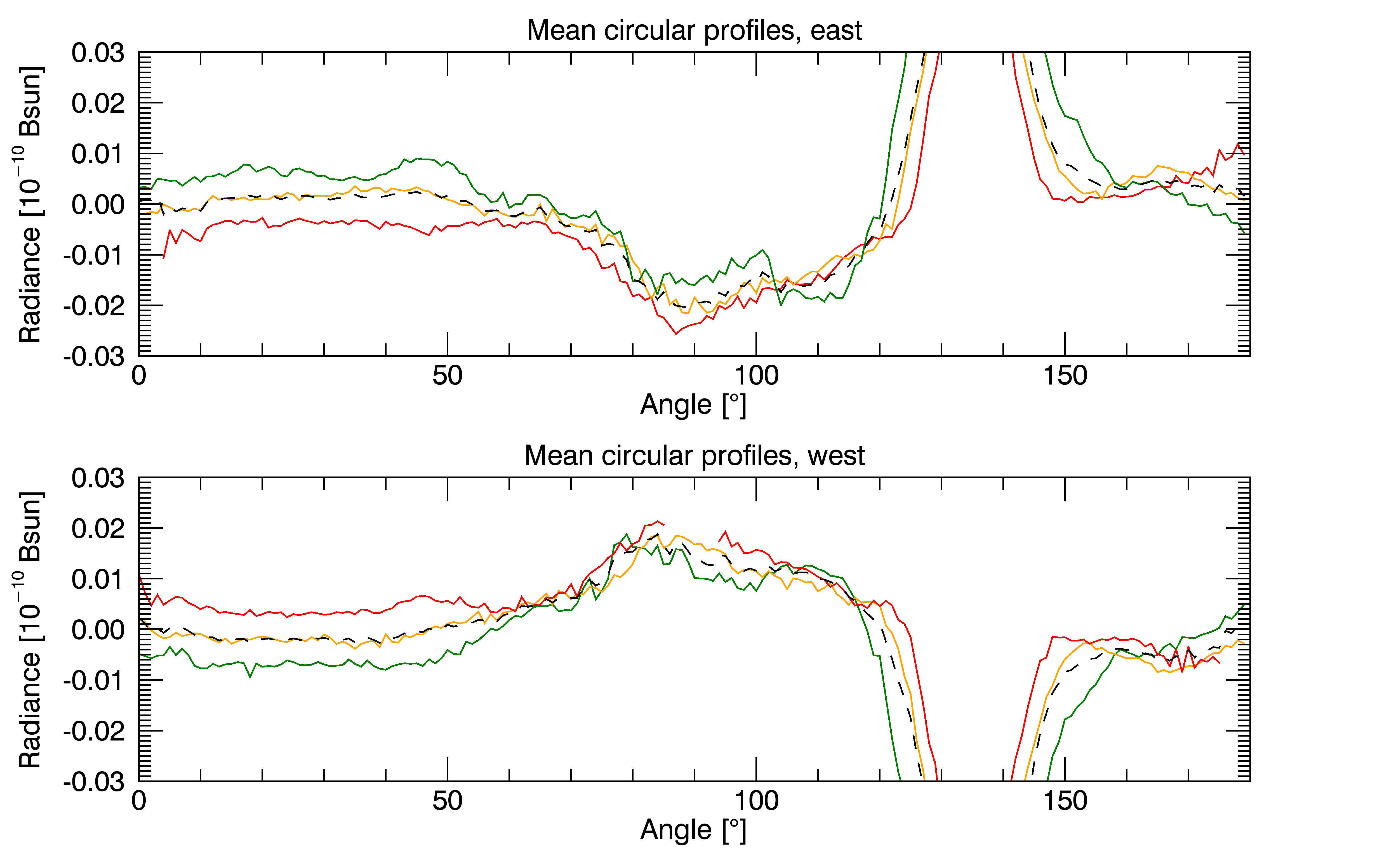}
\caption{Vertical profiles extracted from the resampled polar image displayed in Figure~\ref{fig:ramp_polar}.
For each hemisphere east and west, there are three adjacent profiles whose color sequence red--orange--green starts from the outer edge and proceeds inward.
The dashed black lines represent the average of each group of three profiles and are in excellent agreement with each intermediate (orange) profile.}
\label{fig:EWprofraw}
\end{center} 
\end{figure}

The construction of this EW ramp follows the method developed for the diagonal ramp \citep{Lamy2021} and starts from the difference images obtained before and after 180\deg\ rolls.
Two separate Cartesian to polar transformations were performed, from south to north for the eastern hemisphere and from north to south for the western hemisphere.
The two sub-images were then concatenated so that the EW ramp appears as a broad horizontal structure below the pylon as illustrated in Figure~\ref{fig:ramp_polar}.
The resulting images at the format of 240 pixels (along the x-axis corresponding to the radial direction) times 180 pixels (along the y-axis corresponding to the polar angle) were resampled to a format of 20$\times$180 pixels, thus averaging 12 original columns to improve the signal-over-noise ratio of the ramp.
Figure~\ref{fig:EWprofraw} displays the three outer profiles of each hemisphere as well as their averages.
They coincide with the intermediate (orange) profiles and were retained as baseline profiles to construct the ramp after applying several ``cleaning'' operations.

\begin{figure}[htpb!]
\begin{center}
\includegraphics[width=\textwidth]{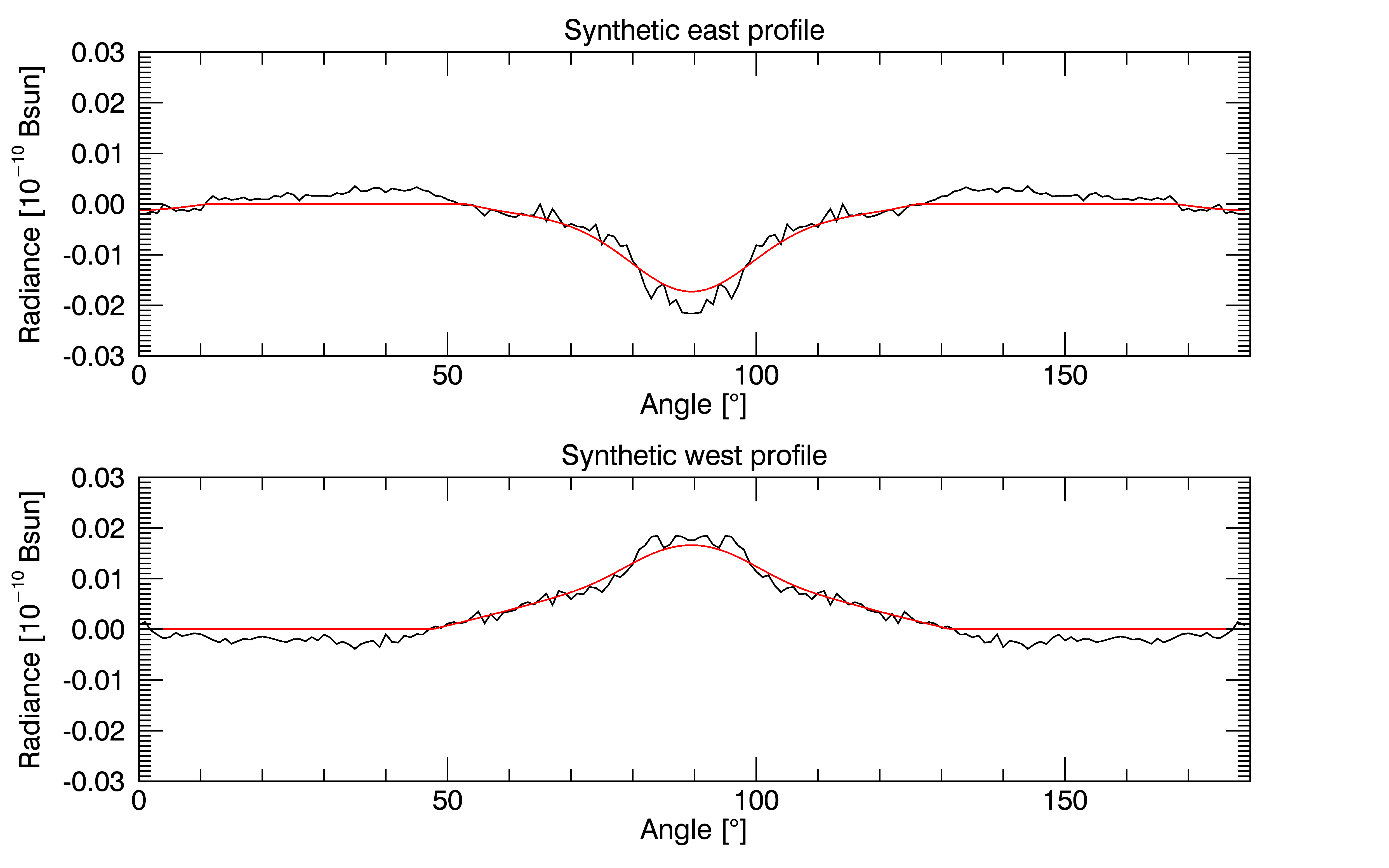}
\caption{Vertical profiles derived from the average profiles of Figure~\ref{fig:EWprofraw}.
The black ones result from the symmetrization applied to remove the contamination by the pylon.
The red ones are the outputs of subsequent operations, smoothing and thresholding, as described in the text.}
\label{fig:EWprofsmo}
\end{center} 
\end{figure}

\begin{figure}[htpb!]
\begin{center}
\includegraphics[width=\textwidth]{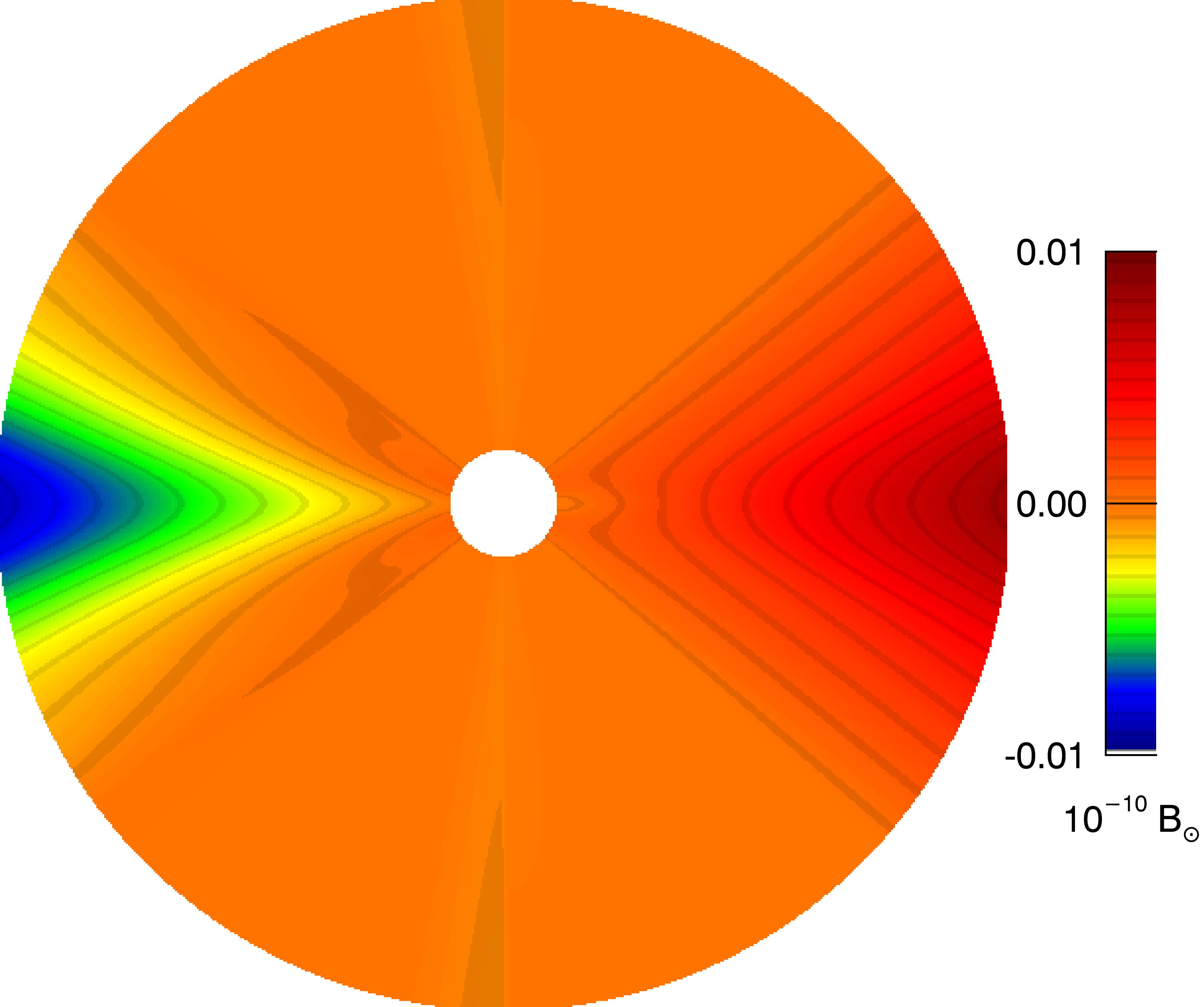}
\caption{Image of the EW ramp displayed on a linear scale given by the color bar.}
\label{fig:EWramp}
\end{center} 
\end{figure}

First, the sections [90\deg -- 180\deg] of these baseline profiles contaminated by the pylon were replaced by the symmetrized sections [0\deg -- 90\deg].
The resulting symmetric profiles were smoothed using a Gaussian filter and the two external branches extending to [0\deg -- 50\deg] and [130\deg -- 180\deg] were forced to zero since their very low levels are not significant (Figure~\ref{fig:EWprofsmo}).
The EW ramp was constructed line by line in the polar frame by applying a linear interpolation between the two (east and west) profiles. 
In practice, a frame of 240$\times$180 pixels was built and these profiles were inserted in the two extreme (first and last) columns of the frame.
Each line of the frame was then calculated by linear interpolation between the extreme values and a polar to Cartesian transformation was applied.
After a division by a factor of two since the difference produces twice the ramp, we obtained a baseline EW ramp whose radiance ranges over $\pm\,10^{-12}$\,\Bsun as displayed in Figure~\ref{fig:EWramp}.

Very much like the case of the diagonal ramp, the absolute scaling of the EW ramp must be performed independently, and this was achieved by a comparison with the calibrated routine unpolarized images which are not affected by this ramp as noted earlier.
These latter images do include some K-corona signal, but at such a negligible level in the outer part of the C3 \fov that it was not a problem at all.
Then a direct comparison with the ``F+SL'' images offered a robust criterion for the absolute scaling. 
This led to adding a constant level of 2.6\,$10^{-12}$\,\Bsun to the baseline ramp, thus ensuring its positivity and therefore its interpretation as an additional component of stray light likewise the diagonal ramp.
To appreciate the very low level of the EW ramp, its maximum value amounts to only 13\,\% of the radiance of the corona at 30\Rsun, the outer limit of the C3 \fov.

\clearpage
\section*{Appendix C: Fine Restoration of the C3 Images of the F-corona}
\label{AppendixResto}
The fine restoration of the C3 images of the F-corona aimed at eliminating the diagonal streak created by the pylon of the occulter as well as faint stray light remnants from the ``F+SL'' images resulting from the polarimetric analysis performed by \cite{Lamy2021}.
On these images, the pylon manifests itself as a very bright diagonal streak oriented along either the south--east (SoHO roll angle of 0\deg) or the north--west (SoHO roll angle of 180\deg) directions.
This compelled us to proceed in two stages, the first one consisting in removing the bulk of these streaks.
This was rather straightforward taking advantage of the quasi symmetry of the F-corona to replace the sector 40\deg\ wide centered on the pylon by the diagonally opposite sector. 

\begin{figure}[htpb!]
\begin{center}
\includegraphics[width=\textwidth]{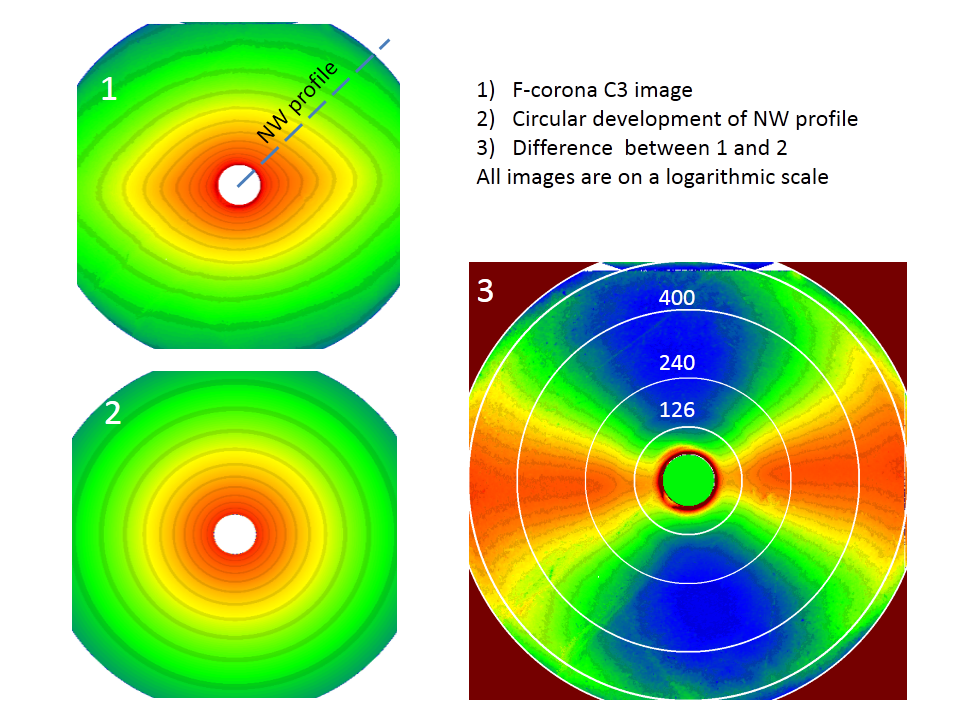}
\caption{Illustration of the construction of a difference image.
Upper left panel: initial F-corona image after removing the bulk of the pylon contribution; the reference NW profile indicated by the dashed line.
Lower left panel: reference axi-symetric image.
Right panel: difference image ``log(image) - log(reference)''.}
\label{fig:C3_Fcor_back}
\end{center} 
\end{figure}

\begin{figure}[htpb!]
\begin{center}
\includegraphics[width=\textwidth]{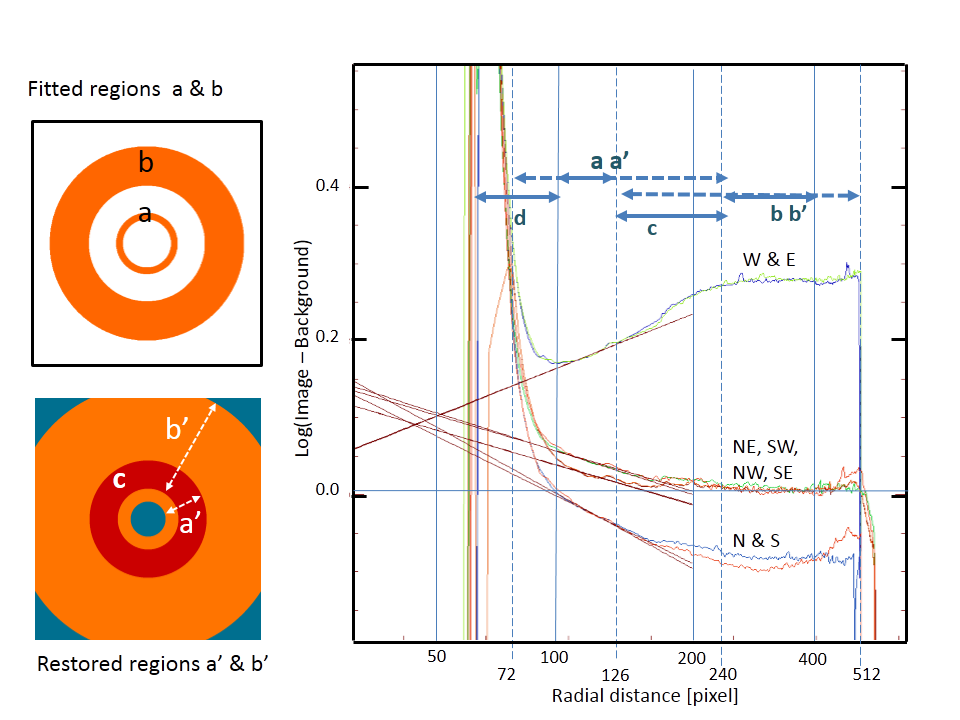}
\caption{Analysis of a difference image ``log(image) - log(reference)''.
The right panel displays the profiles along eight directions as well as the radial extent of the selected regions shown on the two images at left.}
\label{fig:C3_Fcor_prof}
\end{center} 
\end{figure}

The second stage was more complex and consequently, limited to a small  number of images, prominently those obtained at the nodes.
It is reminiscent of the method implemented by \cite{Llebaria2021} to generate the C2-Fcor images and required several steps.
The first one aimed at reducing the dynamical range of the images by taking the logarithm of the radiance and then subtracting a reference image so that the restoration was carried out on the difference images of moderate amplitude.
We analyzed eight radial profiles along the cardinal directions north (N), south (S), east(E), and west(W) and along the diagonal directions NE, SE, NW, and SW.
It turned out that the profiles along the latter directions were the most well-behaved and we finally selected the profiles opposite to the pylon (\eg NW when the pylon was SE), smoothed them, and constructed the reference images using a circular development (Figure~\ref{fig:C3_Fcor_back}).
We then analyzed the difference images ``log(image) - log(reference)'' and distinguished four annular regions characterized by different radial profiles as illustrated in Figure~\ref{fig:C3_Fcor_prof}.
Regions ``a'' from 100 to 126 pixels and ``b'' from 240 to 400 pixels have rather well defined radial variations, linear in the log-lin representation for the former and constant for the latter.
Region ``c'' in-between from 126 to 240 pixels was considered as a transition between regions ``a'' and ``b''.
Finally, the innermost region ``d'' from 56 to 100 pixels is affected by stray light from the diffraction fringe surrounding the occulter.

The second step consisted in constructing models of regions ``a'' and ``b'' of the difference images relying on circular profiles, in practice linear profiles extracted from the polar transformed of the difference images. 
Each quarter of the quasi sinusoidal profiles from a minimum to a maximum was fitted by a polynomial of degree eight, retaining only the even exponents and further imposing the continuity at the extrema (first derivatives equal to zero).
In the case of region ``b'', a single, average profile was constructed in view of the constancy of the radial profiles noted above.
In the case of region ``a'', the procedure was more complex and the polynomials were fitted to each individual belonging to the interval [100\,--\,126] pixels.
The radial variation of each coefficient of the polynomials was then smoothed by fitting low degree polynomials.
Altogether, these two operations performed a two-dimensional filtering of the difference image in region ``a''.

\begin{figure}[htpb!]
\begin{center}
\includegraphics[width=\textwidth]{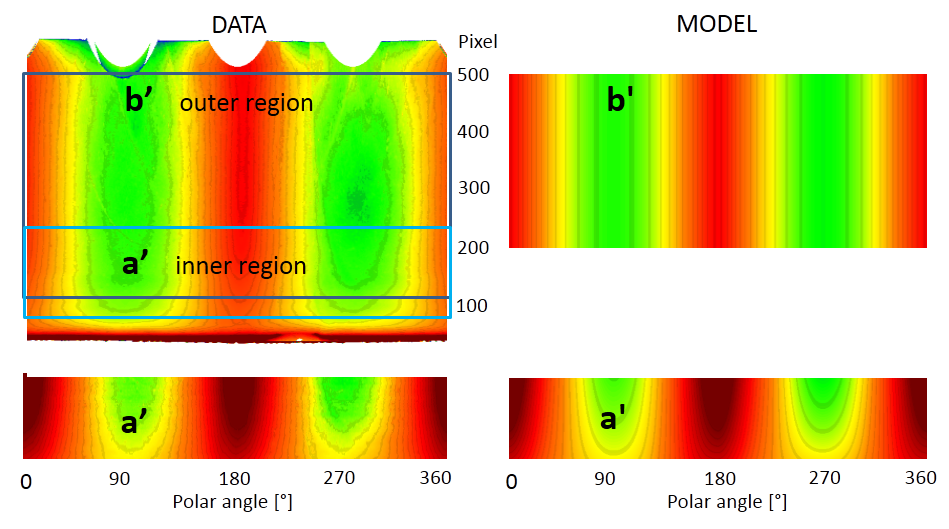}
\caption{Illustration of the restoration of a difference image ``log(image) - log(reference)'' in the regions ``a$^{\prime}$'' and ``b$^{\prime}$'' defined in Figure~\ref{fig:C3_Fcor_prof}.
The polar transformed images of the data are displayed in the left panels and those of the models, in the right panels.
Different color scales are used for the upper and lower panels.}
\label{fig:C3_polar_restit}
\end{center} 
\end{figure}

\begin{figure}[htpb!]
\begin{center}
\includegraphics[width=\textwidth]{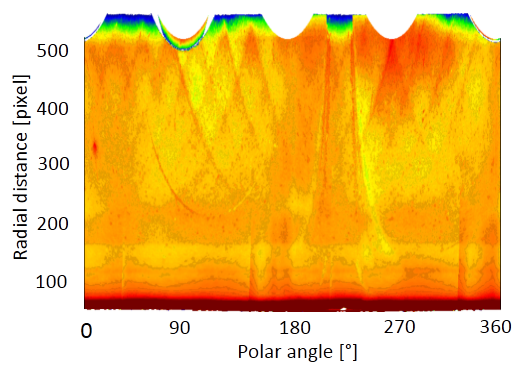}
\caption{Example of a model of the difference image ``log(image) - log(reference)'' in polar coordinates.}
\label{fig:C3_diff_model}
\end{center} 
\end{figure}

In a third step, the models of regions ``a'' and ``b'' were extrapolated to regions ``a$^{\prime}$'' and ``b$^{\prime}$'' as illustrated by Figure~\ref{fig:C3_polar_restit}.
In the transition region ``c'', we replaced the dual determinations by their average weighted according to the radial distance between the outer limit of region ``a'' and the inner limit of region ``b''.
Figure~\ref{fig:C3_diff_model} displays an example of the resulting models of the difference images.
Recombining with the respective reference images and switching back to Cartesian coordinates led to the fully restored C3-Fcor images.

\clearpage
\section*{Appendix D: Tabulated LASCO reference model of the F-corona}
\label{AppendixFdata}

Table~\ref{tab:Fmap} presents a set of numerical values of the radiance sampling the LASCO reference and extended maps of Figure~\ref{fig:Fmap} in longitude $\lambda-\lambda_{\odot}$ and latitude $\beta$. 
It is valid in the spectral range 450--600\,nm and the values can be transformed to \SZL using the formula: 1\,\SZL = 4.5 $\times$ $10^{-16}$\,\Bsunnosp.
On the basis of the applied corrections and the validation performed in Section~\ref{sec:photometry}, the uncertainty is estimated at 5\,\%.

\begin{table}[htpb!]
\caption{LASCO reference map of the F-corona: radiance in units of $10^{-10}$\,\Bsun for an observer at 1\,AU in the plane of symmetry of the zodiacal cloud as a function of longitude $\lambda-\lambda_{\odot}$ and latitude $\beta$ both expressed in degree.} 
\begin{tabular}{cccccccc}
\noalign{\smallskip}\hline\noalign{\smallskip}
\backslashbox[4mm]{$\lambda-\lambda_{\odot} $}{$\beta$}& 0.3 & 0.4 & 0.5 & 0.6 & 0.7 & 0.8 & 0.9  \\
\noalign{\smallskip}\hline\noalign{\smallskip}
     0.3 &   162.93 &    96.24 &    62.26 &    43.33 &    31.65 &    23.85 &    18.74 \\
     0.4 &    94.88 &    65.67 &    47.39 &    35.04 &    26.67 &    20.70 &    16.60 \\
     0.5 &    59.97 &    46.44 &    35.94 &    28.02 &    22.17 &    17.72 &    14.51 \\
     0.6 &    40.53 &    33.43 &    27.31 &    22.26 &    18.24 &    15.00 &    12.54 \\
     0.7 &    28.55 &    24.62 &    20.96 &    17.71 &    14.96 &    12.62 &    10.77 \\
     0.8 &    20.63 &    18.38 &    16.16 &    14.08 &    12.21 &    10.54 &     9.17 \\
     0.9 &    15.53 &    14.17 &    12.76 &    11.37 &    10.07 &     8.87 &     7.85 \\
\noalign{\smallskip}\hline\noalign{\smallskip}
\\
\\
\noalign{\smallskip}\hline\noalign{\smallskip}
\backslashbox[4mm]{$\lambda-\lambda_{\odot} $}{$\beta$}& 1. & 1.5 & 2.5 & 4. & 5. & 6.5 & 7.5  \\
\noalign{\smallskip}\hline\noalign{\smallskip}
     1.0 &    6.032 &    3.454 &    1.639 &    0.702 &    0.450 &    0.259 &    0.190 \\
     1.5 &    2.830 &    2.118 &    1.197 &    0.577 &    0.388 &    0.233 &    0.174 \\
     2.5 &    1.045 &    0.904 &    0.642 &    0.382 &    0.279 &    0.183 &    0.142 \\
     4.0 &    0.365 &    0.344 &    0.290 &    0.210 &    0.169 &    0.122 &        \\
     5.0 &    0.218 &    0.209 &    0.186 &    0.147 &    0.123 &        &        \\
     6.5 &    0.114 &    0.112 &    0.104 &    0.089 &        &        &        \\
     7.5 &    0.079 &    0.077 &    0.074 &        &        &        &        \\
\noalign{\smallskip}\hline\noalign{\smallskip}
\end{tabular}
\label{tab:Fmap}
\end{table}

\clearpage 
\bibliographystyle{aps-nameyear}      
\bibliography{Fcor-Prop-Biblio}                
\nocite{*}

\end{document}